\input harvmac.tex
\input epsf

% Macros plagiarized from P.G.

\def\figin{\epsfcheck\figin}\def\figins{\epsfcheck\figins}
\def\epsfcheck{\ifx\epsfbox\UnDeFiNeD
\message{(NO epsf.tex, FIGURES WILL BE IGNORED)}
\gdef\figin##1{\vskip2in}\gdef\figins##1{\hskip.5in}% blank space instead
\else\message{(FIGURES WILL BE INCLUDED)}%
\gdef\figin##1{##1}\gdef\figins##1{##1}\fi}
\def\DefWarn#1{}
\def\figinsert{\goodbreak\midinsert}
\def\ifig#1#2#3{\DefWarn#1\xdef#1{fig.~\the\figno}
\writedef{#1\leftbracket fig.\noexpand~\the\figno}%
\figinsert\figin{\centerline{#3}}\medskip\centerline{\vbox{\baselineskip12pt
\advance\hsize by -1truein\noindent\footnotefont{\bf Fig.~\the\figno:} #2}}
\bigskip\endinsert\global\advance\figno by1}

%\def\subsubsec#1{$\underline{\rm #1}$}

% Something to deal with sub-sub-sections

\def\unlockat{\catcode`\@=11}
\def\lockat{\catcode`\@=12}

\unlockat
% Something to deal with sub-sub-sections

\def\newsec#1{\global\advance\secno by1\message{(\the\secno. #1)}
\global\subsecno=0\global\subsubsecno=0\eqnres@t\noindent
{\bf\the\secno. #1}
\writetoca{{\secsym} {#1}}\par\nobreak\medskip\nobreak}
\global\newcount\subsecno \global\subsecno=0
\def\subsec#1{\global\advance\subsecno
by1\message{(\secsym\the\subsecno. #1)}
\ifnum\lastpenalty>9000\else\bigbreak\fi\global\subsubsecno=0
\noindent{\it\secsym\the\subsecno. #1}
\writetoca{\string\quad {\secsym\the\subsecno.} {#1}}
\par\nobreak\medskip\nobreak}
\global\newcount\subsubsecno \global\subsubsecno=0
\def\subsubsec#1{\global\advance\subsubsecno by1
\message{(\secsym\the\subsecno.\the\subsubsecno. #1)}
\ifnum\lastpenalty>9000\else\bigbreak\fi
\noindent\quad{\secsym\the\subsecno.\the\subsubsecno.}{#1}
\writetoca{\string\qquad{\secsym\the\subsecno.\the\subsubsecno.}{#1}}
\par\nobreak\medskip\nobreak}

\def\subsubseclab#1{\DefWarn#1\xdef
#1{\noexpand\hyperref{}{subsubsection}%
{\secsym\the\subsecno.\the\subsubsecno}%
{\secsym\the\subsecno.\the\subsubsecno}}%
\writedef{#1\leftbracket#1}\wrlabeL{#1=#1}}% Macros for boxes
\lockat

\def\IL{\relax{\rm I\kern-.18em L}}
\def\IH{\relax{\rm I\kern-.18em H}}
\def\IR{\relax{\rm I\kern-.18em R}}
\def\IC{\relax\hbox{$\inbar\kern-.3em{\rm C}$}}
\def\IZ{\relax\ifmmode\mathchoice
{\hbox{\cmss Z\kern-.4em Z}}{\hbox{\cmss Z\kern-.4em Z}}
{\lower.9pt\hbox{\cmsss Z\kern-.4em Z}}
{\lower1.2pt\hbox{\cmsss Z\kern-.4em Z}}\else{\cmss Z\kern-.4em
Z}\fi}
\def\CM {{\cal M}}
\def\CN {{\cal N}}
\def\CR {{\cal R}}
\def\CD {{\cal D}}
\def\CF {{\cal F}}
\def\CJ {{\cal J}}

\def\CL {{\cal L}}
\def\CV {{\cal V}}
\def\CO {{\cal O}}

\def\CE {{\cal E}}

\def\CH {{\cal H}}
\def\CC {{\cal C}}
\def\CB {{\cal B}}

\def\CA{{\cal A}}
\def\CK{{\cal K}}

%% MORE MACROS
\def\CM {{\cal M}}
\def\CN {{\cal N}}

\def\CO {{\cal O}}

\def\cmtld{\widetilde{\CM} }

\def\CE{{\cal E }}
\def\CV{{\cal V }}

\def\CX {{\cal X }}

\def\ch{{\rm ch}}

\def\CW{{\cal W }}

\def\Gr{{\rm Gr}}

\def\Km{{\rm Km}}

\font\manual=manfnt \def\dbend{\lower3.5pt\hbox{\manual\char127}}

\def\IZ{\relax\ifmmode\mathchoice
{\hbox{\cmss Z\kern-.4em Z}}{\hbox{\cmss Z\kern-.4em Z}}
{\lower.9pt\hbox{\cmsss Z\kern-.4em Z}}
{\lower1.2pt\hbox{\cmsss Z\kern-.4em Z}}\else{\cmss Z\kern-.4em
Z}\fi}
\def\half {{1\over 2}}
\def\sdtimes{\mathbin{\hbox{\hskip2pt\vrule height 4.1pt depth -.3pt
width
.25pt
\hskip-2pt$\times$}}}
\def\p{\partial}
\def\pb{\bar{\partial}}

\def\lsim{
{\ \lower-1.2pt
\vbox{\hbox{\rlap{$>$}\lower5pt\vbox{\hbox{$\sim$}}}}\ }
}
\def\grsim{
{\ \lower-1.2pt\vbox{\hbox{\rlap{$<$}\lower5pt\vbox{\hbox{$\sim$}}}}\ }
}

\def\CJ {{\cal J}}
\def\CM {{\cal M}}
\def\CN {{\cal N}}

\def\CO {{\cal O}}

\def\CE{{\cal E }}
\def\CV{{\cal V }}

\def\ch{{\rm ch}}

% more macros, alphabetically

\def\IZ{\relax\ifmmode\mathchoice
{\hbox{\cmss Z\kern-.4em Z}}{\hbox{\cmss Z\kern-.4em Z}}
{\lower.9pt\hbox{\cmsss Z\kern-.4em Z}}
{\lower1.2pt\hbox{\cmsss Z\kern-.4em Z}}\else{\cmss Z\kern-.4em
Z}\fi}
\def\IB{\relax{\rm I\kern-.18em B}}
\def\IC{{\relax\hbox{$\inbar\kern-.3em{\rm C}$}}}
\def\ID{\relax{\rm I\kern-.18em D}}
\def\IE{\relax{\rm I\kern-.18em E}}
\def\IF{\relax{\rm I\kern-.18em F}}
\def\IG{\relax\hbox{$\inbar\kern-.3em{\rm G}$}}
\def\IGa{\relax\hbox{${\rm I}\kern-.18em\Gamma$}}
\def\IH{\relax{\rm I\kern-.18em H}}
\def\II{\relax{\rm I\kern-.18em I}}
\def\IK{\relax{\rm I\kern-.18em K}}
\def\IP{\relax{\rm I\kern-.18em P}}

\def\IQ{\relax\hbox{$\inbar\kern-.3em{\rm Q}$}}
\def\IP{\relax{\rm I\kern-.18em P}}
\def\im{{\rm Im}}

\def\lieg{{\underline{\bf g}}}
\def\liet{{\underline{\bf t}}}

\def\inbar{\,\vrule height1.5ex width.4pt depth0pt}

\def\mod{{\rm mod}}
\def\p{\partial}
\def\pb{{\bar \p}}

\font\cmss=cmss10 \font\cmsss=cmss10 at 7pt
\def\IR{\relax{\rm I\kern-.18em R}}

\def\re{{\rm Re}}
\def\sdtimes{\mathbin{\hbox{\hskip2pt\vrule
height 4.1pt depth -.3pt width .25pt\hskip-2pt$\times$}}}
\def\Tr{\rm Tr}

%%% New number theory macros
\def\End{ {\rm End}}
\def\Im{{\rm Im}}
\def\Cof{{\rm Cof}}

% Macros for boxes

\def\boxit#1{\vbox{\hrule\hbox{\vrule\kern8pt
\vbox{\hbox{\kern8pt}\hbox{\vbox{#1}}\hbox{\kern8pt}}
\kern8pt\vrule}\hrule}}
\def\mathboxit#1{\vbox{\hrule\hbox{\vrule\kern8pt\vbox{\kern8pt
\hbox{$\displaystyle #1$}\kern8pt}\kern8pt\vrule}\hrule}}

%% ANOTHER SET OF MACROS

\def\lieg{{\underline{\bf g}}}

\def\liet{{\underline{\bf t}}}

\def\inbar{\,\vrule height1.5ex width.4pt depth0pt}
\def\ndt{\noindent}
\def\p{\partial}

\def\pb{{\bar \p}}

\font\cmss=cmss10 \font\cmsss=cmss10 at 7pt
\def\IR{\relax{\rm I\kern-.18em R}}

\def\rank{{\rm rank}}
\def\sdtimes{\mathbin{\hbox{\hskip2pt\vrule
height 4.1pt depth -.3pt width .25pt\hskip-2pt$\times$}}}
\def\Tr{\rm Tr}

%REFERENCES
%

%%%%%%
%%%%%%
%%%%%% REFS FOR ATTRACTORS
%%%%%%
%%%%%%

\lref\andermoore{G. Anderson and G. Moore,
``Rationality in conformal field theory,''
Comm. Math. Phys. {\bf 117}(1988)441}

\lref\adf{L. Andrianopoli, R. D'Auria, and S. Ferrara,
``U-duality and central charges in various dimensions
revisited,'' hep-th/9612105; ``Flat symplectic bundles
of N-extended supergravities, central charges and
black hole entropy,'' hep-th/9707203}

\lref\adfii{L. Andrianopoli, R. D'Auria, and S. Ferrara,
``Duality, Central Charges and Entropy of Extremal BPS Black Holes,''
hep-th/9709113;
``Five Dimensional U-Duality, Black-Hole Entropy and Topological Invariants,''
hep-th/9705024, Phys.Lett. B411 (1997) 39-45;
``U-Duality, attractors and Bekenstein-Hawking
entropy in four and five dimensional supergravities,'';
``U-Invariants, Black-Hole Entropy and Fixed Scalars,''
hep-th/9703156,Phys.Lett. B403 (1997) 12-19;
``Central Extension of Extended Supergravities in Diverse Dimensions,''
hep-th/9608015, Int.J.Mod.Phys. A12 (1997) 3759-3774}

\lref\adfesev{L. Andrianopoli, R. D'Auria,  S. Ferrara,  P. Fre',  M.
Trigiante,
``$E_{7(7)}$ Duality, BPS Black-Hole Evolution and Fixed Scalars,''
hep-th/9707087}

\lref\enntwosugra{
L. Andrianopoli,  M. Bertolini,  A. Ceresole,  R. D'Auria,  S. Ferrara,  P.
Fre',
T. Magri, ``N=2 Supergravity and N=2 Super Yang-Mills Theory on General Scalar
Manifolds: Symplectic Covariance, Gaugings and the Momentum Map,''
hep-th/9605032,J.Geom.Phys. 23 (1997) 111-189 ;
``General Matter Coupled N=2 Supergravity,'' hep-th/9603004,
Nucl.Phys. B476 (1996) 397-417}

\lref\adfl{L. Andrianopoli, R. D'Auria, S. Ferrara, M.A. Lledo,
``Horizon geometry, duality and fixed scalars in six
dimensions,'' hep-th/9802147}

\lref\asplut{P. Aspinwall and C. Lutkin,  ``Geometry of
mirror manifolds,'' Nucl. Phys. {\bf B355}(1991)482}

\lref\agm{P. Aspinwall, B. Greene, and D. Morrison,
``Calabi-Yau Moduli Space, Mirror Manifolds and Spacetime Topology Change in
String Theory,''  hep-th/9309097;
``Measuring Small Distances in N=2 Sigma Models,''
hep-th/9311042}

\lref\aspmorr{P. Aspinwall and D. Morrison,
``String Theory on K3 Surfaces,''  hep-th/9404151}

\lref\aspinwall{P. Aspinwall, ``K3 surfaces and
string duality,'' hep-th/9611137}

\lref\amiii{P. Aspinwall and D. Morrison,
``Point-like Instantons on K3 Orbifolds,''
hep-th/9705104, Nucl.Phys. B503 (1997) 533-564}

\lref\asphm{P. Aspinwall, ``Aspects of the hypermultiplet
moduli space in string duality,'' hep-th/9802194.}

\lref\amrp{P. Aspinwall and D. Morrison,
``Non-simply-connected gauge groups and rational
points on elliptic curves,'' hep-th/9805206}

\lref\balasub{V.J. Balasubramanian,
``How To Count the States of Extremal Black Holes in N=8 Supergravity,''
hep-th/9712215}

\lref\bpv{W. Barth, C. Peters, A. Van de Ven,
{\it Compact Complex Surfaces}, Springer-Verlag 1984}

\lref\batyrev{V.V. Batyrev, ``Variations of the mixed Hodge
structure of affine hypersurfaces in algebraic tori,''
Duke Math. J. {\bf 69}(1993) 349; ``Dual polyhedra and
mirror symmetry for Calabi-Yau hypersurfaces in toric varieties,'' J. Algebraic
Geom. {\bf 3}(1994) 493}

\lref\bbs{K. Becker, M. Becker, and A. Strominger,
``Fivebranes, Membranes and Non-Perturbative String Theory,''
hep-th/9507158;Nucl.Phys. B456 (1995) 130-152}

\lref\cardi{ K. Behrndt, G. Lopes Cardoso, B. de Wit, R. Kallosh, D. Löst, T.
Mohaupt, ``Classical and quantum N=2 supersymmetric black holes,''
hep-th/9610105}

\lref\cardii{K. Behrndt, G. Lopes Cardoso, I. Gaida,
``Quantum N = 2 Supersymmetric Black Holes in the S-T Model,''
hep-th/9704095; K. Behrndt and I. Gaida,
``Subleading contributions from instanton corrections in
N=2 supersymmetric black hole entropy,'' hep-th/9702168}

\lref\bls{K. Behrndt,  D. L\"ust, W.A.  Sabra,
``Stationary solutions of N=2 supergravity,''
hep-th/9705169,Nucl.Phys. B510 (1998) 264-288}

\lref\blsii{K. Behrndt,  D. L\"ust, W.A.  Sabra,
``Moving moduli, Calabi-Yau phase transitions
and massless BPS configurations in type II
superstrings,'' hep-th/9708065;
Phys. Lett. {\bf B418}(1998)303}

\lref\bcdlms{K. Behrndt, G.L. Cardoso, B. de Wit,
D. L\"ust, T. Mohaupt, and W.A. Sabra,
``Higher-order black-hole solutions in N=2
supergravity and Calabi-Yau string backgrounds,''
hep-th/9801081}

\lref\berglund{P. Berglund, P. Candelas, X. de la Ossa,
A. Font, T. Hubsch, D. Jancic, and F. Quevedo,
``Periods for Calabi-Yau and Landau-Ginzburg vacua,''
hep-th/9308005}

\lref\bkmt{P. Berglund, A. Klemm, P. Mayr, and
S. Theisen, ``On type IIB vacua with varying
coupling constant,'' hep-th/9805189}

\lref\berpan{M. Bershadsky, T. Pantev, and V. Sadov,
``F-Theory with quantized fluxes,'' hep-th/9805056}

\lref\birklang{H. Birkenhake and Ch. Lange,
{\it Complex Abelian Varieties}, Springer Verlag 1992}

\lref\borcea{C. Borcea, ``Calabi-Yau threefolds and
complex multiplication,'' in {\it Essays on Mirror
Manifolds}, S.-T. Yau, ed., International Press, 1992.}

\lref\borcha{R. E. Borcherds, ``The monster Lie algebra,'' Adv. Math. {\bf 83}
No. 1 (1990).}

\lref\borchi{R. Borcherds,``Monstrous moonshine
and monstrous Lie superalgebras,'' Invent. Math.
{\bf 109}(1992) 405.}

\lref\borchii{R. Borcherds, ``Automorphic forms
on $O_{s+2,2}(R)$ and infinite products,''
Invent. Math. {\bf 120}(1995) 161.}

\lref\borchiii{R. Borcherds, ``Automorphic forms
on $O_{s+2,2}(R)^+$ and generalized Kac-Moody
algebras,'' contribution to the Proceedings of
the 1994 ICM, Zurich. }

\lref\borchiv{R. Borcherds, ``The moduli space
of Enriques surfaces and the fake monster Lie
superalgebra,''  preprint (1994).}

\lref\borchalg{R. Borcherds, ``Generalized Kac-Moody algebras,'' Journal of
Algebra
{\bf 115} (1988) 501.}

\lref\borevich{Z.I. Borevich and I.R. Shafarevich,
{\it Number Theory} Academic Press 1966}

\lref\bsv{M. Bershadsky, V. Sadov, and C. Vafa,
``D-branes and topological field theories,''
 hep-th/9511222;
Nucl. Phys. {\bf B463}(1996)166}

\lref\buell{D. Buell, {\it Binary quadratic forms}
Springer-Verlag, 1989}

\lref\cdgp{P. Candelas, X. de la Ossa, P. S. Green and L. Parkes,
``A pair of Calabi-Yau manifolds as an exactly soluble
superconformal theory,'' Nucl. Phys. {\bf B359} (1991) 21.}

\lref\twop{P. Candelas X. de la Ossa, A. Font,
S. Katz, and D.R. Morrison, ``Mirror symmetry
for two-parameter models - I,''
hep-th/9308083}

\lref\twopii{ P. Candelas, A. Font,
S. Katz, and D.R. Morrison,
``Mirror Symmetry for Two Parameter Models -- II,''
hep-th/9403187}

\lref\cardcurl{G. L. Cardoso, G. Curio,
D. L\"{u}st, and T. Mohaupt, ``On the Duality between the Heterotic String and
F-Theory in 8 Dimensions,''  hep-th/9609111}

\lref\clm{G. L. Cardoso, D. L\"{u}st and T. Mohaupt,
``Threshold corrections and symmetry enhancement in string
compactifications,'' Nucl. Phys. {\bf B450} (1995) 115,
hep-th/9412209.}

\lref\lust{Gabriel Lopes Cardoso, Gottfried Curio, Dieter Lust, Thomas Mohaupt,
``On the Duality between the Heterotic String and F-Theory in 8 Dimensions,''
hep-th/9609111}

\lref\cassels{J.W.S. Cassels, {\it Rational Quadratic Forms},
Academic Press, 1978}

\lref\cecotti{S. Cecotti, ``$N=2$ supergravity, type IIB
superstrings, and algebraic geometry,''
Commun.Math.Phys.131:517-536,1990}

\lref\cdfvp{A. Ceresole,  R. D'Auria,  S. Ferrara,  A. Van Proeyen,
``Duality Transformations in Supersymmetric Yang-Mills Theories coupled to
Supergravity,''  hep-th/9502072}

\lref\fvdeei{A. Chamseddine, S. Ferrara, G. Gibbons,
and R. Kallosh, ``Enhancement of Supersymmetry Near 5d Black Hole Horizon,''
hep-th/9610155}

\lref\fvdeeii{A. Chou,  R. Kallosh,  J. Rahmfeld, S.-J. Rey,  M. Shmakova,
W.K. Wong,
``Critical Points and Phase Transitions in 5D Compactifications of M-Theory,''
hep-th/9704142}

\lref\clemens{H. Clemens, {\it A scrapbook of
complex curve theory}  Plenum Press, 1980
}

\lref\cohen{H. Cohen,
{\it A Course in Computational Algebraic Number Theory}, Springer GTM}

\lref\pbcohen{P.B. Cohen, ``Humbert surfaces and
transcendence properties of automorphic
functions,'' Rocky Mountain J. Math. {\bf 26}
(1996) 987}

\lref\slg{J.H. Conway and N.J.A. Sloane,
{\it Sphere Packings, Lattices, and Codes},
Springer-Verlag, 1993 }

\lref\csiv{J.H. Conway and N.J.A. Sloane,
``Low-dimensional lattices. IV. The mass
formula,'' Proc. R. Soc. Lond. {\bf A419} (1988)259.}

\lref\cox{D.A. Cox, {\it Primes of the form $x^2 + n y^2$}, John Wiley, 1989.}

\lref\cj{E. Cremmer and B. Julia, ``The $SO(8)$ supergravity''
Nuc. Phys. {\bf B159}(1979)141}

\lref\cvetic{M. Cvetic and  A. A. Tseytlin,
``Solitonic Strings and BPS Saturated Dyonic Black Holes,''
hep-th/9512031; M. Cvetic and D. Youm,
``Dyonic BPS saturated black holes of heterotic string on a six
torus,'' hep-th/9507090; Phys. Rev. {\bf D53}584-588 (1996)
%
%M. Cvetic and D. Youm,
%``General Static Spherically Symmetric Black Holes of Heterotic String on a Six Torus,''
%hep-th/9512127;Nucl.Phys. B472 (1996) 249-267
}

\lref\deligne{P. Deligne, ``La conjecture de Weil pour
les surfaces K3,'' Inv. Math. {\bf 15} (1972) 206}

\lref\delignelcsl{P. Deligne, ``Local behavior of Hodge structures at
infinity,'' in {\it Mirror Symmetry II}, B. Greene and
S.-T. Yau eds. International Press 1991.}

\lref\deser{S. Deser, A. Gomberoff, M. Henneaux,
and C. Teitelboim, ``p-Brane Dyons and
Electric-magnetic Duality,'' hep-th/9712189}

\lref\difrancesco{P. Di Francesco,  P. Mathieu, and
D. S\'en\'echal, {\it Conformal Field Theory} Springer 1997}

\lref\dv{R. Dijkgraaf and E. Verlinde, ``Modular invariance and
the fusion algebras,'' Nucl. Phys. (Proc. Suppl) {\bf 5B} (1988)
110}

\lref\dolgachev{I. Dolgachev,
``Mirror symmetry for lattice polarized K3 surfaces,''
alg-geom/9502005}

\lref\dolgachevi{I. Dolgachev, ``Integral quadratic forms:
applications to algebraic geometry,'' Sem. Bourbaki,
1982, no. 611, p. 251}

\lref\donagi{R. Donagi, ``ICMP lecture on Heterotic/F-theory
duality,'' hep-th/9802093}

\lref\fhsv{
S. Ferrara, J. A. Harvey, A. Strominger, C. Vafa ,
``Second-Quantized Mirror Symmetry, '' Phys. Lett. {\bf B361} (1995) 59;
hep-th/9505162. }

\lref\fgk{S. Ferrara,  G. W. Gibbons,  R. Kallosh,
``Black Holes and Critical Points in Moduli Space,''  hep-th/9702103;
Nucl.Phys. B500 (1997) 75-93}

\lref\fks{S. Ferrara, R. Kallosh, and A. Strominger,
``N=2 Extremal Black Holes,''   hep-th/9508072}

\lref\fk{S. Ferrara and R. Kallosh, ``Universality of Sypersymmetric
Attractors,''   hep-th/9603090;  ``Supersymmetry and Attractors,''
hep-th/9602136; S. Ferrara, ``Bertotti-Robinson Geometry and Supersymmetry,''
hep-th/9701163}

\lref\fklz{S. Ferrara, C. Kounnas, D. L\"{u}st and F. Zwirner,
``Duality-invariant
partition functions and automorphic superpotentials for $(2,2)$ string
compactifications,''  Nucl. Phys.  {\bf B365} (1991) 431. }

%\CveticZQ
\lref\CveticZQ{
M.~Cvetic and C.~M.~Hull,
``Black holes and U-duality,''
Nucl.\ Phys.\ B {\bf 480}, 296 (1996)
[arXiv:hep-th/9606193].
%%CITATION = HEP-TH 9606193;%%
}

\lref\fm{S. Ferrara and J.M. Maldacena,
``Branes, central charges and U-duality invariant BPS conditions,''
hep-th/9706097}

\lref\fg{S. Ferrara and M. G\"unaydin,
``Orbits of Exceptional Groups, Duality and BPS States in String Theory,''
hep-th/9708025}

\lref\font{A. Font, ``Periods and duality symmetries in
Calabi-Yau compactifications,'' hep-th/9203084}

\lref\fre{P. Fr\'e, ``Supersymmetry and First Order Equations for Extremal
States: Monopoles, Hyperinstantons, Black-Holes and p-Branes,''
hep-th/9701054,Nucl.Phys.Proc.Suppl. 57 (1997) 52-64 }

\lref\fmw{R. Friedman, J. Morgan, and E. Witten,
``Vector Bundles And F Theory,'' hep-th/9701162,
Commun.Math.Phys. 187 (1997) 679-743;
``Principal G-bundles over elliptic curves,''
alg-geom/9707004}

\lref\fvdeeiv{I. Gaida,
``N = 2 Supersymmetric Quantum Black Holes in Five Dimensional Heterotic String
Vacua,''
hep-th/980214}

\lref\gmms{I. Gaida, S. Mahapatra, T. Mohaupt, and W. Sabra,
``Black holes and flop transitions in M-theory on
Calabi-Yau threefolds,'' hep-th/9807014}

\lref\gms{O. Ganor, D. Morrison, and N. Seiberg,
``Branes, Calabi-Yau Spaces, and Toroidal Compactification of the N=1 Six-Dimensional $E_8$ Theory,'' hep-th/9610251}

\lref\gauss{C.F. Gauss, {\it Disquisitiones Arithmeticae},
Leipzig, 1801. English translation, Yale University
Press, 1966. }

\lref\gordon{B. Brent Gordon, ``A survey of the
Hodge conjecture for Abelian varieties,''
alg-geom/9709030}

\lref\bhgross{B.H. Gross, ``Groups over $\IZ$,''
Invent. Math. {\bf 124}(1996) 263}

\lref\gzsm{B. Gross and D. Zagier,
``On singular moduli,'' J. reine angew. Math.
{\bf 355} (1985) 191}

\lref\greenekantor{B.R. Greene and
Y. Kanter, ``Small Volumes in Compactified String Theory,''
hep-th/9612181}

\lref\gmsii{B. Greene, D. Morrison, and A. Strominger,
`` Black Hole Condensation and the Unification of String Vacua,''
hep-th/9504145;Nucl. Phys. B451 (1995) 109}

\lref\griffiths{P. Griffiths, et. al.
{\it Topics in transcendental algebraic geometry}, Ann. Math. Studies.
Princeton Univ.
Press, Princeton, 1984}

\lref\gkp{S. Gubser, I. Klebanov, and A. Polykov,
``Gauge Theory Correlators from Non-Critical String Theory,''
hep-th/9802109}

\lref\harris{J. Harris, {\it Algebraic Geometry} Springer 1992}

\lref\hmi{ J. Harvey and G. Moore, ``Algebras, BPS States, and Strings,''
hep-th/9510182;
Nucl.Phys. B463 (1996) 315-368}

\lref\hmalg{J. Harvey and G. Moore,
``On the algebras of BPS states,''
hep-th/9609017}

\lref\hilbert{D. Hilbert, ``Mathematical problems,''
in {\it Mathematical developments arising from
Hilbert problems}, Proc. Symp. Pure. Math.
{\bf 28} vol. 1. AMS 1976}

\lref\hornemoore{J. Horne and G. Moore }

\lref\hulekgrit{K. Hulek and V. Gritsenko}

\lref\husemoller{D. Husemoller, Elliptic curves}

\lref\hkty{ S. Hosono, A. Klemm, S. Theisen and S.-T. Yau,
``Mirror Symmetry, Mirror Map and Applications to Complete Intersection
Calabi-Yau Spaces, ''
hep-th/9406055;Nucl.Phys. B433 (1995) 501-554}

\lref\hkt{S. Hosono, A. Klemm, and S. Theisen,
``Lectures on Mirror Symmetry,'' hep-th/9403096}

\lref\hull{C. Hull and P. Townsend, ``Unity of superstring dualities,''  hep-th/9410167 }

\lref\itzykson{{\it From Number Theory to
Physics}, C. Itzykson, J.-M. Luck, P. Moussa,
and M. Waldschmidt, eds., Springer Verlag,
1992}

\lref\ireland{K. Ireland and M. Rosen, {\it A Classical
Introduction to Modern Number Theory},
Springer 1990}

\lref\kalkol{R. Kallosh and B. Kol,
``E(7) Symmetric Area of the Black Hole Horizon,''
hep-th/9602014}

\lref\klm{A. Klemm, W. Lerche, and P. Mayr,
``K3--Fibrations and Heterotic-Type II String Duality,''
hep-th/950611}

\lref\klemmtheis{A. Klemm and S. Theisen,
``Considerations of One-Modulus Calabi-Yau
Compactifications: Picard-Fuchs equations,
Kahler potentials and mirror maps,''
hep-th/9205041}

\lref\hkmiii{R. Kobayashi and A.N. Todorov,
``Polarized Period Map for Generalized K3 Surfaces and
the Moduli of Einstein Metrics,'' Tohoku Math. J.
{\bf 39} (1987) 341}

\lref\kulikov{V.S. Kulikov and P.F. Kurchanov,
``Complex algebraic varieties: periods of integrals
and Hodge structures,'' in {\it Algebraic Geometry III}
A.N. Parshin and I.R. Shafarevich, eds. Springer 1998}

\lref\langi{S. Lang, {\it Algebra}, Addison-Wesley, 1971}

\lref\langii{S. Lang, {\it Complex Multiplication}
Springer  1983}

\lref\langdg{S. Lang, {\it Survey of Diophantine
Geometry}, Springer Verlag 1997}

\lref\langlands{R.P. Langlands,
``Some contemporary problems with
origins in the Jugendtraum,''
in \hilbert.}

\lref\llw{W. Lerche, D. L\"ust, and N. Warner,
Phys. Lett. {\bf 231B}(1989)417}

\lref\lerchstie{W. Lerche and S. Stieberger,
``Prepotential, Mirror Map, and F-theory on
K3,'' hep-th/9804176}

\lref\lianyau{B.H. Lian and S.T. Yau,
``Arithmetic Properties of Mirror Map and Quantum Coupling,''
hep-th/9411234; Commun.Math.Phys. 176 (1996) 163-192;
``Mirror Maps, Modular Relations and Hypergeometric Series I,
hep-th/9507151}

\lref\luck{{\it Number Theory and Physics},
J.-M. Luck, P. Moussa, and M. Waldschmidt, eds.
Proceedings in Physics {\bf 47}, Springer Verlag
1990}

\lref\lustrev{D. L\"ust, ``String Vacua with
N=2 Supersymmetry in Four Dimensions,''
hep-th/9803072}

\lref\lustline{G.L. Cardoso, G. Curio, and D. L\"ust,
``Perturbative Couplings and Modular Forms in N=2 String Models with a Wilson Line,''
hep-th/9608154}

\lref\msw{J. Maldacena, A. Strominger, and E. Witten,
``Black Hole Entropy in M-Theory,''  hep-th/9711053}

\lref\maldacena{J. Maldacena,``The Large N Limit of Superconformal Field
Theories and Supergravity.''  hep-th/9711200}

\lref\mazur{B. Mazur, ``Arithmetic on curves,''
Bull. AMS {\bf 14}(1986) 207}

\lref\mikhailov{A. Mikhailov, ``Momentum Lattice for CHL String,''
hep-th/9806030 }

\lref\mirandai{R. Miranda and D.R. Morrison,
``The number of embeddings of integral quadratic forms. I,II''
Proc. Japan Acad. {\bf 61}(1985) 217}

\lref\ms{G. Moore and N. Seiberg,
``Naturality in conformal field theory,''
Nucl. Phys. {\bf B313}(1989)16 }

\lref\aaintro{G. Moore, ``Attractors and Arithmetic,''
hep-th/9807056}

\lref\aalong{G. Moore, ``Arithmetic and Attractors,''
hep-th/???}

\lref\hkmii{D.R. Morrison, ``Some remarks on the
moduli of K3 surfaces,'' in K. Ueno ed.
{\it Classification of Algebraic and Analytic
Manifolds}, Prog. in Math. {\bf 39},
Birkh\"auser, 1983}

\lref\morrpicard{D.R. Morrison,
``On $K3$ surfaces with large Picard number,''
Invent..Math.75 (1984), no. 1, 105--121}

\lref\dmlcsl{D.R. Morrison, ``Making enumerative
predictions by means of mirror symmetry,''
  in {\it Mirror Symmetry II}, B. Greene and
S.-T. Yau eds. International Press 1991.}

\lref\dmcomps{D.R. Morrison,
``Compactifications of moduli spaces inspired by mirror symmetry,''
alg-geom/9304007}

\lref\lookingglass{D.R. Morrison, ``Through the
Looking Glass,'' alg-geom/9705028}

\lref\vfmr{D. Morrison and C. Vafa, ``Compactifications of F-theory on
Calabi-Yau threefolds -I,'' hep-th/9602114;``Compactifications of F-theory on

Calabi-Yau threefolds -II,'' hep-th/9603161}

\lref\mumford{D. Mumford, {\it Tata Lectures on
Theta I}, Birkh\"auser 1983}

\lref\nikulin{V. Nikulin, ``Integral symmetric
bilinear forms and some of their applications,''
Math. Izv. {\bf 14} (1980) 103}

\lref\swi{N. Seiberg and E. Witten,
``Monopole Condensation, And Confinement In $N=2$ Supersymmetric Yang-Mills
Theory,''
hep-th/9407087;Nucl. Phys. {\bf B426} (1994) 19.}

\lref\swii{N. Seiberg and E. Witten,
``Monopoles, Duality and Chiral Symmetry Breaking in N=2 Supersymmetric QCD,''
hep-th/9408099;Nucl. Phys. {\bf B431} (1994) 484.}

\lref\shiodaef{T. Shioda, ``On elliptic modular surfaces,''
J. Math. Soc. Japan {\bf 24}(1972) 20}

\lref\shiodamitani{T. Shioda and N. Mitani,
``Singular abelian surfaces and binary quadratic
forms,'' in {\it Classification of algebraic varieties and comopact complex
manifolds} SLN 412 (1974) 259.}

\lref\shioda{T. Shioda and H. Inose, ``On singular
K3 surfaces,'' in Complex analysis and algebraic geometry,
Cambridge University Press, Cambridge, 1977}

\lref\slater{L.J. Slater, {\it Generalised Hypergeometric
Functions}, Cambridge University Press 1966 }

\lref\stark{H. Stark, ``Galois theory, algebraic
number theory and zeta functions,''
in {\it From Number Theory
to Physics} C. Itzykson et. al. eds. op. cit. }

\lref\susskind{L. Susskind and E. Witten,
``The Holographic Bound in Anti-de Sitter Space,''  hep-th/9805114}

\lref\psshaf{I.I. Piatetski-Shapiro and I.R. Shafarevich,
``A Torelli theorem for algebraic surfaces of type
$K3$,'' Math. USSR Izvestia {\bf 5}(1971)547}

\lref\sabra{W.A. Sabra, ``Black holes in N=2 supergravity theories and harmonic
functions,''  hep-th/9704147, Nucl.Phys. B510 (1998) 247-263}

\lref\fvdeeiii{W. Sabra,
``General BPS Black Holes In Five Dimensions,''
hep-th/9708103}

\lref\schmakova{M. Schmakova, ``Calabi-Yau black holes,''
hep-th/9612076}

\lref\sen{A. Sen, ``Stable Non-BPS Bound States of BPS D-branes,''
hep-th/9805019}

\lref\serre{J.-P. Serre, {\it Statement of results},
in {\it Seminar on Complex Multiplication},
A. Borel et. al. eds., SLN 21, 1966}

\lref\shiga{H. Shiga and J. Wolfart,
``Criteria for complex multiplication and
transcendence properties of automorphic
functions,'' J. reine angew. Math. {\bf 463} (1995) 1}

\lref\shimtan{G. Shimura and Y. Taniyama,
{\it Complex Multiplication of Abelian Varieties
and its applications to number theory}, Math. Soc. of Japan,
1961.}

\lref\gshimura{G. Shimura, {\it Abelian
Varieties with Complex Multiplication and Modular Functions}, Princeton
University Press, Princeton, 1998}

\lref\silvermanag{J. Silverman, {\it Arithmetic Geometry},
G. Cornell and J.H. Silverman, eds. Springer Verlag 1986}

\lref\silveraec{J. Silverman, {\it The Arithmetic of Elliptic
Curves}, Springer Verlag GTM 106 1986}

\lref\silveradvtop{J. Silverman, {\it Advanced Topics in the
Arithmetic of Elliptic Curves} Springer Verlag GTM 151, 1994}

\lref\stromsg{A. Strominger, ``Special Geometry,'' Commun. Math.
Phys. {\bf 133}(1990)163}

\lref\conifold{A. Strominger, ``Massless Black Holes
and Conifolds in String Theory
,'' hep-th/9504090 }

\lref\sv{A. Strominger and C. Vafa,
``Microscopic Origin of the Bekenstein-Hawking Entropy,''
hep-th/9601029; Phys.Lett. B379 (1996) 99-104 }

\lref\stromi{A. Strominger, ``Macroscopic Entropy of $N=2$ Extremal Black
Holes,''  hep-th/9602111}

\lref\syz{A. Strominger, S.-T. Yau, and E. Zaslow,
``Mirror symmetry is T-duality,'' hep-th/9606040,
Nucl.Phys. B479 (1996) 243-259}

\lref\takeuchi{K. Takeuchi, ``Arithmetic triangle
groups,'' J. Math. Soc. Japan {\bf 29} (1977) 91}

\lref\taormina{A. Taormina and S.M.J. Wilson,
``Virasoro character identities and Artin L-functions,''
physics/9706004}

\lref\thomas{R. Thomas, ``A holomorphic Casson
invariant for Calabi-Yau 3-folds and bundles on
K3 fibrations,'' IAS preprint}

\lref\hkmi{A. Todorov, ``Applications of K\"ahler-Einstein-
Calabi-Yau Metric to Moduli of K3 surfaces,''
Inv. Math. {\bf 61} (1980) 251}

\lref\ueno{K. Ueno, ``On fibre spaces of normally
polarized abelian varieties of dimension 2,'' J. Fac. Sci.
Tokyo, {\bf 18}(1971) 37}

\lref\vafa{C. Vafa, ``Evidence for F-theory,'' hep-th/9602022}

\lref\vafacyi{C. Vafa, ``Black Holes and Calabi-Yau Threefolds,''
hep-th/9711067}

\lref\vafacyii{C. Vafa, ``Extending Mirror Conjecture to Calabi-Yau with
Bundles,''
hep-th/9804131}

\lref\viehweg{E. Viehweg, {\it
Quasi-projective moduli for polarized manifolds}
Springer-Verlag, 1995}

\lref\vladut{S.G. Vladut, {\it Kronecker's Jugendtraum and modular functions},
Gordon and Breach, 1991.}

\lref\weil{A. Weil, {\it Elliptic functions according
to Eisenstein and Kronecker}, Springer-Verlag 1976 }

\lref\weilii{A. Weil, ``The field of definition of a
variety,'' Amer. J. Math. {\bf 78}(1956) 509}

\lref\grssmm{E. Witten, ``Quantum field theory,
grassmannians and algebraic curves,'' Commun.Math.Phys.113:529,1988}

\lref\wittenphase{E. Witten, ``Phases of N=2 theories in
two dimensions,'' hep-th/9301042; Nucl. Phys. {\bf B403}(1993)159}

\lref\WittVar{E. Witten, ``String theory in various dimensions,''
hep-th/9503124}

\lref\wittenholog{E. Witten, ``Anti De Sitter Space And Holography,''
hep-th/9802150}

\lref\youm{D.Youm, ``Black Holes and Solitons in String Theory,''
hep-th/971004}

%\deWitPX
\lref\deWitPX{
  B.~de Wit, P.~G.~Lauwers and A.~Van Proeyen,
  ``Lagrangians Of N=2 Supergravity - Matter Systems,''
  Nucl.\ Phys.\  B {\bf 255}, 569 (1985).
  %%CITATION = NUPHA,B255,569;%%
}

\Title{\vbox{\baselineskip12pt
\hbox{hep-th/9807087}
\hbox{YCTP-P17-98 }
}}
{\vbox{\centerline{Arithmetic and Attractors}
 }}

\bigskip
\centerline{Gregory Moore}
\bigskip
\centerline{\sl Department of Physics, Yale University}
\centerline{\sl New Haven, CT  06511}
\centerline{ \it moore@castalia.physics.yale.edu }

\bigskip
\bigskip
\noindent
We study relations between some topics in
number theory and supersymmetric black holes.
These relations are based on the ``attractor
mechanism'' of $\CN=2$ supergravity. In
IIB string compactification this mechanism
singles out certain ``attractor varieties.''
We show that these attractor varieties are
constructed from
products of elliptic curves with complex
multiplication
for $\CN=4,8$ compactifications. The heterotic
dual theories are related to rational conformal field
theories. In the case of $\CN=4$ theories
$U$-duality inequivalent backgrounds with
the same horizon area are counted by the
class number of a quadratic imaginary field.
The attractor varieties are defined over
fields closely related to class fields of
the quadratic
imaginary field. We discuss some extensions
to more general Calabi-Yau compactifications
and explore further connections to
arithmetic including connections to
Kronecker's Jugendtraum and the theory
of modular heights. The paper also includes
a short review of the attractor mechanism.
A much shorter version of the paper
summarizing the main points is the companion
note entitled ``Attractors and Arithmetic,''
hep-th/9807056.

\Date{July 10, 1998}
%\draft

\newsec{Introduction}

Many people  learning modern string theory
and supersymmetric gauge theory are struck by the
fact that much of the necessary mathematical
background is best  learned from textbooks on
number theory.  For example, modular forms,
congruence subgroups, and elliptic curves,
are all mathematical objects of central concern both to
string theorists and to number theorists.
This suggests the obvious idea that there might
be a deeper relationship between the two subjects.
Such a relation, if truly valid, would clearly
have a beneficial effect on both subjects.

Of course, one should be cautious about speculations
of this nature. For example, various partial
differential equations occur in widely disparate
fields of physics and engineering.
The mere appearance of a technical tool in two disparate subjects does not necessarily imply a deeper unity
(except insofar as the same equation appears).
Indeed, upon closer examination, one is often
disappointed to find that the precise
questions of the number theorist and the string
theorist  generally seem to be orthogonal.

One example will illustrate this orthogonality
of the world-views of the string- and number- theorist.
In string theory and supersymmetric gauge theory
one often meets elliptic curves. These play a central
role in conformal field theory, string perturbation
theory, supersymmetric gauge theory, string duality, and F-theory.
Yet, in all these applications,
there has never been  any compelling reason to
restrict attention to elliptic curves defined over
$\IQ$ (or any other  number field).  On the other hand,
it is the special properties of arithmetic elliptic
curves  which often take center stage in number
theory.

The point of this paper is that the
``attractor mechanism''  (explained below)
used in the construction of
supersymmetric  black holes
and black strings {\it does} provide a
compelling reason to focus on certain varieties,
and, at least in the examples
where we can solve the   equations
exactly, these  attractor varieties turn out to be
arithmetic. We believe this observation opens
up some opportunities for fruitful interactions
between string theory and number theory,
and the present paper is an attempt to explore
some of those relations.
The nonexpert reader should be warned that
the author knows very little number theory.

\lref\attractortalk{ http://online.kitp.ucsb.edu/online/strings98/moore/ }

Here is an outline of the paper:
A short introduction to this paper can be
found in the separate text  \aaintro. A talk on the subject
can be heard at \attractortalk. Section two contains a
review of the attractor mechanism. It is
primarily written for the mathematician
who wishes to learn something about the
subject.
Section three illustrates the first connection
between arithmetic and supersymmetric black holes.
This connection is related to questions about the orbits
of the arithmetic U-duality groups. Sections
4, 5, and 6, discuss some exact attractor varieties
in special compactifications with a high
degree of  symmetry: compactifications of
type II string theory with $\CN=4,8$
supersymmetry, and the FHSV model.
Section 7 constitutes the second, and main,
connection to arithmetic. It
attempts to explain why the attractor
varieties in the $\CN=4,8$ examples  are
arithmetic. The essence of the matter is that
the attractor varieties are related to curves with
``complex multiplication.''  Section 8 discusses some
extensions to larger classes of Calabi-Yau
3-folds and states the main conjecture
of the paper. Section 9 examines what can be
said about attractors near a point of
maximal unipotent monodromy/large radius.
Section 10 explains a relation to heterotic
compactification on rational conformal field
theories. Section 11 explores an arithmetic
property of the K3 mirror map. Section 12
explains the relation of the conjectures of
section 8 to Kronecker's Jugendtraum
and Hilbert's twelfth problem.
Section 13 mentions some more speculative
ideas including a relation to the absolute
Galois group and to heights of arithmetic
varieties.  In the conclusions we
list our principle results and some of the
main speculations.
Three appendices
briefly cover some background material
and some technical proofs.

In an effort to make
this unreadable paper readable we have
explicitly marked digressions, remarks, and
examples. No harm is done if they are ignored.
A list of some of our notation appears in
appendix D.

Finally, the references in this paper are
incomplete. There is a surprisingly
substantial literature
on the relation of arithmetic
and physics. The reader might wish to consult
the proceedings of a Les Houches school
\itzykson\luck\ for an introduction to some
aspects of the subject.

\newsec{Review of the attractor mechanism}

The attractor mechanism is an interesting
phenomenon discovered by Ferrara, Kallosh,
and Strominger in their work on dyonic
black holes in supergravity \fks. The attractor
equations \stromi\fk\  are central to the
ideas of this paper. They were interpreted in
terms of the minimization of the BPS mass by
Ferrara, Gibbons and Kallosh in
\fgk. In this section we review some of this
work. There is an extensive literature on the subject.
Some recent reviews include \adf\lustrev\youm, which
the reader should consult for more complete
references.

\subsec{Electric-magnetic duality and the
Gaillard-Zumino construction}

A common theme of modern string and gauge theory
is the study of a families of abelian gauge theories
with no natural electric/magnetic
splitting of the fieldstrengths. We now review the
standard formalism for describing such theories.

Let $M_4$ be a four-dimensional Minkowski
signature spactetime. Let
$\lieg=\IR^r$ be the Lie algebra of the
gauge group. Then   the total 2-form
fieldstrength $\CF$
(electric plus magnetic) is valued in
$\Omega^2(M_4;\IR) \otimes V$ where
$V\equiv \lieg\oplus\lieg^*$ is a real symplectic
vector space with symplectic
product $\langle \cdot, \cdot \rangle : V \times V \rightarrow \IR$.

In the physical problems the family of theories
is often labeled by a complex symplectic structure
on $V$, i.e., a linear transformation
 $\CJ: V \rightarrow V$ with $\CJ^2=-1$ and
$\langle \CJ \cdot v_1 , \CJ \cdot v_2 \rangle =
\langle   v_1 ,   v_2 \rangle$.  Given such a complex
structure there are two natural constructions
we can make.

First, we may define a symmetric
bilinear form on $V$:
\eqn\symmform{
(v_1, v_2)_{\CJ} \equiv
\langle v_1, \CJ \cdot v_2 \rangle
= (v_2, v_1)_{\CJ}
}
which will be
used to write the Hamiltonian of the theory.

Second, we can use $\CJ$ to define the correct
number of degrees of freedom of the theory.
Since $*_4^2=-1$ on
$\Omega^2(M_4)$ the operator  $*_T\equiv *_4 \otimes \CJ$
satisfies $*_T^2=+1$ and
we can therefore impose the all-important
anti-self-duality constraint:
\eqn\sdconst{
\CF = - *_T \CF
}
on {\it real} fieldstrengths $\CF$.
\foot{We choose the $(-)$ sign to agree with
several standard conventions below.} The
 equation of motion and Bianchi identity
of the electromagnetic theory are combined
in the single equation:
\eqn\eombi{
d \CF = 0
}
If we choose a space/time
splitting $M_4=M_3\times \IR$,
 the electric and magnetic fields $\vec E, \vec B\in \Omega^3(M_3) \otimes
\lieg$ are not
functionally independent because of \sdconst.
Moreover, the  formula for the energy density in terms of
the spatial components $\vec \CF$ is simply:
\eqn\enerdens{
\CH = (\vec \CF, \vec \CF)_{\CJ}
}
Equations \sdconst\eombi\enerdens\ constitute
a manifestly dual
formulation of the abelian theory.
It is impossible to write a local Lorentz invariant and symplectically
invariant action.

In order to connect to more standard treatments of the subject (see,
e.g. \deWitPX\adf\enntwosugra) we proceed as follows. If we
complexify we can simultaneously diagonalize the operators $*_4$ and
$\CJ$: \eqn\diagonal{ \eqalign{ \Omega^2(M_4;\IC) & = \Omega^{2,+}
\oplus \Omega^{2,-} \cr
*_4 & = -i \Pi^+ \oplus i \Pi^-\cr
V \otimes \IC & = V^{1,0} \oplus V^{0,1} \cr
\CJ & = +i \Pi^{1,0} \oplus -i \Pi^{0,1} \cr}
}
Here $\Pi$ are projection operators and
we use the notation $\xi^- = \half (\xi-i *_4 \xi)$ for any
two-form $\xi$.
Note that $*_4 \xi^-=+i \xi^-$.
If we now choose a symplectic (Darboux)
basis $\hat \alpha_I, \hat \beta^I$, $I=1, \dots, r$ for
$V$ with
 $\langle \alpha_I, \hat \beta^J \rangle = \delta_I^{~J}$
then we may always choose a basis $\{ f_I \}$ for
$V^{0,1}$ with $f_I = \hat \alpha_I + \tau_{IJ} \hat \beta^J$.
Let $\bar f_I$ be the complex conjugate basis.
By symplectic invariance of $\CJ$ it follows that
the period matrix $\tau_{IJ}= \tau_{JI}$ is symmetric.
Equivalently, the symplectic form $\langle \cdot, \cdot \rangle$
is of type $(1,1)$ with respect to the complex
structure $\CJ$.

The components of the total fieldstrength are, by
definition,
\eqn\defcompst{
\CF = F^I \hat \alpha_I - G_I \hat \beta^I \quad .
}
On the other hand, by the self-duality constraint
\sdconst\ we have
\eqn\dcsp{
\CF = F^{I,+} f_I + F^{I,-} \bar f_I
}
and combining these we arrive at
\eqn\geetoeff{
\eqalign{
G_I^- & = -\bar \tau_{IJ} F^{J,-} \cr
G_I & = (\im \tau_{IJ}) *_4 F^{J} -  (\re \tau_{IJ})F^J \cr}
}
and hence we recognize \eombi\ as the standard
Bianchi identities and equation of motion. The energy
is, after a short calculation:
\eqn\energy{
\CH =  \im \tau_{IJ} \bigl( \vec E^I \cdot \vec E^J +
\vec B^I \cdot \vec B^J \bigr)
}
where $\vec E^I, \vec B^I$ are the standard
spatial components of $F^I$.
Thus, physically, we require $\im \tau_{IJ}>0$,
that is, $\tau\in \CH_r$ where $\CH_r$ is the
Siegel upper half plane for $r \times r $ matrices. Hence, if
$\Lambda$ is the integral span of $\hat \alpha_I, \hat \beta^I$
for a symplectic basis then $V/\Lambda$ is a
principally polarized abelian variety.

As a final remark one might attempt to form a
local, Lorentz invariant, and symplectic invariant
action using  $\int_{M_4} (\CF, \CF)_{\CJ}$.
A short  calculation reveals this to be zero.
(Although the symmetric form \symmform\
is positive definite, $\Omega^2(M_4)$  has nilpotents.) If one chooses
a symplectic basis then $\int_{M_4} (\CF, \CF)_{\CJ}$
is naturally written as a sum of two cancelling terms,
either one of which provides an action:
\eqn\action{
\int_{M_4} G_I \wedge *_4 F^I = +2 \int_{M_4} \im \biggl(
\bar \tau_{IJ} F^{I,-} \wedge F^{J,-} \biggr)
}

\subsec{Low energy supergravity for Calabi-Yau
compactification of IIB supergravity}

In order to explain the attractor phenomenon
we will focus attention on the compactification
of IIB string theory on Calabi-Yau 3-folds.
In this section we review a few of the relevant
details of the resulting low-energy $d=4, \CN=2$ supergravity
needed to describe the dyonic black holes.
In the present discussion a key role is played by
the abelian gauge  fields in the theory.

Gauge fields in the four-dimensional theory
arise from the 5-form fieldstrength $G$
of IIB supergravity.
The equations of motion and Bianchi
identity follow from the anti-self-duality
constraint: $G= -*_{10} G$.
Let $X_{}$ be a Calabi-Yau 3-fold, that is, a
compact, complex 3-fold with Ricci flat K\"ahler
metric. If $b_1(X_{})=0$ this is
the only source of abelian gauge fields and
consequently the total fieldstrength is
\eqn\iibredi{
\CF \in \Omega^2(M_4) \otimes H^3(X_{};\IR).
}
We are now exactly in the general
setup of the previous section since $V=H^3(X;\IR)$
has symplectic form:
\eqn\symplfrm{
\langle \hat \gamma_1, \hat \gamma_2 \rangle
\equiv \int_X \hat \gamma_1 \wedge \hat \gamma_2
}
and a metric on $X_{}$ defines a complex structure
$\CJ= *_X: H^3 \rightarrow H^3$. The
selfduality constraint \sdconst\ is just that inherited
from $G$. The principally polarized variety
$V/\Lambda$ is known as the Weil Jacobian.
In the supergravity literature the period matrix
is denoted as $N_{IJ} = - \tau_{IJ}$.

\subsubsec{Hodge structures}

We next want to decompose the
gauge fields \iibredi\ into those coming from
the different   supersymmetric multiplets namely
the graviphoton in the gravity multiplet and the
remaining vectors in the vectormultiplets.
In order to do this we need to use a
little of the theory of variation of
Hodge structures, see, e.g.,  \griffiths\cecotti\kulikov\stromsg.
Consider the universal family
$\pi: \CX \rightarrow \CM $. The
fiber at
$s\in \CM$ is
$X_s$,  the Calabi-Yau with complex
structure $s$ (we fix a Kahler class and use Yau's
theorem). The complex structure at $s$ determines a
Hodge decomposition:
\eqn\hdec{
H^3(X_s;\IC) = H^{3,0}_s \oplus H^{2,1}_s
 \oplus H^{1,2}_s \oplus H^{0,3}_s
}
in terms of which the Weil complex structure
$\CJ= *_X$ is diagonal:
\foot{For simplicity we are assuming $b_1(X)=0$ here, otherwise
we need to distinguish the primitive from the nonprimitive cohomology.}
\eqn\weilcplx{
\CJ= *_X= -i \Pi^{3,0} \oplus i \Pi^{2,1} \oplus -i \Pi^{1,2} \oplus+i
\Pi^{0,3}
}
Here $\Pi^{p,q}(v)$ is the component of
$v$ in $H^{p,q}$. We also use the notation
$v^{p,q}\equiv \Pi^{p,q}(v)$.
Accordingly, the anti-self-duality constraint is solved by:
\eqn\sdconsti{
\CF= (\Pi^{2,1} \oplus \Pi^{0,3})(\CF^-) +
(\Pi^{1,2} \oplus \Pi^{3,0})(\CF^+)
}

Now choose a neighborhood $\CU\subset \CM$ and
a holomorphic family $\Omega^{3,0}(s)$ of nonwhere
zero holomorphic $(3,0)$ forms.
\foot{In a more precise description we consider
  the Hodge line bundle over $\CM$:
$\CL = R\pi_*\omega_{\CX/\CM}$ where
$\omega_{\CX/\CM}$ is the relative dualizing sheaf.
The fiber at $s$ is $H^{3,0}(X_s)$. We choose a
local holomorphic section of $\CL$.}
Using $\Omega$ and Kodaira-Spencer theory we have
the isomorphism: $T_s^{1,0}\CM \cong H^{2,1}(X_s)$, and
moreover, if we choose local holomorphic coordinates
$z^i$, $i=1,\dots, h^{2,1}(X)$,  we have a basis for $H^{2,1}$:
\eqn\definechis{
\chi_i \equiv e^{K/2} \Pi^{2,1}(\p_i \Omega)
=e^{K/2} \biggl( \p_i \Omega - {\langle \p_i \Omega, \bar \Omega\rangle \over
\langle \Omega, \bar \Omega \rangle} \Omega\biggr) .
}
Here $K$ is a K\"ahler potential for the Weil-Peterson-Zamolodchikov
(WPZ) metric:
\eqn\wpz{
e^{-K} = i \langle \Omega, \bar \Omega \rangle.
}
A short calculation shows that
$g_{i \bar j} = - i \langle \chi_i, \bar \chi_{\bar j} \rangle$.

\subsubsec{Supersymmetry transformations}

The massless multiplets are the gravity
multiplet, vectormultiplet, and
hypermultiplet. We generally follow
the notation and conventions of
\enntwosugra\ and denote the gravity multiplet
as $(g_{\mu\nu} , \psi_{\mu A} , A_\mu^0)$,
where the subscript
$A$ is an $sl(2)$ R-symmetry
index and $A^0_\mu$ is the graviphoton.
The vectormultiplets are
denoted $(z^i, \lambda^{Ai}, A^i_\mu)$,
$i=1,\dots, n_V$.
The vectormultiplet scalars are coordinates
for a nonlinear  sigma model with   target the
moduli of  complex structures on a polarized 3-fold $X$:
\eqn\vmscalr{
z: M_4 \rightarrow \CM(X_{}).
}
The kinetic energy follows from the WPZ metric.

Variations $\delta z^i$ of the vectormultiplet scalars
are related to tangents to $\CM$, and are also
related by $\CN=2$ supersymmetry to the fieldstrengths
of the vectormultiplets. Hence we define the
vectormultiplet fieldstrengths by:
\eqn\vmfldstrngth{
G^{i,-} \otimes \chi_i \equiv -\half \Pi^{2,1}(\CF^-)
}
The supersymmetry transformations for the
associated gauginos must contain two terms
corresponding to raising or lowering helicity:
\eqn\gaugin{
\delta \lambda^{Ai} = i \dsl z^i  \epsilon^A + G^{i,-}_{\mu\nu} \gamma^{\mu\nu}
\varepsilon^{AB} \epsilon_B
}
Here $\epsilon^A, \epsilon_B$ are supersymmetry
parameters of opposite chirality. $\varepsilon^{AB}=i \sigma^2$
is a numerical matrix.

Since $b_3(X)= 2+ 2h^{2,1}(X)$ there is one remaining
gauge field. This  gauge field is the graviphoton, whose
fieldstrength is defined by the projection of $\CF^-$
onto $H^{0,3}$:
\eqn\grvphtstrngth{
T^-   \equiv e^{K/2} \langle \Omega, \CF^- \rangle
}
The corresponding susy transformation law is:
\eqn\gauginii{
\eqalign{
\delta \psi_{\mu A} & = \CD_\mu  \epsilon_A  + T^-_{\mu\nu} \gamma^{ \nu} \varepsilon_{AB}
\epsilon^B\cr
\CD_\mu \epsilon_A & = \bigl( \p_\mu - {1 \over  4} \omega_\mu^{ab} \gamma_{ab}
+ {i \over  2 } Q_\mu\bigr)\epsilon_A \cr}
}
where the covariant derivative is the standard
spinor and K\"ahler covariant derivative.

Neglecting hypermultiplets, the  bosonic part of the
action is accordingly:
\eqn\bosonact{
I_{\rm boson} = \int_{M_4} - \half e R +
\parallel \nabla z \parallel^2 - {1 \over  8 \pi} \im[ \bar \tau_{IJ} F^{I,-}
F^{J,-}  ]
}
where $\tau_{IJ}$ is the period matrix of the Weil Jacobian.

\subsec{Charge lattices}

%
%The main tool in recent years in understanding
%the nonperturbative spectrum of string theories
%consists of the Olive-Witten observation that
%dyonic BPS representations of supersymmetry
%are rigid. Accordingly, in order to have a clear
%understanding of dyonic objects one must have
%a clear understanding of the charge lattices
%

Supersymmetric
black holes and strings in
$D=4,5,6 $-dimensional   compactifications of
string/M/F theory  are charged under certain
one-form or two-form gauge fields.
The set of charges are valued in a
lattice   $\Lambda$. In the examples
related to the attractor phenomenon the relevant
charge lattices are given in Table 1.

\def\tablerule{\omit&\multispan{6}{\tabskip=0pt\hrulefill}&\cr}
\def\tablepad{\omit&height3pt&&&&&&&\cr}
$$\vbox{\offinterlineskip\tabskip=0pt\halign{
\strut$#$\quad&\vrule#&\quad\hfil $#$ \hfil\quad &\vrule #&\quad \hfil $#$
\hfil \quad&\vrule #& \quad $#$ \hfil\ &\vrule#&\quad $#$\cr
&\omit&\hbox{  $T^d$, $ \CN=32$  }&\omit&
\hbox{ $K3 \times T^d$, $\CN=16$  }&\omit&
\hbox{ $X_{}$, $\CN=8$  }&\omit&\cr
\tablerule\tablepad
&& II^{5,5} &&  II^{21,5} &&  H^{1,1}(B;\IZ) &&\cr
\tablerule\tablepad
&& II^{6,6} && II^{21,5}   &&
H_2(X_{};\IZ)  &&\cr
\tablerule\tablepad
&& II^{6,6} && II^{21,5}   &&
 H_4(X_{};\IZ)  &&\cr
\tablerule\tablepad
&&   \IZ^{28}_{el}\oplus
\IZ^{28}_{mag}  && II^{22,6}_{\rm el} \oplus
II^{22,6}_{\rm mag}  &&
  H_3(X_{};\IZ),  H^{\rm ev}(\tilde X_{};\IZ)   && \cr
\tablerule
\noalign{\bigskip}
\noalign{\narrower\noindent{Table 1. In the first line we have listed
charge lattices of 6D strings in various compactifications
of type IIB string theory and $F$-theory. Here $B$ is the
base of an elliptic fibration
$\pi: X_{} \rightarrow B$. In the second and third lines
we have listed the charge lattices of 5D black holes and
strings, respectively. In the final line we have listed
the electric/magnetic charge lattices of 4D supersymmetric
black holes.  }
 }
 }}
$$

A full explanation of all entries of this table is
beyond the scope of this short review. We
content ourselves with some explanation of
the entry on the lower right corner, namely
$IIB$ on $X$.

We have shown that the Lie algebra is such
that $\lieg \oplus \lieg^* \cong H^3(X_{};\IR)$.
To specify the physical theory we must
specify the corresponding Lie {\it group }
$G= \IR^r \mod 2 \pi L$ where $L\subset \IR^r$ is
a rank $r$ ``lattice.''
\foot{See the remark  below on the use of the word
``lattice.'' } The electric charge
lattice is the lattice of unitary
irreps of $G$ and is just  $L^*$ while the magnetic
charge lattice is the lattice of chern classes
of gauge bundles on a sphere at infinity and is
just $L$. Together we form the symplectic
rank $2r$ charge lattice $\Lambda = L \oplus L^*$.
In this form it has a symplectic splitting.

Let us determine $\Lambda$ for the compactification
of $IIB$ theory on $X_{}$.
The quantization of the abelian charges is justified
by the existence of D-branes.
For example, suppose
a D3-brane wraps a real
3-cycle $\Sigma_3$ in nine-dimensional
space $M_9$ in $IIB$ theory. Suppose $\Sigma_5$ is
a linking $5$-cycle in $M_9$. Then:
\eqn\quantiz{
\int_{\Sigma_5} {G \over  2 \pi}
}
measures the ``number of enclosed
$D3$-branes.''  If there are no fractional D3-branes
\quantiz\  is necessarily integral. In the compactification
of $IIB$ on $M_9=M_4 \times X_{}$ described above
supersymmetric configurations arise when
a D3-brane wraps a supersymmetric 3-cycle
$\Sigma_3 \subset X_{}$ \bbs. If $M_4$ is an
asymptotically Minkowskian spacetime and
we consider the compactification $M_4 \times X_{}$
then we can choose the linking 5-cycle to
be $\Sigma_5 = S^2_{\infty} \times \Sigma_3$
and it follows that
\eqn\quantizch{
\int_{S^2_{\infty}  \times \Sigma_3} {G \over  2 \pi}
= \int_{S^2_{\infty}} \langle \widehat{[ \Sigma_3] } , c_1(\CF) \rangle
}
where $\widehat{[ \Sigma_3]}$ is the Poincar\'e dual
to the homology class $[ \Sigma_3]$ and
 $c_1(\CF) = {1 \over  2 \pi} [\CF]$ is the
Chern class of the $G$-bundle over $S^2_{\infty}$
defining a topological sector of configuration space.
Thus, the magnetic charges are quantized and hence
so are the electric charges. If the fundamental
D3 brane has charge 1 then:
\eqn\iiblatt{
\Lambda = H^3(X_{};\IZ)
}
in $IIB$ theory with the natural symplectic structure.

%
%\item{1.} If $D3$ branes are treated as external
%Dirac-like branes they enter into the supergravity
%path integral via the phase-factor
%\eqn\phasefactor{
%\exp\biggl( i \int_{\Sigma_3 \times \IR} C^{(4)}\biggr)
%}
%Hence from the quantization of RR charge we see that
%the ``gerbe gauge group'' of the RR 4-form gauge potential
%includes ``large gauge transformations''
%$C^{(4)} \rightarrow C^{(4)} + 2 \pi \omega$ where
%$\omega\in H^4(M_{10};\IZ)$.
%

\bigskip
\noindent
{\bf Remark.} In 6D the charge lattice $\Lambda$ of
strings is, on general principles a {\it lattice} in
the sense that there is an integral symmetric
bilinear form: $(\cdot, \cdot): \Lambda \times \Lambda \rightarrow \IZ$
\deser. In 4D the general principles only
guarantee the existence of a nondegenerate integral
symplectic structure on the electric/magnetic
charge lattice.
In the above discussion the word ``lattice'' means a rank
$r$ $\IZ$-submodule of $\IR^r$. In fact, in the
theories under discussion $L$
turns out to have an integral quadratic form
and hence is a lattice in the usual sense. This does
not follow from any general physical principles (as
far as the author is aware) and is a deep consequence of
mirror symmetry. Choosing a mirror map between a
point of maximal unipotent monodromy and a large
radius limit we have an isomorphism \asplut\dmlcsl\syz:
\eqn\lattisom{
\mu: H^{\rm odd}(X;\IZ) \rightarrow H^{\rm even}(\tilde X;\IZ)
}
where $\tilde X$ is the mirror CY manifold. Hence the
charge lattices $\Lambda$ have a natural antisymmetric
symplectic structure {\it and} a natural symmetric
quadratic form. To choose a simple example,
$K3 \times T^2$ is self-mirror. By the Kunneth theorem:
\eqn\kunneth{
H^3(K3 \times T^2;\IZ) \cong H^2(K3;\IZ) \otimes H^1(T^2;\IZ)
}
and, if we make a choice of $a,b$ cycles on $T^2$,
 we can identify
\eqn\kunnethii{
H^3(K3 \times T^2;\IZ) \cong II^{19,3} \oplus II^{19,3}
}
where $II^{19,3}$ is the even unimodular lattice of
signature $((-1)^{19},(+1)^3)$.

\subsec{Construction of certain dyonic black holes in
$IIB$ supergravity compactifications on CY 3-folds}

We now construct black holes in
$d=4$, $\CN=2$  compactification of $IIB$ string
theory on a  CY 3-fold $X_{}$. We assume they are:

1. Static, spherically symmetric, asymptotically
Minkowskian spacetimes $M_4$.

2. Dyonic of  charge $\hat \gamma\in H^3(X_{};\IZ)$.

3. BPS, that is, satisfy
$\delta \lambda = \delta \psi =0 $ for a 4 real dimensional
space of spinors. (``Preserve $1/2$ the supersymmetry. '')

{}From conditions one and three we obtain the metric
ansatz:
\eqn\metanstz{
ds^2 = - e^{2U(r)} dt^2 + e^{-2 U(r)} (d \vec x)^2
}
where $r^2 = \vec x^2$ is the Euclidean norm,
and $e^{2U} \rightarrow 1+ \CO(1/r)$ for $r \rightarrow \infty$.
The complex structure moduli $z^i$ are functions
only of $r$ and the electric and magnetic fields are
radial and only depend on $r$.

{}From condition 2 we obtain the Chern-class:
\eqn\cherncls{
\int_{S^2_{\infty}} {\CF \over  2 \pi} = \hat \gamma
}
One now makes an ansatz to split the spacetime from
internal degrees of freedom as much as is consistent
with the self-duality constraint \sdconsti. Introduce
the complex 2-form \fre:
\eqn\eemin{
E^- \equiv \sin \theta d \theta d \phi - i {e^{2U(r)} \over  r^2} dt \wedge dr
}
which satisfies $*_4 E^- = i E^-$ in the metric
\metanstz. The ansatz for the electromagnetic
fieldstrengths is that the total fieldstrength
$\CF$ is:
\eqn\effanstz{
\CF = \re\biggl[ E^- \otimes \bigl(  \hat \gamma^{2,1} +
 \hat \gamma^{0,3}  \bigr) \biggr]
}

We now impose the BPS condition:
$\delta \psi = \delta \lambda = 0$ for some
Killing spinor $\epsilon$ on $M_{10} = M_4 \times X_{}$.
This defines   a dynamical
system on $\CM_{VM}$ with flow parameter $r$.
In order to write it we will need to pass to the universal cover
$\widetilde{\CM} \rightarrow
\CM = \Gamma\backslash \widetilde{\CM}  $, where
$\Gamma$ is the duality group $\Gamma\subset Sp(b_3;\IZ)$.
Thus   $\pi: \widetilde{\CX}  \rightarrow \cmtld$, is a
family of marked CY 3-folds. The
Hodge bundle $\CL= R\pi_*\omega_{\CX/\CM}$
pulls back to $\widetilde{\CL}$.

 For $\gamma\in H_3(X;\IZ),
\Omega\in \widetilde{\CL}$ we
define a function on the total space of
$\widetilde{\CL}
\equiv R\pi_*\Omega^{3,0}(\widetilde{\CX}) \rightarrow \widetilde{\CM}$:
\eqn\defzee{
Z(\Omega;\gamma) \equiv e^{K/2} \int_\gamma \Omega  .
}
We will write $Z$ when $\Omega, \gamma$ are understood.
Note that   $\Pi^{3,0}(\hat \gamma) = i \bar Z e^{K } \Omega$. Note also
that
\eqn\bpsmass{
\vert Z(\Omega;\gamma) \vert^2 \equiv {\vert \int_\gamma \Omega \vert^2 \over
i \int \Omega \wedge \bar \Omega}
}
 is a nonnegative
well defined real function on $ \widetilde{\CM}$
(but not on $\CM$), so we sometimes
denote this as $\vert Z(z;\gamma) \vert^2 $.

\ifig\flowii{Radial evolution from infinity to the
horizon corresponds to the flow of a
dynamical system in moduli space $\CM$.   }
{\epsfxsize2.5in\epsfbox{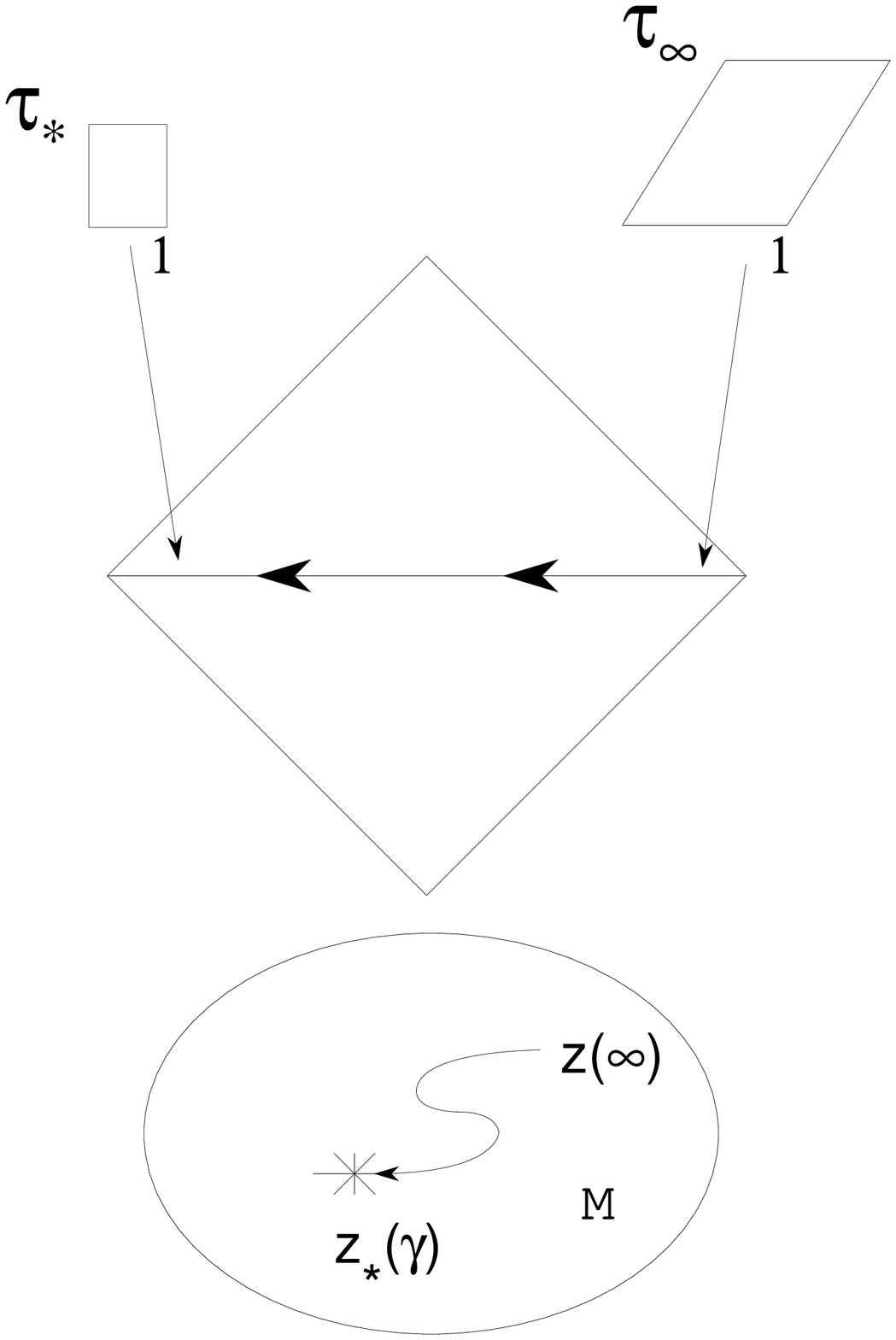}}

The dynamical system is obtained by solving
$\delta \psi_{A\mu}=0, \delta \lambda^{i A}=0$
using \gaugin\gauginii. Starting with the time-component
equation $\delta \psi_{A 0}=0$ we find that the
supersymmetry transformation parameters
are related by:
\foot{There is an unfortunate clash of notation.
$\gamma$ refers to a representation of a Clifford
algebra, not a charge in this equation. Also,
the $\gamma$-matrix has a flat space index.}
\eqn\susypars{
\gamma_{\underline{0}} \epsilon_A = i { Z \over  \vert Z \vert}
\varepsilon_{AB}\epsilon^B
}
Since
$\psi_A, \psi^A$ transform as sections of
$\CL^{\pm 1/2} \rightarrow \widetilde{\CM}$ the
relative phase of $Z$ is necessary. Equation
\susypars\ is the only condition we impose on
the supersymmetry parameters, so the black
holes we are constructing preserve $d=4,\CN=1$ supersymmetry as long as $Z \not=0$.

The next step involves  solving $\delta \lambda=0$ to get
a first order differential equation for the
$r$-dependence of the complex structure moduli.
In this way one derives a dynamical system
on $\IR^* \times \widetilde{\CM}$:
\eqn\dynsys{
\eqalign{
  { d \over  dr} (e^{-U})  & =
- { \vert Z(\Omega; \gamma)\vert \over  r^2}  \cr
     \Pi^{2,1}\bigl(e^{K/2} {d \Omega \over  dr} \bigr)
& = i { e^{U}  \over  r^2} {Z \over  \vert Z \vert }
\hat \gamma^{2,1} \cr}
}
The phases in \dynsys\ are very important
and have physical significance.

To construct a supersymmetric
black hole we  choose  a boundary
 condition at $r=\infty$,
$z^i(r=\infty) = z^i_\infty, e^{U(r=\infty)} = 1$
and use \dynsys\ to evolve $U(r), z^i(r)$
inwards to $r<\infty$ until we meet a horizon or a
singularity.  Ordinary differential
equations with Lipschitz vector fields
always have solutions for some range of
$r$. Thus we can always construct
dyonic black hole solutions for any charge
$\gamma$ and any initial conditions $z^i_\infty$. The
resulting spacetime might or might not
be singular. What actually happens  depends on
the evolution of the complicated nonlinear
dynamical system \dynsys.

Under ``good conditions'' on $\gamma$
and $z^i_\infty$ the
equations evolve smoothly to a fixed point
in moduli space $z=z_*(\gamma)$ at $r=0$, where
there is a horizon. This is the
attractor mechanism of \fks\fk.
We will give some (incomplete) discussion of what
constitute ``good conditions'' below. For the moment
we focus on the fixed point condition.

Suppose the flow \dynsys\ has a fixed point  with
$Z\not=0$. Then
 $\Omega(r)$ is constant so $\Pi^{2,1}(\hat \gamma)=0$.
Since $\hat \gamma$ is real it has Hodge decomposition:
\eqn\hodgetype{
\hat \gamma = \hat \gamma^{3,0} + \hat \gamma^{0,3}.
}
This equation is a condition on the Hodge structure of
$X_{}$ known as the ``attractor equation.''
The fixed points are also sometimes referred to as
``fixed scalars.'' In the literature, the
attractor equations are usually written
somewhat differently.
Since
$h^{3,0}(X)=1$ the condition on the Hodge decomposition
is equivalent to the statement that there is a complex
number $C$ such that $\hat \gamma^{3,0} = - i \bar{C} \Omega$,
hence:
\eqn\hodgede{
2 {\rm Im} (\bar C \Omega^{3,0}) = \hat \gamma
\in H^3(X;\IZ)
}
Now, choosing a symplectic basis
$\hat \alpha_I, \hat \beta^I$ for $H^3(X;\IZ)$
we write the equations as
\eqn\hodgedeii{
2 {\rm Im} (\bar C \Omega^{3,0})
= p^I \hat \alpha_I - q_I \hat \beta^I
}
with $p^I,q_I$ integral. In terms of the
Poincar\'e dual basis $\alpha^I, \beta_I$
we can define periods
  $X^I = \int_{\alpha^I} \Omega, F_I =\int_{\beta_I}
\Omega $ and write the equations
in the familiar form:
\eqn\attreqs{
\eqalign{
\bar C X^I - C \bar X^I & = i p^I \cr
\bar C F_I -  C \bar F_I & = i q_I \cr}
}
The central charge at the attractor point is
defined by $C= e^{K/2} Z(\Omega;\gamma)$
with
\foot{Our conventions for Poincar\'e duality are:
$\int_X \alpha\wedge\eta(\gamma)    = \int_\gamma \alpha$,
$\eta(\alpha^I) \equiv \hat \beta^I$, $\eta(\beta_I) \equiv \hat \alpha_I$, so
that $\hat \gamma = - \eta(\gamma)$. }
\eqn\gmmapq{
\gamma = \gamma(p,q) \equiv q_I \alpha^I - p^I \beta_I
}

The equations \attreqs\ constitute $b_3$ real equations
for $b_3$ real variables (if we do not fix the gauge
of $\Omega$). Therefore, one can hope that
generically the equations \attreqs\ will determine
isolated points $z_*(\gamma)$ in
$\cmtld$.

If the initial condition $z^i_{\infty}$ is an
attractor point the
  geometry of the black hole  is easily described.
Let $Z_*$ be the fixed point value of
$Z(\Omega_*;\gamma)$. Since $Z_*$ is independent
of $r$ we have:
\eqn\radius{
 e^{-U}=1 + \vert Z_*\vert /r
}
In particular the near horizon geometry is $AdS_2 \times S^2$:
\eqn\nearhor{
ds^2 = - {r^2 \over  \vert Z_*\vert^2} dt^2 + \vert Z_*\vert^2 { dr^2 \over  r^2} +
\vert Z_*\vert^2 d\Omega^2
}
with a horizon at $r=0$.
The   Ricci curvature is
\eqn\ricci{
\CR^\mu_{~\nu}  = \pmatrix{1& 0 \cr
0 & -{1 \over  3} {\bf 1_{3 \times 3} } \cr} { \vert Z_* \vert^2 \over
(r+ \vert Z_* \vert)^4 }
}
and hence the condition for the validity of supergravity
outside the horizon is $ \vert Z_* \vert \gg 1$.

\bigskip
\noindent
{\bf Remarks.}

\item{1.}  The equations \dynsys\ were also written
as gradient flow equations in \fgk:
\eqn\gradflow{
\eqalign{
{dU \over  d \rho} & = - e^{U} \vert Z \vert \cr
{d z^i \over  d \rho} &
= -2 g^{i \bar j} e^{U} \p_{\bar j} \vert Z \vert\cr}
}
where $\rho=1/r$.  Equation \gradflow\ is
 a gradient flow on $\IR^* \times \widetilde{\CM}$
with metric $2dU^2 +   ds^2_{WPZ}$ and potential function
$W = e^U \vert Z(z ;\gamma) \vert$.
The derivation of \gradflow\ from \dynsys\ is a
straightforward application of special geometry using:
\eqn\dergrdfl{
 \p_{\bar j} \vert Z \vert   = \half { Z \over  \vert Z \vert} e^{K/2}\int_{\gamma} \p_{\bar j} \bar \Omega + \half \vert Z \vert \p_{\bar j} K   = \half   { Z \over  \vert Z \vert} \int_\gamma
\bar \chi_{\bar j}
}
Note that away from the locus $Z(\Omega;\gamma)=0$, $U$ is a strictly
decreasing function of $\rho$. It is convenient to define a new
variable $\mu \equiv e^{-U}$ in terms of which the equations
\gradflow\ take the suggestive form:
\eqn\gradflowii{
\eqalign{
\mu { d z^i \over d \mu}
& = - g^{i \bar j} \p_{\bar j} \log\vert Z\vert^2\cr
{d \mu \over d \rho} & = \vert Z \vert \cr}
}
which turns out to be more convenient for investigating
the flows. Note in particular that the trajectory
on $\widetilde{\CM}$ is itself a gradient flow
with potential $\log\vert Z\vert^2$.

\item{2.}
The flow equations \dynsys\ were first
written in \fks.
The condition \hodgede\ for a fixed point
was first written by Strominger in \stromi.
The attractor equations have been widely
discussed in the literature.
See, for examples, \fre\sabra\bls\bcdlms.
The latter   references even
claim to give  a general
solution to \dynsys, and even generalize it
to include rotating and multicenter solutions.
(The derivation of these solutions requires
a choice of gauge for $\Omega$
which is not manifestly a consistent choice.
This issue deserves some further study.)
Finally, similar phenomena have been found for
5-dimensional extremal black holes
and for 5- and 6-dimensional black holes and
strings. A partial list of references includes
\fvdeei\fvdeeii\fvdeeiii\fvdeeiv\adfl.

\subsec{The minimization principle}

The dyonic black holes of the previous section
are semiclassical descriptions of BPS states
(when they exist as solutions valid in the
supergravity approximation). Assuming that
$M_4 \times X_{}$ is a true nonperturbative
vacuum of $IIB$ string theory the Hilbert space
of the theory decomposes into superselection
sectors:
\eqn\hilbspce{
\CH = \oplus_{\gamma\in \Lambda}
\CH^\gamma
}
In each sector there is a representation of the
$d=4, \CN=2$ supersymmetry algebra with central
charge. If $Q_{\alpha A}$ are supercharges then
\eqn\centchrc{
\{Q_{\alpha A}, Q_{\beta B} \} = \epsilon_{\alpha \beta} \epsilon_{AB}
Z(\Omega; \gamma) .
}
Therefore, the space of BPS states is also
graded by the lattice
$\Lambda$:
\eqn\bpsstate{
\CH_{\rm BPS} = \oplus_{\gamma\in \Lambda}
\CH_{\rm BPS}^\gamma
}
In this equation $\CH_{\rm BPS}^\gamma$ can be
the zero vector space.
We say that ``$\gamma$ supports a   BPS  state''
  if $\CH_{\rm BPS}^\gamma$ is not the
zero vector space.  In this case
the mass in Planck units, $M^2(z;\gamma)/M_{\rm Pl}^2$,
of the BPS states
is given by \bpsmass\ \cdfvp.

Not much is known  about the BPS states in
the general Calabi-Yau compactification.
We are particularly interested in mutliplicities
 $\dim \CH_{\rm BPS}^\gamma$ (and their
asymptotics for ``large'' $\gamma$), as well
as the mass-spectrum.
Moreover, it has been conjectured in \hmalg\
that there is an interesting algebraic
structure on  $\CH_{\rm BPS}$ generalizing
generalized Kac-Moody algebras. Thus,
any information we can gain on the
existence and nature of BPS states is of
interest. The attractor equations are
relevant to these questions as we discuss
in detail in the next section. The
discussion is based in part on the following
key result \fgk:

\bigskip
\ndt
{\bf Theorem 2.5.1}.

\medskip
\ndt
a.) $\vert Z(z;\gamma) \vert^2 $
has a  stationary point at $z= z_*(\gamma) \in \widetilde{\CM}$,
with fixed point value $Z_* \not=0$ iff $\hat \gamma$
has Hodge decomposition \hodgetype,
that is: $\hat \gamma = \hat \gamma^{3,0} + \hat \gamma^{0,3}$.

\medskip
\ndt
b.) If such a stationary point exists in the interior of
$\cmtld$ then
it is a local minimum of $\vert Z(\Omega;\gamma) \vert^2 $.

\bigskip
\noindent
{\it Proof:} Choose a local
holomorphic family $\Omega(s)$ for $s\in \CU\subset
\CM$ and local holomorphic coordinates $z^i$.
Straightforward computation gives:
\eqn\compi{
\p_i \vert Z \vert^2 = \bigl( \int_\gamma \chi_i \bigr) \bar Z
}
and since $\chi_i$ span $H^{2,1}(X_s) \cong T_s^{1,0}\CM$
(a) follows immediately.

Similarly, using the WPZ metric we use identities of
special geometry to compute
\eqn\compiii{
\eqalign{
D_i D_j \vert Z \vert^2 & = \p_i(\p_j \vert Z \vert^2 ) + \Gamma^k_{ij} \p_k  \vert Z \vert^2 \cr
& = \int_\gamma\biggl( \p_i \chi_j + \half \p_i K \chi_j\biggr) \bar Z + \int_\gamma \Gamma^k_{ij} \chi_k \bar Z \cr
& = \int_\gamma (D_i \chi_j) \bar Z  = i C_{ijk} g^{k \bar k} \bigl( \int_\gamma \bar
\chi_{\bar k}  \bigr) \bar Z\cr}
}
where $C_{ijk}$ are Yukawa couplings.
Now, using the stationary point condition  we see that
$\p_i\p_j \vert Z \vert^2$ vanishes. Moreover
$\p_i \pb_{\bar j} \log \vert Z \vert^2 = g_{i \bar j} $ so
the Hessian at a stationary point is positive definite.
$\spadesuit$

\bigskip
\noindent
{\bf Remark.}

\item{1.} The connection between the minimization
theorem and the flow equations for supersymmetry
is provided by spherically symmetric static
dimensional reduction of the supergravity
Lagrangian \fgk. The   real part of the bosonic Euclidean
action is then
\eqn\sqmact{
\int d \rho \biggl[ \bigl( {dU \over  d \rho} \bigr)^2
+ \parallel \nabla z \parallel^2 + e^{2U} (\hat \gamma, \hat \gamma)_{\CJ}\biggr]
}
where $\rho=1/r$ and the   last term, which comes from the electromagnetic
action, uses the symmetric bilinear form \symmform\
associated to the Weil intermediate Jacobian.
It would be nice to interpret the
system \sqmact\ in terms of a supersymmetric quantum mechanics
on the total space of the Hodge bundle
$\widetilde{\CL} \rightarrow \widetilde{\CM}$
with holomorphic
superpotential $\CW = \int_\gamma \Omega$.
\foot{We thank N. Nekrasov for some interesting discussion
on this.}

\ifig\zgrow{Three generic types of behavior of
central charges.   }
{\epsfxsize2.5in\epsfbox{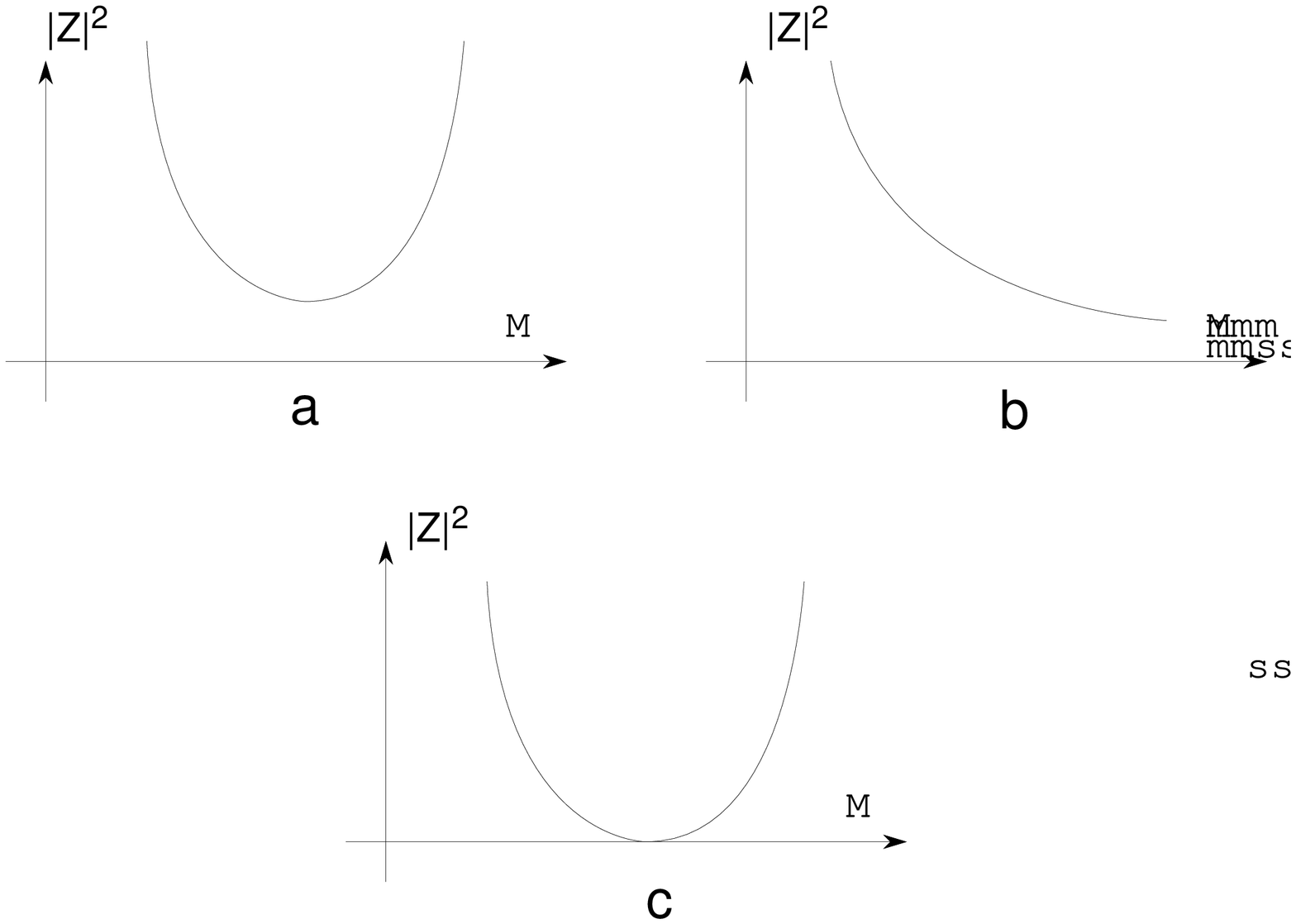}}

\subsec{Criteria for the existence of BPS states}

As mentioned above, it is of great interest
to determine which charges $\gamma$ support
BPS states. This can be answered in part using
the attractor mechanism.
Theorem 2.5.1 suggests that there
are roughly three different types of behavior
for the function $\vert Z(\Omega;\gamma) \vert^2 $
on $\widetilde{\CM}$, depending on the
nature of $\gamma$, and illustrated in \zgrow.
\foot{We are ignoring
global issues, such as the existence of multiple
minima. See sec. 9.2 below for further discussion.
Some global aspects
of BPS masses on $\CM $ have also
been discussed in
\greenekantor.}

\item{Type a:}  $\vert Z(\Omega;\gamma) \vert^2 $ has a nonvanishing local
minimum. In this case we expect to have a BPS
state in the theory.

\item{Type b:} It can also happen that
$\vert Z(\Omega;\gamma) \vert^2 $ has no stationary
point in $\widetilde{\CM}$. It might or might not
vanish at the boundary. In this case the supergravity
approximation breaks down and we cannot decide
whether $\gamma$ supports a BPS state without
further information.

\item{Type c:} For some vectors $\gamma$ it can
happen that $Z(\Omega;\gamma) =0$ for some
complex structures in the interior of $\widetilde{\CM}$.
This is a holomorphic
equation on the complex structure and hence, if nonempty,
the solution set is  a
nontrivial divisor in the interior of  $\widetilde{\CM}$.
Such charges $\gamma$ do not support BPS states.

%\DenefNB
\lref\DenefNB{
F.~Denef,
``Supergravity flows and D-brane stability,''
JHEP {\bf 0008}, 050 (2000)
[arXiv:hep-th/0005049].
%%CITATION = HEP-TH 0005049;%%
}
%\DenefXN
\lref\DenefXN{
F.~Denef, B.~Greene and M.~Raugas,
``Split attractor flows and the spectrum of BPS D-branes on the quintic,''
JHEP {\bf 0105}, 012 (2001)
[arXiv:hep-th/0101135].
%%CITATION = HEP-TH 0101135;%%
}

We now explain these criteria  for BPS states in more detail.
\foot{It turns out that it is important to include non-spherically
symmetric black holes to account for all BPS states. This is
especially relevant for criterion (c). For a discussion of these
points see \DenefNB\DenefXN.}

\subsubsec{Type a: BPS states exist}

In this case there is a complete flow with a smooth nonsingular
supergravity solution outside the horizon. Indeed near the
stationary point we may choose holomorphic
coordinates $z^i$ so that
\eqn\neara{
\vert Z(\Omega;\gamma)\vert^2 \cong \vert Z_*\vert^2(1+ g_{i\bar j} z^i \bar z^{\bar j}+\cdots)
}
where $g_{i\bar j}$ is the nonsingular WPZ metric at the stationary point
and since the charge is type a, $Z_*\not=0$. Analysis of the flow equations
\gradflow\ near the fixed point shows that this flow is rapidly
driven to the solution
\eqn\nearaii{
\eqalign{
e^{-U} & \sim 1+ \vert Z_* \vert \rho \cr
z^i & \sim k^i/\rho \cr}
}
as $\rho \rightarrow \infty$. Here $k^i$ are constants.
As in \radius\ricci\ the curvature
and its derivatives are small, the supergravity solution is
accurate, the semiclassical state exists, and hence the quantum
state exists.

\bigskip\noindent
{\bf Remark}.
As several people have noticed,
there is a close analogy between
renormalization group
flow and the attractor flow which can probably
be made precise using the ideas of
\maldacena\susskind. Regarding the evolution
$1/r = \rho \rightarrow \infty$ as a flow to the
infra-red, the above result
shows that at the fixed points there are no
relevant   directions. Indeed by \nearaii\ there
is only one ``critical exponent'' for the approach
to the fixed point. The hypermultiplets
constitute exactly marginal directions. In this
analogy $\log \vert Z\vert^2$ corresponds to a
Zamolodchikov $c$-function.

\subsubsec{Type c: BPS states do not exist}

We now consider charges such that $\int_\gamma \Omega=0$
defines a nontrivial divisor $\CD_\gamma$ in the
{\it interior} of $\widetilde{\CM}$.
If such charges $\gamma$ supported BPS
states then there would be a singularity in the
low energy Lagrangian in the interior of moduli
space. This is physically unreasonable, and hence
we do not expect a BPS state for such
vectors.

This conclusion raises the following question:  The
equations for supersymmetry \dynsys\ or \gradflow\
are ODE's, and ``ODE's always have a solution.'' One might
wonder   what goes wrong if $\gamma$ is a
charge that {\it does not} support a BPS state.

In order to investigate this we consider a flow with
initial condition such that
$z_\infty$ begins near to the divisor $\CD_\gamma$.
Working locally we assume that $\CD_\gamma$ has
transverse coordinate $z$, and we denote the
other coordinates by $\xi^\alpha$. For simplicity we
assume that
we may choose $z$ so that the Kahler potential
is of the form $K = z \bar z + \hat K(\xi, \bar \xi)$.
The central charge near $z=0$ is given by
\eqn\nearc{
Z(\Omega;\gamma)= e^{K/2} k z \sim k z
}
where $k$ is a constant.
For small $z$ the flow equations are well approximated
by
\eqn\nearcii{
\eqalign{
{dU \over d \rho} & = - e^U \vert k z \vert \cr
{d \vert z\vert^2 \over d \rho} & = 2 {dU \over d \rho}\cr}
}
the phase of $z(\rho)$ becomes constant. Thus we find
that $\vert z\vert = \sqrt{2 U + \vert z_{\infty}\vert^2}$,
so that $U$ is rapidly driven to a fixed point
$U_* = - \half \vert z_\infty\vert^2$. Similarly, $\xi^\alpha$
are driven to fixed point values.

For small values of $z_\infty$ the flow is driven to
the fixed point values at the spacetime radius
\eqn\nearciii{
\rho_* \cong \vert z_\infty/k\vert e^{-\half \vert z_\infty\vert^2}
}
Thus, the flow {\it stops} at a fixed radius $\rho = \rho_*$
where $z=0$ and $U= U_*$.
One checks that the curvature and all its derivatives
remain smooth near $\rho = \rho_*$
(e.g. $\CR^{\mu}_{\nu}\sim \vert z_\infty\vert^4$)
so the supergravity approximation remains valid.
 Technically, the flow cannot be continued
because the vectorfield defining \gradflow\ is no
longer Lipschitz. The Killing spinors solving
$\delta \lambda = \delta \psi_{A \mu}=0$ at $Z=0$
are not consistent with those obtained from
the limit of \susypars.

It might well be that charges of type c support
stable {\it nonsupersymmetric} states such as
are described in \sen. Such partially supersymmetric
spacetimes might be of considerable interest.

\subsubsec{Type b: BPS states might or might not exist}

This is the most inolved case, and actually comprises
several distinct cases since there are several
different boundaries of $\CM$, characterized in part
by the Jordan form of the monodromy matrix on the
periods. We limit the discussion to  a few illustrative
examples of what can happen. We also restrict attention
to  charges with $Z(\Omega;\gamma)\rightarrow 0$
on the boundary. The candidate BPS states for
such charges include perturbative string states and
 the BPS states analogous to the massless monopole of  Seiberg-Witten
theory \swi\swii, e.g.  the ``massless black hole''
of the conifold transition \conifold\gmsii.
Charges with no stationary point and
$Z(\Omega;\gamma)\not= 0$ are also interesting
and are probably associated with topology-changing
transitions.

Let us first consider a one-parameter family
near a LCSL point. We take the inhomogeneous
prepotential to be $\CF \sim - {\kappa \over 6}t^3$
with the boundary at $\im t \rightarrow +\infty$.
We consider central charges which vanish in
the LCSL.
(Perturbative string states are of this type.)
The central charge is then of the form:
\eqn\nearbi{
Z \sim {q_0 + q_1 t \over \sqrt{(\im t)^3}}
}
There are now two subcases, $q_1=0$ and $q_1\not=0$.
If $q_1=0$ then for $r\rightarrow 0$ (i.e., large $\rho$)
 we have
$\im t \sim k_1 r^{-1/2}$ and $e^U \sim k_2 r^{1/4}$
where $k_1, k_2$ are positive constants.
Similarly, if $q_1\not=0$ then
$\im t \sim k_1 r^{-2/9}$ while
$e^{U}\sim k_2 r^{8/9}$ as $r \rightarrow 0$.
In general, if $e^{U(r)}\sim r^\alpha$ with
$\alpha<1$ and $\alpha \not=0$ then a curvature
singularity develops. Thus, we cannot trust
the supergravity approximation in these cases.

Another boundary of $\CM$ of considerable interest is
the discriminant locus. An example of a famous
BPS state associated with a charge $Z \rightarrow 0$
on the discriminant locus is the conifold
hypermultiplet of \conifold\gmsii. Since the
internal CY metric degenerates we cannot
necessarily trust the supergravity approximation.
Nevertheless, for completeness, we include a
description of the flows near the conifold
point of the quintic analyzed in \cdgp.

Following \cdgp\ we take the
period vector $\Pi=\pmatrix{F_1 & F_2 & X^1 & X^2\cr}$
such that monodromy around the conifold point
($\psi=1$, in the notation of \cdgp)
is $F_2 \rightarrow F_2 + X^2$.
Choosing the vanishing period $z=X^2/X^1$ as a local
coordinate near the conifold point  the inhomogeneous
prepotential and K\"ahler potential take the form:
\eqn\nearbi{
\eqalign{
\CF = & {1 \over 4 \pi i} z^2 \log z + b_0 + b_1 z + b_2 z^2+\cdots\cr
e^{-K}& = \vert X^1\vert^2\biggl(-4 \im b_0
+ {1 \over 2 \pi} \vert z\vert^2 \log\vert z\vert^2
-2 (\im b_1) ( z + \bar z) + \CO(z^2) \biggr)\cr}
}
Here $b_i$ are complex constants. Since the
hypergeometric series for the periods given in
\cdgp\ actually converge at $\psi=1$ these
constants can be evaluated numerically to
give, approximately, $b_0\cong 0.263-0.012i$,
$b_1\cong -0.04 - 0.04 i$. The metric is well-approximated
by $g_{z\bar z} \cong \log(z\bar z)/(8\pi \im b_0)$.
Central charges vanishing at the conifold point
must have the form
\eqn\nearbii{
\vert Z \vert^2 \cong {\vert q_2 z\vert^2 \over 4 \vert \im b_0\vert}
\equiv e^{-K_0} \vert q_2 z\vert^2
}
where $q_2$ is the hypermultiplet charge. One again finds
that the flow reaches the fixed point at a finite value
of $\rho=\rho_*$.
If the initial condition $z_\infty$ at $\rho=0$
is small the flow is approximately given by:
\eqn\nearbiv{
\eqalign{
4\pi e^{-K_0} (U_*-U)& \sim
\vert z\vert^2 \log \vert z\vert^2 -\vert z\vert^2\cr
e^{U_*}\vert q_2\vert e^{-K_0/2}(\rho-\rho_*)&\sim
\vert z\vert \log \vert z\vert -\vert z\vert\cr}
}
The boundary condition deterines $U_*$ in terms of
$z_{\infty}$ and the second equation
in \nearbiv\ then fixes the critical
radius  to be
\eqn\nearbiii{
\rho_* \cong - {e^{-K_0/2} \over   \vert q_2\vert}
\vert z_\infty\vert\log \vert z_\infty\vert
}

The spacetime geometry near the critical radius can
be obained from the approximate solution of the
parametric equations \nearbiv
from which one finds the approximate behavior
\eqn\nearbv{
U_*-U \sim  {e^{2 U_*} q_2^2 \over 2 \pi}
{(\rho-\rho_*)^2 \over \log(\rho_*-\rho)+ \CO(\log(\vert \log(\rho_*-\rho)\vert)) }
}
for $\rho\nearrow\rho_*$. While the curvature remains
finite and well-defined, the higher covariant derivatives
of the curvature in fact diverge at $\rho=\rho_*$ so,
again, the supergravity approximation breaks down, and
one cannot decide about the existence of the BPS state.

Of course, from \conifold\gmsii\ we expect that in the
exact quantum theory there is a state for
$\vert q_2\vert =1$ and no state for $\vert q_2\vert > 1$.

\subsec{Example:  The diagonal torus}

A simple example of the three types of behavior
described above
is supplied by the case where $X$ is the ``diagonal torus.''
Let
\eqn\ellip{
E_\tau \equiv \IC/(\IZ + \tau \IZ)
}
denote the elliptic curve with modular parameter
$\tau$. We consider the 3-fold
$X_{} = E_\tau \times E_\tau \times E_\tau$.
We may choose a cycle $\gamma$ depending
on   4 integral charges
so that the central charge is:
\eqn\specchrge{
\vert Z(\Omega;\gamma)  \vert^2 = { \vert q_0 + 3 q \tau + 3 p \tau^2 - p^0 \tau^3\vert^2 \over
 8 (\Im \tau)^3}
}
The cubic in $\tau$ in the numerator has 3 roots.
Let $\Delta = 16 (e_1-e_2)^2 (e_2-e_3)^2 (e_1-e_3)^2$ be the discriminant of
this cubic. One checks that
\eqn\discr{
\CD \equiv      12 p^2 q^2 - (3 pq + p^0 q_0)^2 + 4 (p^0 q^3-q_0 p^3)= {(p^0)^4
\over  27} {\Delta\over  16}
}

Because the coefficients of the cubic are real (in fact,
integral)  we have $\CD>0$ iff there are 3 real roots.
In this case
there is a unique minimum in the upper half plane
given by the solution of the attractor equations.

When $\CD=0$ there are two coincident real roots.
It is easy to check that $ \vert Z \vert^2$ then
takes its mimimum on the boundary of the
upper half plane and vanishes there. These are
$U$-dual to perturbative string states.

When $\CD<0$ there are two complex roots. Thus
\eqn\speccplx{
\vert Z(\Omega;\gamma)  \vert^2
 = { \vert p^0 (\tau- \alpha) (\tau - \bar \alpha) (\tau- r)
\vert^2 \over  8 (\Im \tau)^3}
}
where $\alpha$ is complex, and $r$ is real. Now we see
that $Z$ vanishes at a unique point within the upper
half plane and this is the unique minimum. In
section 6.2 below  we will see that $\CD$ is just
the $E_{7,7}$ invariant $I_4$, discussed in \kalkol.
These charges do not support BPS states.
Therefore, this analysis is consistent with the results of
\CveticZQ\fm.

\subsec{Some mathematical predictions}

If we combine the discussion of section 2.6
 with the microscopic description
of BPS states as wrapped D-branes we arrive
at some interesting predictions for mathematics.

\bigskip

\item{1.} If $\hat \gamma$ is of type $(a)$   then, for
all complex structures $X_s$ in the basin of attraction of
$z_*(\gamma)$ the homology class $\gamma$
must support a supersymmetric 3-cycle in $X_s$.

\bigskip
\item{2.}  Let $\mu$ be the mirror symmetry isomorphism
of \lattisom. Define Chern classes of a sheaf on the
mirror $\CE \rightarrow \tilde X$ via the
generalized Mukai vector \hmalg:
\eqn\genmuk{
\mu(\hat \gamma) = \ch \CE \sqrt{{\rm Td}( T^{1,0} \tilde X)}
}
Given the description of BPS states advocated in, e.g.,
\bsv\hmalg,
  if the vector $\hat \gamma$ is of type $(a)$
we expect the moduli of sheaves
$\CE$ on $\tilde X$ with Chern classes
\genmuk\ to be nonempty. Indeed,
following the general reasoning of
\sv\msw\ we would expect
the Euler character to grow like
$\chi(\CM) \sim e^{\pi \vert Z_* \vert^2}$.
(See, however \vafacyi.)

{\bf Remark.} We expect that the space of
BPS vectormultiplets and hypermultiplets
 will jump in dimension
when the moduli cross real codimension
one walls: this is the phenomenon of
marginal stability. The mathematical reflection
of this on the IIA side is that the moduli space
of sheaves should jump across real codimension
one walls in the moduli of complexified Kahler
classes. It is known that the moduli space of
sheaves jumps discontinuously across walls
due to the constraint of Mumford-Takemoto
stability. Hence it is natural to suppose that
these walls coincide. In the case of marginal
stability the dimension
$\dim \CH_{\rm BPS}^{vm} -\dim \CH_{\rm BPS}^{hm}$
remains constant. Mathematically, this suggests
that although the moduli space jumps discontinuously
the Euler character does not.
\foot{We thank R. Thomas for a useful discussion
about this. For some related discussion see
\thomas\vafacyii.}

\newsec{The discriminant of a BPS state}

In this section we define a $U$-duality invariant
of charges $\gamma$ which support BPS states.  We call
this invariant  the discriminant of $\gamma$,
and denote it by $D(\gamma)$. The discriminant
 is essentially the same as  the
``topological invariants'' discussed in
 \adfii\CveticZQ\fm\fg.
In \CveticZQ\fm\fg\ the relation between
$U$-duality and the discriminant was
investigated for the duality group
$U(\IR)$ over $\IR$. The justification
for working over $\IR$ and not $\IZ$ is
  that the large charges appropriate to
supergravity solutions are in some sense
continuous. In this section we investigate some
of the finer arithmetic points that arise when one
takes into account the integral structure of the
$U(\IZ)$-duality group.

\bigskip
\ndt
{\bf Definition}: Suppose $\gamma\in H_3(X;\IZ)$ defines an attractor point $z_*(\gamma)\in \widetilde{\CM} $.
 Then
\eqn\discrdef{
\vert Z(z_*(\gamma);\gamma) \vert^2\equiv \sqrt{-D(\gamma)}
\qquad D(\gamma)\leq 0
}
is the
{\it ``discriminant  of a BPS state labelled by $\gamma$. ''}
There are attractor equations
for theories with more than 8 supersymmetries
 (see below) which determine the
complex structure (for IIB theory). So we can
also speak of the discriminant in these cases too.

Let us give some examples.

\item{1.} $IIB/X_{}$,  $\hat \gamma \in H^3(X;\IZ)$:
\eqn\expli{
D(\gamma) = - \Biggl( \int_{X_{}} \hat \gamma \wedge * \hat \gamma \Biggr)^2 = -
(\hat \gamma, \hat \gamma)_{\CJ_*}^2
}
The quadratic form is the natural one on the Weil
Jacobian.
Note that since $*^2=-1$ this scales quartically with the
charges.
\expli\  is just a rewriting of the equations
from section two.

\item{2.}  $II/K3 \times T^2$. In a similar way we have:
\eqn\kiitii{
\hat \gamma = (p,q) \in
H^3(K3 \times T^2;\IZ)   \cong II^{19,3} \oplus II^{19,3}
}
using the isomorphism \kunnethii.
More generally one can take
$\hat \gamma=(p,q)\in
\Lambda = II^{22,6}_{e} \oplus II^{22,6}_m$.
The solution of
the attractor equations  will
 be described below. In order to
write the discriminant it is useful to introduce
the matrix
\eqn\bqform{
Q_{p,q} \equiv \half
\pmatrix{ p^2 & -p\cdot q \cr - p\cdot q & q^2 \cr}
}
in terms of which one finds  \cvetic\fk:
\eqn\expldisc{
D(\gamma)  = -4 \det Q_{p,q} = (p\cdot q)^2 - p^2 q^2.
}
Sometimes we denote this quantity by $D_{p,q}$.
In the supergravity approximation one
finds the conditions $p^2>0$, $Q_{p,q} >0$
for a well-defined solution.

\item{3.}  In the FHSV model \fhsv\
  $X_{}= (S \times T^2)/G$, where
$S $ is a double-cover of an  Enriques surface and
 $G\cong \IZ/2\IZ$ acts as $(\sigma_E, -1)$
where  $\sigma_E$ is the
fixed-point free holomorphic involution on $S$.
The charge lattice is:
\eqn\fshvch{
\hat \gamma = (p,q) \in  \Lambda
= II^{10,2}_{e}(2) \oplus II^{10,2}_m(2)  \cong
H^3(X_{};\IZ)
}
where $II^{s,t}$ denotes the even unimodular
lattice of signature $(-1^s,+1^t)$, and
$II^{s,t}(a)$ means the quadratic form is multiplied
by $a$. The discriminant is now given by $D_{p,q}/4$
with $D_{p,q}$ defined in \expldisc. The
extra factor of 1/2 in $\vert Z_*\vert^2$ comes
from the $\IZ_2$ quotient.

\item{4.}  $IIB/T^6$. Using the
$E_{7,7}(\IZ)$ $U$-duality group we may take:
\eqn\torcase{
\hat \gamma\in H^3(T^6;\IZ)\subset
\Lambda
}
$\Lambda\cong \IZ^{56}$ is a module for $E_{7,7}(\IZ)$
defined by integral symplectic transformations
preserving a quartic form $I_4$ on $\Lambda$
described in \kalkol\fg\balasub. One finds:
$D(\gamma) = - I_4(\gamma)$ \kalkol.

\item{5.}  {\it 6D strings from F-theory.}
Let $\pi: X_{}\rightarrow B$ be an elliptically
fibered CY 3-fold. One can consider the
$F$-theory compactification of $IIB$ theory.
This compactification has $n_T = h^{1,1}(B)$
tensor multiplets coupling to charged
strings with charge
$\gamma     \in H^{1,1}(B)$. The $U$-duality
group is a subgroup of $Aut(H^{1,1}(B))$
presumably isomorphic to all of $O(1,n_T-1;\IZ)$. The
discriminant is  $D(\gamma) = -\gamma^2$ and
a well-defined solution requires $\gamma^2>0$
\adfl.

\subsec{  $U$-duality inequivalent black holes with
the same near horizon metric}

In the SUGRA approximation the near-horizon metric
of black holes and strings only
depends on the discriminant. For examples
\eqn\expls{
\eqalign{
D=4: \qquad
{ A(\gamma) \over  4 \pi} & = \vert Z_*\vert^2 = \sqrt{-D(\gamma)},  \cr
D=5: \qquad
{ A(\gamma) \over  2\pi^2} & =  ({Z_* \over  3})^{3/2}= \sqrt{-D(\gamma)}  \cr
D=6: \qquad
{ A(\gamma) \over  2\pi^2} &
\sim (\vert Z_* \vert  )^{3/2} = (-D(\gamma))^{3/4}  \cr}
}

While $D(\gamma)$ controls the horizon area in
the supergravity approximation we stress that it
 is defined for {\it all} charges supporting BPS
states, and
is a second $U(\IZ)$-duality invariant in addition
to the number of BPS states. The
discriminant is expected
to be related
to the asymptotics of the number of BPS states,
as in the Strominger-Vafa calculation \sv:
\eqn\stromvaf{
 \log \dim \CH_{BPS}(\gamma) =
\pi \sqrt{-D(\gamma)} + \cdots
}
This distinction raises the question of central importance
to this section.
While $D(\gamma)$ is invariant under
$U(\IZ)$, it might be that $U$-inequivalent
$\gamma$'s have the same $D(\gamma)$.
(i.e., the same near-horizon metric).
We define
\eqn\defenn{
\CN(D) \equiv \# \{ [\gamma]_U : D(\gamma) = D \}
}
and, in the remainder of this section, we examine
$\CN(D)$.

\subsec{The discriminant for $K3 \times T^2$ compactifications}

We now focus on $X_{} = K3 \times T^2$. The
$U$-duality group is  a product: $U(\IZ) = SL(2,\IZ) \times O(6,22;\IZ)$.
Suppose the Euclidean lattice $L_{p,q}\equiv \langle p,q\rangle_{\IZ}
\hookrightarrow II^{3,19}$ is primitive. There is then
a unique primitive embedding by the Nikulin embedding
theorem \nikulin.
Consequently, if $L_{p,q}$  is primitive, then
$(p',q') \sim (p,q)$ under $U$-duality
iff there exists $s\in SL(2,\IZ)$ such that:
\eqn\tranfsfm{
s Q_{p,q} s^{tr} = Q_{p',q'} .
}

\subsubsec{The class number}

An integral binary quadratic form is a matrix
\eqn\abc{
Q = \pmatrix{ a & b/2\cr b/2 & c \cr}
}
where $a,b,c\in \IZ$. We also sometimes denote it
as $(a,b,c)$. Two forms $Q$ and $Q'$ are said to be
(properly) equivalent if there is an element
$s\in SL(2,\IZ)$ such that:
\eqn\equivfrm{
s \pmatrix{a & b/2 \cr b/2 & c \cr} s^{tr}
= \pmatrix{a' & b'/2 \cr b'/2 & c' \cr}
}
The study of the
equivalence of integral binary quadratic forms
is an old and venerable problem in number theory \gauss.
We summarize a few of the main facts.
First and foremost, the number of equivalence
classes is finite and it is bigger than one with
the exception of 13 values of $D$.
If the form is primitive, i.e., if ${\rm g.c.d.}(a,b,c)=1$,
then the number of classes is denoted $h(D)$
where $D=b^2-4 a c$ is the discriminant, and is
called the class number.

In standard number theory texts (see, for examples,
\buell\borevich\ireland\cox\stark) it is shown  the set of classes $C(D)$ forms
an abelian group of order $h(D)$
which is naturally identified with
 the group of ideal classes in the quadratic
imaginary field $K_D\equiv
\IQ[i \sqrt{\vert D \vert} ]$.
%
%\eqn\quadfield{
%K_D\equiv
%\IQ[i \sqrt{\vert D \vert} ] \equiv \{ a + i b \sqrt{\vert D \vert} :
%a,b \in \IQ \}.
%}
We collect a few of the relevant definitions from the theory of
quadratic imaginary fields in appendix A.

It is convenient   to label the  classes in $C(D)$
by  points $[\tau_i ] \in \CF= \CH_1/PSL(2,\IZ)$, the
fundamental domain for the action of
$SL(2,\IZ)$ on the upper half-plane. To any
binary form we associate $\tau\in \CH_1$ via:
\eqn\uhp{
a x^2 + b xy  + cy^2 \equiv a \vert x - \tau y \vert^2 }
Thus, to a quadratic form $Q$ we associate
\eqn\tauque{
\tau_Q \equiv  {-b + \sqrt{D} \over  2 a}  \qquad \im \tau_Q>0 .
}
The $SL(2,\IZ)$ action on $Q$ becomes
the standard fractional linear action  on $\tau$.

\bigskip
{\bf Example.}
Perhaps the simplest example of a nontrivial
class group is provided by $D=-20$. There are two
inequivalent classes with reduced forms:
\foot{A form is reduced iff $\tau_Q\in\CF$. }
\eqn\explclss{
\eqalign{
\pmatrix{1 & 0 \cr 0 & 5 \cr} \qquad
&
 x^2 + 5 y^2 \qquad\qquad  \tau_1=i\sqrt{5} \cr
 \pmatrix{2 & 1\cr 1& 3\cr} \qquad
&
 2 x^2   + 2 xy + 3 y^2 \qquad \tau_2={-1 + i\sqrt{5}\over  2}  \cr}
}
Note that, since $\tau_1, \tau_2$ are both in
the fundamental domain we can immediately
conclude that these two forms are inequivalent.
In fact, the class group $C(-20)\cong \IZ/2\IZ$,
and $[\tau_1]$ is the unit so
$[\tau_2] * [\tau_2] = [\tau_1]$.

\subsubsec{Result for $\CN(D)$}

Returning to black holes in $K3 \times T^2$,
it follows that if we restrict to charges such
that $L_{p,q}$ is
primitive and  $Q_{p,q}$ primitive,
then
\eqn\clssnumi{
\CN(D) = h(D).}
More generally, it is easy to show that any form
$(a,b,c)$ may be realized as $Q_{p,q}$ for a primitive
sublattice $L_{p,q}\hookrightarrow II^{3,19}$, and
hence
\eqn\classnum{
\CN(D) = \sum_{m} h(D/m^2),
}
where the sum is over
$m$ such that $  D/m^2 = 0,1 \mod 4$ (note: $D=0,1\mod 4$).
\foot{Warning: the meaning of the adjective
``primitive'' is inequivalent
for $\hat \gamma$, $L_{p,q}$, and $Q_{p,q}$ !}

One interesting consequence of \classnum\
is that  the number of classes {\it grows} with $\vert D \vert$.
It follows from work of Brauer, Landau, and Siegel
that $\forall \epsilon>0, \exists C(\epsilon)$ with
\eqn\clssgrwth{
\CN(D) > C(\epsilon) \vert D \vert^{1/2 -\epsilon}
}
Therefore, we arrive at the result:

{\it At large entropy the number of $U$-duality
inequivalent black holes with fixed area $A$ grows
like $A$. }

\bigskip
{\bf Remark.}  The asymptotics
\clssgrwth\ can also be written
$\log h(D) \sim \log \vert D \vert^{1/2}$.
Removing the $\log$ is subtle and
depends on how $D$ approaches
infinity.
It is possible to be more explicit
for some families of charges going to infinity.
Let $1<s<t$ be a pair of   relevatively prime
integers, and consider $D=-4 s^2 t^2$.
An example of a primitive form with
discriminant $D$ is:
\eqn\explfrm{
\pmatrix{ s^2 & 0 \cr 0 & t^2 \cr}
}
Using \cox, 7.28, we can compute the class number
as a product over prime divisors:
\eqn\comptclss{
h(-4 s^2 t^2) = \half st \prod_{p \vert st} (1- {(-1)^{(p-1)/2} \over  p})
}
If, moreover, $s,t$ are primes then
$h(-4 s^2 t^2) = \half (s\pm 1) (t \pm 1)$ where
the sign is determined by the Legendre symbol:
We choose $s+1$ for $s=3 \mod 4$ etc.

\subsec{The discriminant for the FHSV model}

If we consider the $FHSV$ model
then the criterion for uniqueness in
the Nikulin embedding theorem fails.
Let $L_{p,q}$ be the rank 2 lattice spanned
by $p,q$.
There can be inequivalent primitive embeddings
of $L_{p,q} \hookrightarrow II^{10,2}$.
For example, taking the orthogonal complement
we see that
\eqn\embed{
L_{p,q}^{0,2}   \perp K^{10,0} \subset II^{10,2}
}
where $K$ is a definite lattice of discriminant $\vert D \vert$.
Inequivalent lattices $K$ lead to inequivalent
embeddings. There can also be inequivalent
choices of ``glue vectors'' for a fixed choice of
$K$ \nikulin\slg\mirandai.
\foot{We collect a few
  definitions relevant to this section in appendix B.
See also \mikhailov\ for a recent discussion of
related mathematics.}
If we take into account the different embeddings then
the  number of $U$-duality inequivalent brane
configurations with the same near-horizon geometry will
grow at least as fast as
\eqn\grwth{
\CN(D) \lsim C \vert D \vert^{9/2 }
}
where $C$ is a constant.
This is proved in appendix B.
Estimating the actual growth appears to be a
subtle problem and depends on
the arithmetic nature of $D$. From the
considerations of  appendix
B one can make a crude guess that
\eqn\grwthest{
\CN(D) \sim  C \vert D \vert^{5 }.
}

\subsec{The discriminant for $6D$ strings}

In a very similar way, we can consider
$6D$ strings from $F$-theory
compactifications. $\CN(D)$ then counts
inequivalent lightlike vectors of fixed $D=-\gamma^2$.
Again using embedding theory and the
mass formula, as in appendix B,  we can establish a lower
bound like \grwth,  and, moreover,
give an estimate like \grwthest\
  for compactifications with
$n_T$ tensor multiplets:
\eqn\grwthii{
\CN(D) \sim  \vert D \vert^{(n_T-1)/2}.
}

\newsec{A description of the attractor varieties for
$IIB/K3 \times T^2$ }

We now consider compactification of IIB theory on
$S\times T^2$ where $S$ is a K3 surface.

\subsec{The $\CN=4$ attractor equations}

In this case we face the complication that the
moduli space is
not a product $\CM_{vm} \times \CM_{hm}$,
but is given by:
\eqn\ennfour{
\CM = O(II^{22,6})\backslash {\rm Gr}_+(6,II^{22,6}\otimes \IR) \times
SL(2,\IZ)\backslash SL(2,\IR)/SO(2) .
}
By relating the $R$-symmetries to an $\CN=2$ embedding
we can locally decompose \ennfour\ into a product of
scalar manifolds for different $\CN=2$
representations. The dimensions work as follows:
$132+2 = 88_{HM} + 44_{VM} + 2_{TM} $.
In this case the attractor equations have been written in
\adf\ with the following result.
A point $\Upsilon\in {\rm Gr}_+(6,II^{22,6}\otimes \IR)$ determines  orthogonal
projections: $p= p^{22,0 } + p^{0,6} = p^L + p^R$.
The attractor equations are:
\eqn\ennfour{
p^L= p^{22,0} = 0 \qquad\qquad q^L= q^{22,0} = 0
}
and hence define an $88$-dimensional
subvariety in moduli
space:
\eqn\ennfouri{
\CV(p,q) = \{ \Upsilon\in {\rm Gr}_+(6,II^{22,6}\otimes \IR):
p,q \in \Upsilon \}  .
}

We now give a {\it geometrical interpretation}
in terms of the compactification data.
Using $U$-duality we can take $p,q\in H^2(S;\IZ)$.
Then the attractor equations impose conditions
 on the geometrical data:
\eqna\interpi
$$
\eqalignno{
\bar Z e^{K/2} \Omega &
={p^2 \over  2 \sqrt{-D} }  (q- \bar \tau p)\wedge dz
&  \interpi a \cr
B^{2,0} & = (C^{(2)})^{2,0} = (\int_{T^2} C^{(4)})^{2,0} = 0
&  \interpi b \cr}
$$
We give a proof in appendix C using the
geometrical interpretation of Narain moduli
spaces \'a la Aspinwall \& Morrison.
Note that
\eqn\slvtau{
\int_S \Omega^{0,2} \wedge \Omega^{0,2} = 0
\quad \Rightarrow\quad  p^2 \tau^2 - 2 p\cdot q \tau + q^2 =0
}
so we arrive at the important result (known before)
that the torus has complex structure:
\eqn\solvtau{
\tau=\tau(p,q) \equiv {p\cdot q +   \sqrt{D_{p,q}} \over p^2}.
}

\subsec{Attractive $K3$ surfaces}

Equation   \interpi{a}\ is simply the
solution of the
attractor equation \hodgetype\
for the complex structure of
$S \times T^2$ for $\gamma = p\oplus q$.
One can give a direct
solution to the
equations for the complex structure
by choosing a symplectic basis
$a \times \gamma^I$, $b \times \gamma_I$
for $S \times T2$, where
$\gamma^I,I=1,\dots, 22$ is an integral basis
for $H_2(S;\IZ)$ and $\gamma_I$ is a dual basis.
The attractor equations are then just the system:
\eqn\dirsolv{
\eqalign{
2 \Im \bar C \int_{a \times \gamma^I} dz \wedge \Omega^{2,0} & = 2 \Im \bar C
\int_{ \gamma^I}   \Omega^{2,0} = p^I \cr
2 \Im \bar C \int_{b \times \gamma_I} dz \wedge \Omega^{2,0} & = 2 \Im \bar C
\tau \int_{ \gamma_I}   \Omega^{2,0} = q_I \cr}
}
It is straightforward to solve \dirsolv\ and  equivalent
computations have appeared in several papers in the
literature.
The novel point in this paper is the interpretation
of the solution in terms of a
condition on the complex structure expressed through
a condition on the Neron-Severi lattice. This we
now describe.

By the Torelli theorem, the complex structure
of the $K3$ surface
is determined by $\Omega^{2,0}=C( q - \bar \tau p)$.
Recall that the
Neron-Severi lattice is the kernel
of the period map:  $NS(S) \equiv  \ker\{ \gamma \rightarrow \int_\gamma
\Omega\}$.
The transcendental lattice is the orthogonal
complement: $T_S \equiv  (NS(S))^\perp$.
Evidently, \interpi{a}\ implies that the
K3 surface  has
 $NS(S) = \langle p, q \rangle^\perp \subset H^2(K3;\IZ)$.
The lattice has    rank $  \rho(S) = 20$ and signature
$(+1, (-1)^{19})$. Equivalently, we have:
\eqn\attrkthree{
H^{2,0}(S) \oplus H^{0,2}(S) = T_S \otimes \IC
\qquad {\rm rank}(T_S)=2
}
These conditions
define what we call
{\it attractive K3 surfaces}.
\foot{In the literature they are usually called
``singular K3 surfaces,'' because their arithmetic
behavior is singularly interesting. However, we
object to this term since, as complex surfaces,
they are not singular. They are also sometimes
referred to as ``special Kummer'' surfaces, although
they are not necessarily Kummer. They are also sometimes
called ``exceptional K3 surfaces'' and they
{\it are} related to exceptional groups, but then,
isn't everything? }
Attractive K3 surfaces
are ``maximally algebraic'' and
form a dense set in the moduli space of algebraic $K3$'s.

\bigskip
{\bf Remark.} Up to an overall constant
the periods are quadratic imaginary integers:
\eqn\periods{
\int_\gamma \Omega \in \CO(K_D) \qquad \gamma\in H_2(S;\IZ) }
This will play an important role in section 10.

\subsec{The Shioda-Inose theorem}

We have now established a relation between
black holes and integral binary quadratic forms,
as well as between black holes and attractive K3
surfaces. It stands to reason that there should
be a direct relation between attractive K3
surfaces and binary quadratic forms. Indeed
there is, and it is a direct consequence of the
global Torelli theorem.

In this section we review the
results of \shiodamitani  \shioda.
It is convenient to begin by describing the
situation for exceptional abelian surfaces
$A$ defined by $\rho=\rank NS(A) = 4$.

\bigskip
\noindent {\bf Theorem 4.3.1 \shiodamitani }
There is
a 1-1 correspondence between
exceptional abelian surfaces
and  $PSL(2,\IZ)$ equivalence
classes of positive even
binary quadratic forms.

To motivate this theorem let us consider
a product of elliptic curves
\eqn\sthmi{
A = E_{\tau_1} \times E_{\tau_2}
}
Each curve has a  holomorphic 1-form:
$dz^i = dx^i + \tau_i dy^i$, $0 \leq x^i,y^i\leq 1 $.
For general $\tau_i$ the Picard lattice will
be generated by:
\eqn\sthmii{
\eqalign{
dx^1 \wedge dy^1 & = {i\over 2} {dz^1 \wedge d \bar z^1 \over
Im \tau_1} \cr
dx^2 \wedge dy^2 & = {i\over 2} {dz^2 \wedge d \bar z^2 \over
Im \tau_2} \cr}
}
and hence $\rho=2$. The generic transcendental lattice is:
\eqn\prdtrans{
T_A \cong \langle dx^1 dx^2, dy^1 dy^2 \rangle_{\IZ} \perp
\langle dx^1 dy^2, dy^1 dx^2 \rangle_{\IZ} \cong II^{1,1} \perp
II^{1,1}
}

However, if
\eqn\sthmiii{
\eqalign{
\tau_1 & = \alpha_1 + \beta_1 \sqrt{D} \cr
\tau_2 & = \alpha_2 + \beta_2 \sqrt{D} \cr}
}
with $\alpha_i, \beta_i \in \IQ$ and $D<0$
then $\rho$ jumps to $4$. Indeed, we have two new
rational $(1,1)$ forms:
\eqn\sthmiiii{
\eqalign{
Re(dz^1 \wedge d \bar z^2) & = dx^1 dx^2 + \alpha_1 dy^1 dx^2 + \alpha_2 dx^1
dy^2 + (\alpha_1 \alpha_2 - \beta_1 \beta_2 D )dy^1 dy^2 \cr
{1 \over  \sqrt{-D}} Im (dz^1 \wedge d\bar z^2)
& = \beta_1 dy^1 dx^2 - \beta_2 dx^1 dy^2 + (\beta_1 \alpha_2-\alpha_1 \beta_2)
dy^1 dy^2\cr}
}
Multiplying by suitable integers we see that
$H^{1,1}(X)$ contains a four-dimensional
sublattice of $H^2(X;\IZ)$.

In fact, \sthmiii\ is a somewhat redundant way of
parametrizing the exceptional abelian surfaces.
 Indeed, it suffices to consider the special case:
\eqn\shodii{
\tau_1 = {-b + \sqrt{D} \over  2 a} \qquad \tau_2 = {b + \sqrt{D} \over  2 }
}
with $D=b^2 - 4 a c <0 $ for some integer $c$.
Now \sthmiiii\ simplifies to
\eqn\sthmv{
\eqalign{
2 a Re(dz^1 \wedge d \bar z^2) & = 2 a dx^1 dx^2 -b
 dy^1 dx^2 + ab dx^1 dy^2 +  (2 a c - b^2) dy^1 dy^2 \cr
{2  a \over  \sqrt{-D}} Im (dz^1 \wedge d\bar z^2)
& =   dy^1 dx^2 -a dx^1 dy^2 + b dy^1 dy^2\cr}
}
while the holomorphic $(2,0)$ form $dz^1\wedge dz^2$ is given by:
\eqn\holotwo{
\eqalign{
\Omega & = (dx^1 dx^2 + b dx^1 dy^2 - c dy^1 dy^2)
+ \tau_1 (dy^1 dx^2 + a dx^1 dy^2) \cr
& = t_2 + \tau_1 t_1\cr}
}
The two integral 2-forms  in \holotwo\  are an
integral basis for the
transcendental lattice $T_A$.
The matrix of inner products determines an even
binary quadratic form:
\eqn\shiod{
T_A = \langle t_1, t_2 \rangle_{\IZ} \rightarrow
 \pmatrix{t_1^2 & t_1\cdot t_2 \cr
t_1 \cdot t_2 & t_2^2\cr} = \pmatrix{ 2a & b \cr b & 2c \cr}
}

Conversely,  given a
binary quadratic form $Q=2(a,b,c)$
there is a corresponding
abelian variety $A_Q$. Indeed, given a form we
simply construct $A$ as in \sthmi\ using \shodii.
The nontrivial fact, which follows from the Torelli
theorem is that isomorphism classes of
$A_Q$ with $\rho=4$
are in $1-1$ correspondence with $PSL(2,\IZ)$ classes
of $Q$ \shiodamitani.

Let us now turn our attention to $K3$ surfaces.
We have:

\bigskip
\noindent {\bf Theorem 4.3.2}\shioda. There is
a 1-1 correspondence between
attractive K3 surfaces
and  $PSL(2,\IZ)$ equivalence
classes of positive even binary quadratic forms.

The idea of the proof is the following:
The map from  an attractive K3 surface
$S$ to even binary quadratic
forms is straightforward:
\eqn\mapsform{
T_S = \langle t_1, t_2 \rangle_{\IZ} \quad \Rightarrow \quad Q
= \pmatrix{ t_1^2& t_1 \cdot
t_2 \cr t_1 \cdot t_2 & t_2^2 \cr}
}
where $t_1,t_2$ is any integral basis.

The map from forms to a surface $S$ requires
more thought. Consider again
$A_Q = E_{\tau_1} \times E_{\tau_2}$  with
\eqn\taus{
\tau_1 = {-b + \sqrt{D} \over  2 a} \qquad \tau_2 = {b + \sqrt{D} \over  2 }
}
One begins by forming the Kummer surface,
$\Km(A_Q)$, i.e., the smooth K3 surface resolving
$A_Q/\langle -1 \rangle$.
However, the transcendental lattice is then always
divisible by $2$, and it turns out that not all attractive
K3 surfaces are Kummer. Reference \shioda\  remedies this
by constructing a diagram:
\eqn\shiodinse{
\matrix{    &    &   A_Q  &  \cr
                  &    &      &  \cr
                  &    &  \downarrow   & \cr
                  &    &      &  \cr
           S_Q       & {\buildrel 2:1 \over  \rightarrow}    &  {\rm Km}(A_Q)
&  \cr }
}
based on a clever elliptic fibration of $\Km(A_Q)$.
It turns out that $S_Q$ is also an attractive
K3 surface and $T_{S_Q}\cong T_{A_Q}$.
(The Shioda-Inose theorem has been generalized
to surfaces with $\rho<20$ in
\morrpicard.)

\bigskip
{\bf Example}
As an example, consider the even quadratic form:
\eqn\siexpli{
Q = \pmatrix{2 & 0 \cr 0 & 2\cr}
}
$S_Q$ is  the
resolution of $E_i \times E_i$ by the
order four action $g(z_1, z_2) = (i z_1, -i z_2)$
\shioda.
It double-covers $Km[E_i \times E_i]$.
On the other hand the form
\eqn\siexplii{
Q = \pmatrix{4 & 0 \cr 0 & 4\cr}
}
corresponds to   the elliptic modular
surface of level 4,
$\pi: B(4) \rightarrow  \Gamma(4)\backslash \CH$.
 In turn, $B(4)$ is a double-cover of the
K3 surface corresponding to
\eqn\siexpliii{
Q = \pmatrix{8 & 0 \cr 0 & 8\cr}
}
which just gives  the Fermat quartic:
\eqn\fquart{
x_0^4 - x_1^4 = x_2^4 - x_3^4
}
A proof is in \psshaf, pp. 583-586.
The double-cover of $B(4)$ over
the Fermat quartic can be written very
explicitly and beautifully in terms of
theta functions essentially as the
map:
\eqn\thetamap{
(z,\tau) \mod \Gamma^*  \rightarrow \biggl( \vartheta_3(2 z \vert \tau),
\vartheta_4(2 z \vert \tau) ,
\vartheta_2(2 z \vert \tau) ,
\vartheta_1(2 z \vert \tau)  \biggr)\in \IP^3
}
Here $\Gamma^*$ is a certain discrete group
described in \mumford,
such that $(\CH \times \IC)/\Gamma^* = B(4)$.
 \fquart\ follows from the Riemann relations.
See  \mumford, pp. 53-60 for further details.

\subsec{Application of the Shioda-Inose theorem}

{}From the Shioda-Inose theorem we can draw the immediate:

\bigskip
\noindent
{\bf Corollary 4.4.1.} Suppose
$p,q\in II^{22,6}$ span a primitive rank two sublattice
 $L_{p,q} = \langle p, q \rangle_{\IZ} \subset H^2(K3;\IZ)$.
Then the attractor variety $X_{p,q}$ determined
by $\hat \gamma= (p,q)$
is
\eqn\eei{
 S_{2Q_{p,q}} \times E_{\tau(p,q)}
}
where $\tau(p,q)$ is given by \solvtau\
and $S_{2Q_{p,q}}$ is the Shioda-Inose K3 surface associated
to the even quadratic form $2 Q_{p,q}$ defined
by \bqform.

Note that the attractor variety is thus closely
related to a product of 3 elliptic curves:
\eqn\threeelip{
X_{p,q}=
S_{2 Q_{p,q}} \times E_{\tau(p,q)}
\quad {\buildrel 2:1 \over  \rightarrow}\quad   Km\Biggl( E_{\tau(p,q)} \times
E_{\tau'(p,q)}\Biggr)  \times E_{\tau(p,q)}
}
with
\eqn\auxtau{
\tau'(p,q)= {-p\cdot q + i \sqrt{-D} \over  2 }
}

{\bf Remark.} If $L_{p,q}$ is not primitive
things are more complicated.
Note that under scaling $(p,q) \rightarrow (t p, tq)$,
$\tau(p,q)$ is invariant but $\tau'(p,q)$ and $D$ change,
$\tau'(tp, tq) = t^2 \tau'(p, q)$. If $D_{p,q}/4$ is divisible
by a square then it might or might not happen that
$\langle p, q \rangle \subset T_S$ is primitive.
As a simple
example suppose
\eqn\expli{
Q_{p,q} = \pmatrix{ t_1^2 & 0 \cr 0 & t_2^2 \cr}
}
If $L_{p,q}$ is a primitive lattice with this Gram matrix
then $X_{p,q}$ is given by
\threeelip.
Suppose instead that $p = t_1 \tilde p, q = t_2 \tilde q$ with
$\tilde p^2=\tilde q^2=2$.  Then
$X_{p,q} = X_{\tilde p , \tilde q}  = S_{2 \tilde Q}\times E_{\tau(p,q)}$
with
\eqn\explii{
\tilde Q = \pmatrix{ 1 & 0 \cr 0 & 1 \cr}.
}
Using \shioda\ Corollary p.129, we can say
that $X_{p,q}$ and $X_{\tilde p,\tilde q}$
are based on isogenous exceptional abelian varieties
and are related by finite degree rational maps.

\newsec{BPS mass spectrum in the FHSV model at attractor
points}

The special K\"ahler geometry of the FHSV
model \fhsv\ is uncorrected by worldsheet
instantons and is therefore the homogeneous
metric on
\eqn\fhsvmod{
SL(2,\IR)/SO(2)
\times SO(10,2)/SO(10)\times SO(2).
}
Thus, the solution of the attractor equations
is very similar to the solution for the complex
structures in the $\CN=4$ case described in the
previous section.
The novelty in studying the FHSV  model is
that we can also examine the BPS mass spectrum
exactly. In the $\CN=4$ case the BPS
mass spectrum depends on the hypermultiplet
moduli. In the FHSV model, there is only
$\CN=2$ supersymmetry, and the spectrum is
a function only of the vectormultiplet moduli.

Suppose
the charge vector $(p_0,q_0)$
defines an attractor point $z_*(p_0,q_0)$ in moduli
space.  Let us consider a background with constant moduli
fixed at this point. We may then   ask about the central
charges and mass spectrum of all the BPS
states at that point. The central charges in the
sector $\hat \gamma = (p,q)$ are given
by
\eqn\cenchrges{
\eqalign{
Z_0(z_*(p_0,q_0);p,q)
& = {p_0^2 \over  \sqrt{2} \vert D_0\vert^{3/4}} (q-\tau_0p )\cdot (q_0-\bar
\tau_0 p_0) \cr
 = {1 \over   \sqrt{2} \vert D_0\vert^{3/4}} &
\biggl[ p_0^2 q_0\cdot q + q_0^2 p_0\cdot p
- p_0\cdot q_0 \bigl(q_0\cdot p + p_0\cdot q\bigr)
+ i \sqrt{\vert D_0 \vert } \bigl( p_0\cdot q - q_0 \cdot p \bigr)
\biggr]\cr}
}
where $D_0 = D_{p_0,q_0}$.
Thus, up to a simple overall factor, all the central
charges are in the ring of integers $\CO_{K_{D_0}} $ of $K_{D_0}$.
For general points in moduli space the
spectrum of masses is a complicated set of
real numbers. However, from \cenchrges\ it
follows that:  {\it at an attractor point,
the set   $\{ 2 \vert D_0\vert^{3/2} M_\alpha^2\}$,
where $M_\alpha$ are BPS masses,
is a set of rational integers.}

In fact, more is true. The integers
represented by $f(x,y) = a x^2 + b xy + c y^2$
are just the norms of the ideals in the corresponding
ideal class in $\CO(D)$. These norms
can be written as $a \vert x - \tau y \vert^2$ for
$x,y$ integral.
To the quadratic form $Q_{p_0,q_0}$ we
may associate an  ideal:
\eqn\coridel{
\underline{\bf a}_{p_0,q_0}  \equiv
\half p_0^2 \IZ + {p_0 \cdot q_0 + \sqrt{D_0} \over  2} \IZ
}
which has norm $\half p_0^2$. So now we write the
mass spectrum as:
\eqn\writemass{
p_0^2 (-D_0)^{3/2} \vert Z(z_*(p_0,q_0);p,q)\vert^2
= \half p_0^2 \vert A - \tau_0 B \vert^2
}
where
\eqn\asbes{
\eqalign{
A & = p_0^2 q\cdot q_0 + q_0^2 p\cdot p_0 - 2 p_0\cdot q_0 p_0 \cdot q \cr
B& = p_0^2 (q_0 \cdot p - p_0 \cdot q) \cr}
}
are rational
integers. Hence \writemass\ are the norms of ideals in the ideal class of
\coridel.  In particular, the BPS
square-masses at the
attractor point $p_0,q_0$, scaled by
$p_0^2 \vert D_0\vert^{3/2}$,  are the norms of certain
ideals in the ideal class corresponding to
$Q_{p_0,q_0}$. In particular the BPS mass-squared spectrum
for $U$-inequivalent attractor points
$z_*(p_0,q_0)$ and $z_*(p_0',q_0')$ with
$D_{p_0,q_0}=D_{p_0',q_0'}$ are different
sets of integers.

\ifig\bpsmass{The BPS mass spectrum in the
FHSV model for attractor points defined by
the discriminant $D=-20$.
There are two such points corresponding to
$\tau_1 = i \sqrt{5}$, $\tau_2 = (1+ i \sqrt{5})/2$.   }
{\epsfxsize2.5in\epsfbox{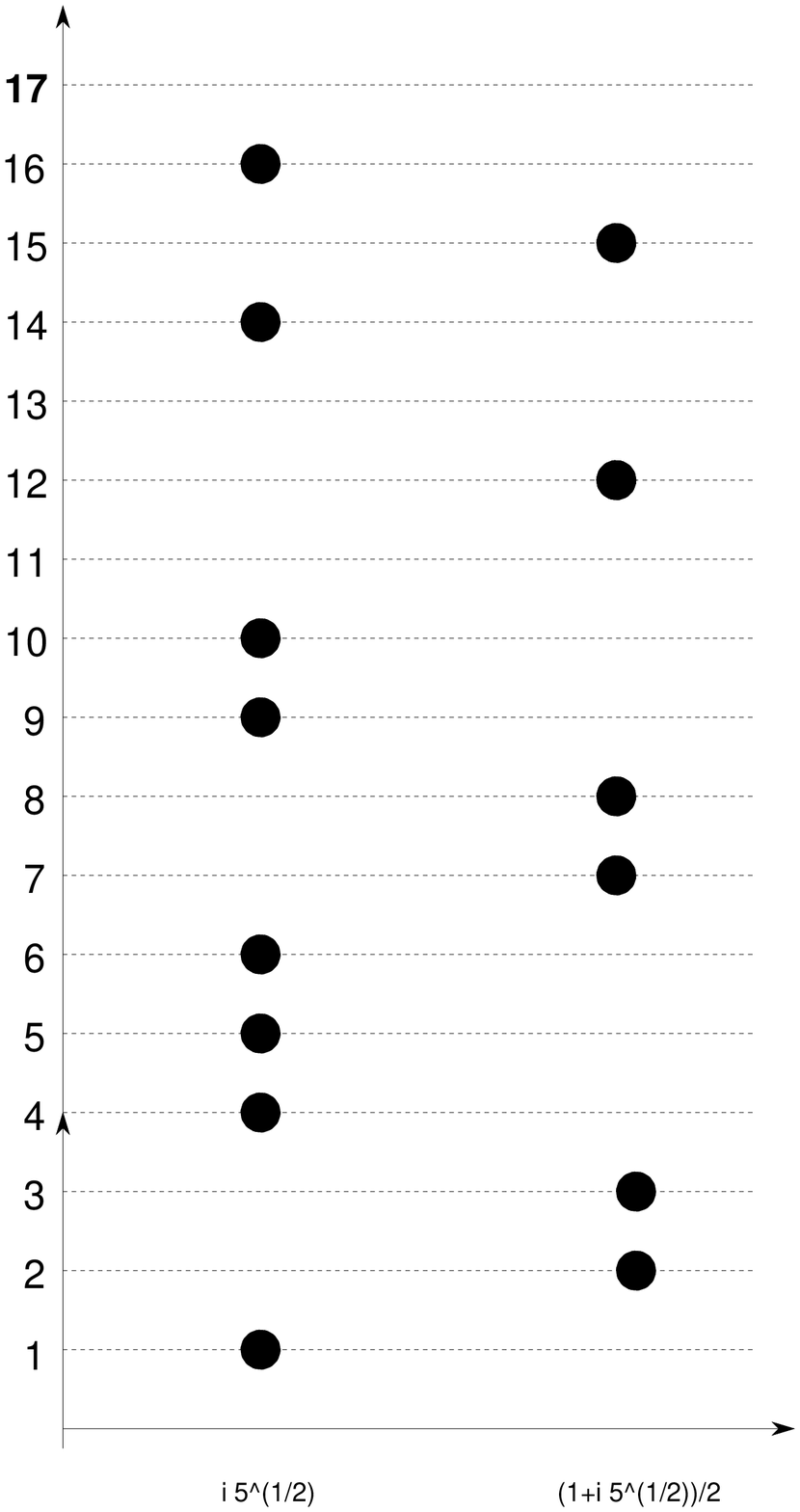}}

\bigskip
\noindent
{\bf Remarks.}

\item{1.} It is interesting to ask
{\it which} integers $n$
actually do occur in the BPS mass spectrum
at an attractor point.
If the ideal ${\bf a}$ has prime
decomposition ${\bf a}= \prod_i {\bf p_i}^{r_i}$
then $N({\bf a})= \prod_i N({\bf p_i})^{r_i}$.
Thus, the question of which $n$'s are BPS
mass-squares
boils down to questions about the prime factorization
of $n$ and how the primes of $\IQ$ split in $K_D$.
These are exactly the kinds of questions number theory
was designed to answer! For example, consider
our example \explclss\ with  $D=-20$.
The primes $p=2,5$ are ramified:
\eqn\rmfed{
\eqalign{
-5 & = (\sqrt{-5})^2 \cr
2 & = (2, -1 + i \sqrt{5})^2\cr}
}
and hence are norms of ideals.
Primes $p=1,9\mod 20$ are represented by the
ideal class of $[\tau_1]$ while primes $p=3,7\mod 20$
are represented by the ideal class of $[\tau_2]$.
The primes $p=11,13,17,19\mod 20$ do not split
and are not represented by either class.
Thus, if an integer $n$ contains such a
prime to an odd power in its factorization in $\IQ$,
then it does not occur in the spectrum. All this is
illustrated in \bpsmass.

\item{2.} An immediate consequence of our
result is that  mass-generating functions for
the BPS states
(such as the ``topological free energy'' of
\fklz )
are natural generalizations of
 $L$-functions
and $\zeta$-functions of the associated number
field $K_D$. However, because of the issue of degeneracies
of different BPS states the relation is not straightforward.
We hope to address it elsewhere.

\newsec{A description of the attractor varieties for $IIB/T^6$}

\subsec{Analysis of the attractor equations}

As in the case of $\CN=4$ compactifications,
 one must decompose the moduli space
into the space of vectormultiplets and hypermultiplets
using an $\CN=2$ embedding \adfesev.
Again, we must give a geometrical interpretation
to the resulting conditions.
Choose $\hat \gamma \in H^3(T^6;\IZ)$.
Then the  complex structure satisfies the usual
equation:
\eqn\attreqstor{
Im(2\bar C\Omega^{3,0}) = \hat \gamma \in H^3(X;\IZ).
}
As we will see presently, \attreqstor\ uniquely determines
the complex structure.
In this complex structure,
$B, C^{(2)} ,   *_6 C^{(4)} $ are of type $(1,1)$.
These conditions fix all the vectormultiplet moduli.

In order to describe the complex structure determined
by the attractor equations we proceed as follows.
We can choose analytic coordinates for
the complex torus $\IC^3/\Lambda$ such that
the holomorphic 1-forms are defined by
$dz^i = dx^i + \tau^{ij} dy^j$ where $\tau$ is
the period matrix, $0\leq x^i, y^i \leq 1$ are defined
$\mod 1$, and $i=1,2,3$.
We choose the gauge
$\Omega^{3,0}=   dz^1 dz^2 dz^3$,
 an orientation
$\int dx^1 dy^1 dx^2 dy^2 dx^3 dy^3=+1$, and
the following symplectic splitting of
$H^3(X;\IZ)$:
\eqn\difforms{
\eqalign{
\hat \alpha_0 & = dx^1 dx^2 dx^3\cr
\hat \alpha_{i,j} & = \half \epsilon_{i l m}
dx^l dx^m dy^j \qquad\qquad 1\leq i,j\leq 3 \cr
\hat \beta^{i,j} & = \half \epsilon_{jlm} dx^i dy^l dy^m \qquad\qquad 1\leq
i,j\leq 3 \cr
\hat \beta^0 & = - dy^1 dy^2 dy^3\cr}
}
so that $\int \hat \alpha_I \wedge \hat \beta^J = + \delta_I^{~ J}$.
One then checks that
\eqn\expdome{
\Omega = \hat \alpha_0 + \hat \alpha_{ij} \tau^{ij} + \hat \beta^{ij}
(\Cof \tau)_{ij} - \hat \beta^0 (\det \tau)
}
where  $\Cof(A)$ is the cofactor matrix,
$\Cof(A)=(\det A) \cdot A^{-1,tr}$.
The   inhomogeneous prepotential is purely cubic
 $\CF = - \det \tau$.

Decomposing the charge vector $\gamma$ with respect to the basis
\difforms\ the central charge can be written as:
\eqn\centrlchrg{
Z(\Omega;\gamma) = e^{K/2}  \biggl[ q_0 + Q_{ij} \tau^{ij} + P^{ij} (\Cof
\tau)_{ij} - p^0 \det \tau\biggr].
}
In this basis the
 equation $\Im 2 \bar C \Omega  = p^I \hat \alpha_I - q_I \hat \beta^I $
 becomes the system:
\eqn\torieqs{
\eqalign{
\Im 2 \bar C & = p^0  \cr
\Im (2\bar C \tau^{ij}) & = P^{ij}   \qquad 1\leq i,j\leq 3 \cr
\Im (2\bar C \Cof(\tau)_{ij} ) & = - Q_{ij}  \qquad 1\leq i,j\leq 3\cr
\Im (2 \bar C \det \tau) & =q_0 \cr}
}

The general solution of the system \torieqs\
is the following.
Define:
\eqn\mtrxdfs{
\eqalign{
R  & \equiv \Cof P + p^0 Q \cr
\CM& \equiv 2 \det P+ p^0(p^0 q_0 + \tr(PQ) )  \cr
\CD & = 2\bigl[ (\tr PQ)^2 - \tr (PQ)^2\bigr] - (p^0 q_0 + \tr PQ)^2 +
4\bigl[ p^0 \det Q - q_0 \det P\bigr] \cr}
}
Then a solution of the equations
\torieqs\ exists for $\det R \not=0$, $\CD>0$.

If, in addition, we want the torus $\IC^3/\Lambda$
to be a principally polarized Abelian variety,
then we require $\tau\in \CH_3$. For this to be the
case we need:
\eqn\condits{
\eqalign{
P = P^{tr} \qquad  & \qquad Q = Q^{tr} \cr
 R > 0 \qquad & \qquad \CD>0  \cr}
}
The solution is then:
\eqn\treqsiv{
\eqalign{
\tau & = \Biggl[
  \biggl( 2PQ -  (p^0 q_0 + \tr(PQ))\cdot 1  \biggr)
+ i \half \sqrt{\CD} \Biggr] \cdot (2 R)^{-1} \cr
2\bar C & = {\CM \over  \sqrt{\CD}} + i p^0 \cr}
}

In our conventions
conventions the extremal mass and entropy
are given by:
\eqn\extrmlmss{
 {S \over  \pi}= M^2   = 8 \vert \bar C \vert^2 \cdot \det \Im \tau
}
In particular, using the solution \treqsiv\
and the identity
$(p^0)^2 \CD  = (4 \det R - \CM^2) $ we find
\eqn\torientr{
S/\pi = M^2 = \sqrt{\CD} .
}

\subsubsec{Proof of $(6.8)$ }

Solving \torieqs\ is straightforward, and is
in fact a special case of Schmakova's calculation
reviewed in
section 9.1 below. It is a little more explicit
in this case so we present it here.
Let  $2 \bar C = \xi^0 + i p^0$ where
$\xi^0$ is real.  Assume
$p^0\not=0$. (The case $p^0=0$ can be obtained
by taking a limit. All the expressions above
are valid for both cases.)
Let $\tau= X + i Y$ be the real and imaginary
parts of $\tau$.
The second equation has general solution:
\eqn\secndeq{
\tau = {1 \over  p^0} (P - 2 C Y )
}
as long as $P^{tr}=P$ is symmetric.
For a nondegenerate torus $\det Y \not=0$, so
the third equation implies:
\eqn\thrdeq{
\Cof(Y) = {1 \over  (\xi^0)^2+ (p^0)^2} (\Cof P + p^0 Q)
}
and therefore, if we define
\eqn\definearr{
R  \equiv \Cof P + p^0 Q
}
then a necessary condition for a nondegenerate
solution is $\det R \not=0$, and for a polarized
Abelian variety $R= R^{tr} >0$. Finally, the
fourth equation determines $\xi^0 = \CM/\sqrt{\CD}$
and requires $\CD>0$. To do the calculation it
is helpful to note that for any $3\times 3$ matrix
$S$:
\eqn\ideness{
\det(1+S) = 1 + \tr S + \half ((\tr S)^2- \tr S^2) + \det S
}

\subsec{Relation to $E_{7,7}$ invariants}

The expression for $\CD$ in \mtrxdfs\
can be identified with the quartic
$E_7$ invariant on the charges formed from two $28$'s
$X^{IJ}$, $Y_{IJ}$ of $SL(8,\IR) \hookrightarrow Sp(56;\IR)$.
The $E_7$ invariant is defined by \cj\kalkol\balasub:
\eqn\esevinvt{
-I_4 = \Tr(XY)^2 - { 1 \over  4} (\Tr(XY))^2 + 4 Pfaff(X) + 4 Pfaff(Y)
}
We identify the charges with the $28$ via:
\eqn\chrgmatch{
\eqalign{
X^{IJ} & =
\pmatrix{ 0 & P & 0 & 0 \cr
- P^{tr}& 0 & 0 & 0 \cr
0 & 0 &0 & q_0 \cr
0  & 0 & - q_0 & 0 \cr} \cr
Y_{IJ} & =
\pmatrix{ 0 & -Q^{tr} & 0 & 0 \cr
Q& 0 & 0 & 0 \cr
0 & 0 &0 & p^0 \cr
0  & 0 & - p^0 & 0 \cr} \cr}
}
and a short calculation shows that $\CD = I_4(\gamma)$.

\subsec{Isogenies and CM type}

Some useful reference material for this
section is
\gordon\birklang\shimtan\langii\gshimura.
Let $Z$ be a complex torus, that is,
$Z=\IC^g/\Lambda$ where $\Lambda$ is a
rank $2g$ $\IZ$-module in $\IC^g$.

\bigskip
\noindent
{\bf Definition}: A homomorphism of
complex tori  is a holomorphic map
$Z \rightarrow Z'$ which is a homomorphism of
groups.  A homomorphism $Z \rightarrow Z$ is
an endomorphism.

Concretely, if  $Z=\IC^n/( \IZ^n + \tau \IZ^n)$,
$Z'=\IC^m/( \IZ^m + \tau' \IZ^m)$
we have $z \rightarrow w = M \cdot  z$  where $M\in Mat_{n\times m}(\IC)$ is
such that there is a matrix $\rho(M)\in Mat_{2m \times 2m}(\IZ)$
with
\eqn\endomorph{
M \pmatrix{1 & \tau \cr} = \pmatrix{1 & \tau' \cr} \rho(M)
}
For $\tau=\tau'$
$\End(Z)$ is the algebra over $\IZ$ of such matrices $M$.
%
%We also let $\End_0(Z)\equiv \End(Z) \otimes_{\IZ} \IQ$.
%If $Z$ is an
%abelian variety then $\End_0(Z)$ is a division
%algebra, and the possible division algebras which
%occur have been classified \birklang\gordon.
%

\bigskip
\noindent
{\bf Definition}: A homomorphism
 of abelian  varieties
$\phi: A \rightarrow B$   is
an {\it isogeny} if it is surjective and
has a  finite kernel. The order of the
kernel is the degree of the isogeny.

Concretely, if $A=\IC^n/\Lambda$ and
$B=\IC^n/\Lambda'$ then
$M\cdot \Lambda \subset \Lambda'$ is a
sublattice of finite index.
A useful standard example is obtained
by supposing that $\Lambda'\subset \Lambda$
is a sublattice of finite index. Then the map
\eqn\isogeny{
\eqalign{
A'= \IC^g/\Lambda '  & \rightarrow A =
 \IC^g/\Lambda \cr
z\  \mod \Lambda' & \mapsto z\  \mod \Lambda \cr}
}
is an isogeny. It is of degree $[\Lambda:\Lambda']$.
Note that if $\phi: A \rightarrow B$ is an isogeny
then there is an isogeny $\phi': B \rightarrow A$
so that $ \phi' \phi: z \rightarrow n z$ where
$n = \deg \phi$.

The result \treqsiv\ shows that the attractor
variety is an abelian 3-fold of CM type in the
sense of \shimtan\langii\gshimura. However it is
of a very special type.
\foot{ In the language of
\gshimura\langii\ the ``reflex field'' is quadratic
imaginary.}
Indeed
the attractor variety $\IC^3/(\IZ^3 + \tau \IZ^3)$
is isogenous to
a product of three elliptic curves
$E_\gamma  \times E_\gamma\times E_\gamma$,
where $E_\gamma \equiv E_{\tau(\gamma)}$ for
$\tau(\gamma)=   i \sqrt{I_4(\gamma) } $. This is
most easily seen from \treqsiv\ which gives
the covering:
\eqn\isogdiag{
\eqalign{
\phi: E_\gamma \times E_\gamma \times E_\gamma
 = \IC^3/\Lambda' & \rightarrow \IC^3/\Lambda \cr
\Lambda' \equiv \IZ^3 \oplus \tau\cdot (2R)\IZ^3
&\subset \Lambda = \IZ^3 \oplus \tau\cdot \IZ^3 \cr}
}
In this was
we arrive at:

\bigskip
\noindent
{\bf Proposition 6.3.1.} The $\CN=8$ attractor as a complex
variety is a polarized abelian variety isogenous to
$E_{\tau(\gamma)} \times E_{\tau(\gamma)} \times E_{\tau(\gamma)} $
with $\tau(\gamma) = i \sqrt{I_4(\gamma)}$
by an isogeny of degree    $8 \det R$.

\newsec{$\CN=4,8$ Attractors are Arithmetic}

\subsec{Complex multiplication}

The elliptic curves occuring in   sections 4 and 6
have ``complex multiplication.''
The theory of complex multiplication is a popular
subject with an extensive
literature.  Some useful references
include
\cox\shimtan\langii\gshimura\serre\vladut\weil\silveradvtop.

An elliptic curve $E=\IC/(\IZ + \tau \IZ)$
is a group, and as in the previous section  we may study
its {\it endomorphisms}, that is, the homomorphisms
$E \rightarrow E$. These must be of the form
$z \rightarrow \lambda z$. On the other hand, for this
to be well-defined we require
$\lambda (\IZ + \tau \IZ) \subset \IZ + \tau \IZ$.
This condition can be written as:
\eqn\cmplxi{
\eqalign{
\lambda \pmatrix{1 &  \tau\cr}& =\pmatrix{1 &  \tau\cr}
\pmatrix{ N & -C \cr A & M \cr}   \cr
& = \pmatrix{1 &  \tau\cr}\cdot \rho(\lambda)^{tr}  \cr}
}
where $N,A,C, M \in \IZ$ define a matrix
$\rho(\lambda)$. All elliptic curves have such
endomorphisms for $\lambda\in \IZ$. However,
curves which admit a larger endomorphism algebra are
special, and referred to as curves with complex multiplication.
We can easily determine these curves as follows.
Since $\lambda$ is an eigenvalue in \cmplxi\ it follows that
$\lambda$ is a quadratic imaginary integer:
\eqn\cmplxii{
\lambda = {N+M \pm \sqrt{(N-M)^2 - 4 AC} \over  2}
}
and $\tau$ is determined to be:
\eqn\cmplxiii{
\tau = {M-N +  \sqrt{(N-M)^2 - 4 AC} \over  2A}
={-b  +  \sqrt{b^2 - 4 a c} \over  2a}
}
where we removed the greatest common factor $\ell$:
$(A, N-M, C) = \ell(a, b, c) $. Let   $D\equiv b^2 - 4 ac<0$
and define:
\eqn\omeg{
\omega \equiv {D + \sqrt{D} \over  2}.
}
Then we find that
\eqn\cmplxiii{
\lambda = M + \ell( 2 a c - \half b (b-1)) + \ell \omega
}
Hence the ring of endomorphisms of
$E_\tau$ is the order $\CO(D) = \IZ + \IZ \omega$.
Multiplication by $n_1 + n_2 \omega$ is represented by :
\eqn\multom{n_1 {\bf 1} + n_2
\pmatrix{ \half(D+b) & a \cr -c & \half(D-b) \cr}
}

\subsubsec{Special values of $j(\tau)$}

We   now quote the first main theorem
of complex multiplication. Details can
be found in the references. Part $(iii)$ below
refers to ``class field theory.''
A readable account
of class field theory, with references to more
rigorous treatments,  can be found in
\cox\stark.

\bigskip
\noindent
{\bf Theorem 7.1.1 } Suppose $\tau$ satisfies the quadratic
equation $a \tau^2 + b \tau + c=0$ with $g.c.d.(a,b,c) = 1$.
Let $D$ be the discriminant of the associated
primitive quadratic form. Then,

i.) $j(\tau)$ is an algebraic integer of degree
$h(D)$.

ii.) If $[\tau_i]$ correspond to the distinct
ideal classes in the order $\CO(D)$ via
the map \tauque, the minimal polynomial of $j(\tau_i) $ is
\eqn\clsseqtn{
p(x) = \prod_{k=1}^{h(D)} (x-j(\tau_k)) \in \IZ[x]
}

iii.) $\widehat{K_D} \equiv
 K_D(j(\tau_i))$ is Galois over $K_D$ and
is independent of $i=1,\dots, h(D)$. In fact,
$\widehat{K_D}$ is
the ring class field of the order $\CO(D)$ in the
ring of integers $\CO_{K_D}$ in $K_D$.

\bigskip
\noindent
{\bf Example.} Continuing our example from
\explclss\  we illustrate parts $(i),(ii)$
of the theorem with:
\eqn\explii{
\eqalign{
\pmatrix{1 & 0 \cr 0 & 5 \cr}
\qquad \leftrightarrow \qquad\quad\quad
j(i \sqrt{5}) & = (50 + 26 \sqrt{5})^3 \cr
& \cr
\pmatrix{2 & 1\cr 1& 3\cr} \qquad
\leftrightarrow\qquad
 j({1+ i \sqrt{5} \over 2} )
& = (50 - 26 \sqrt{5})^3 \cr}}
and the minimal polynomial is:
\eqn\explpoly{
p(x)   =  x^2 -    1264000\ x  -681472000
}
The reader should be warned that the
algebraic numbers involved get ``very complicated''
``very fast.'' For example, using the modular equation
it is not hard to show that $j(6i)$ is given by:
\eqn\jaysixi{
\eqalign{
5894625992142600 + 3403263903336192\,{\sqrt{3}} +\qquad\qquad \cr
  2352\,{\sqrt{2\,\left( 6281131340524109220108468 +
         3626412870266989391644263\,{\sqrt{3}} \right) }}\cr}
}

\bigskip
\noindent
{\bf Remark.}
There is a curious ``converse'' known as
Schneider's theorem: if $\tau$ is algebraic and
{\it not} quadratic imaginary, then $j(\tau)$ is
transcendental! For this theorem and modern
generalizations see \pbcohen\shiga,
and references therein.

\subsec{Arithmetic varieties}

When an elliptic curve has complex multiplication  the
corresponding elliptic  curve  has very special arithmetic properties. Usually
we write
$$
E_\tau \cong \{ (x,y): y^2 = 4 x^3 - g_2 x - g_3 \}
$$
with the map given by the Weierstrass function
$(x,y)=(\wp(z,\tau), \wp'(z,\tau))$.
However,  for arithmetic questions one must be careful
to choose an appropriate lattice in the
homothety class of $\IZ + \tau \IZ$.
 Assume for simplicity that
$j \not=0, (12)^3$ and let $c=27j/(j-1728)$.
Then the curve
\eqn\weiermodel{
\eqalign{
y^2 & = 4 x^3 - c(x+1) \cr
c& = {27 j \over j- (12)^3} \cr}
}
has invariant $j$. We can always achieve this form
by an appropriate rescaling $x \rightarrow \lambda^2 x,
y \rightarrow \lambda^3 y$.
With this rescaling undersood it follows from
the first main theorem Theorem 7.1.1 above, that
there is an arithmetic Weierstrass
model for $E_\tau$  defined
over $\widehat{K_D} \equiv K_D(j(\tau_i))$.
This fact is the key ingredient in establishing that
the attractor varieties are arithmetic.

In order to show that the $\CN=4$ attractor
varieties are arithmetic we must consider the
 two steps in the diagram \shiodinse\
used to construct
 the Shioda-Inose surface $S_Q$. First we must
take a quotient by $\IZ_2$. Then we must
take a branched double cover.

Quite generally,
if $X$ is defined over a field $K$ and a finite group $G$
acts on $X$ then the quotient $X/G$ is defined over
$K$.  An argument for this proceeds along the following
lines.
\foot{The statement is obvious to all experts.
We sketch the proof for the benefit of
amateurs, like the author.}
Suppose an ample line bundle
$\CL \rightarrow X$ embeds $X$
as a hypersurface in  projective space:
\eqn\embed{
p \rightarrow [(Z^0,\dots, Z^N)]=[(\sigma_0(p), \dots, \sigma_N(p))]
}
Here $\sigma_\alpha(p)$ is a basis of sections
of $H^0(X;\CL)$. These sections satisfy some polynomial
relations $W_\alpha(Z^i) =0$ where $W_\alpha$ are
defined over $K$.   We may
assume that   $G$ acts on
$X$ so that the action in projective coordinates
$Z^i$ is arithmetic:
\eqn\groupact{
g\cdot p \rightarrow \rho(g)\cdot [(\sigma_0(p), \dots, \sigma_{N-1} (p))].
}
where $\rho(g)$ is in
$PGL(N,K)$,
and   the $W_\alpha$ form a representation
of $G$.
The crux of the matter is to show that
it is possible to find a subspace
$\CW\subset H^0(X; \CL^{\otimes m})^G
\subset H^0(X; \CL^{\otimes m})$
for some $m$ such that $\CW$
is spanned by invariant sections $\Sigma_\alpha(p)$:
\eqn\bigembd{
[(\Sigma_0(g \cdot p) , \dots, \Sigma_{N'}(g \cdot p))]
=
[(\Sigma_0(  p) , \dots, \Sigma_{N'}(  p))]
}
and such that the linear system $\CW$  maps $X/G$
to $\IP^{N'}$ as an {\it embedding.}
This can be proved using
\harris, pp. 124-129.
In our case we conclude that
$Km(A_Q)$ is defined over $\widehat{K_D}$.

In order to cover the case of attractive
K3 surfaces which are not Kummer we
 now must take the branched cover
$\pi: S_Q \rightarrow Km(A_Q)$. At the very
least the field of definition of $\pi$ will
involve the field of definition of the torsion points.
We believe this field extension is sufficient
and that a proof can be given using the
general results of \weilii, but we have
not checked any details. Experts assure us
this field is indeed sufficiently large, so we
leave it at that, for now.

Since we do need to extend the field of definition
to include that of the torsion points in
$A_Q$ we must now invoke the {\it second main theorem
of complex multiplication}, which we state
in the following form:
\foot{See, e.g., \cox, Theorem 11.39, or
\silveradvtop\ for the real thing. One will wonder
what happens if one includes values of the $y$
coordinates. This gives abelian extensions of
the classfield $K_D(j(\tau))$ itself,
 but in general the extension of
$K_D$ is not abelian \silveradvtop, as
described in Theorem 2.3 and Example 5.8. }

\bigskip
\noindent
{\bf Theorem 7.2.1 }

a.) The torsion points $(x,y)_{a,b,N}$ on the
curve \weiermodel\ corresponding
to $z= {a + b \tau \over  N}$ are arithmetic.
The values of the $x$-coordinate, \foot{More precisely, of the Weber function.}
$x({a + b \tau \over  N})$,
generate finite abelian extensions of $K_D$
\eqn\torspots{
\widehat{K}_{a,b,N,D} = K_D\bigl( j(\tau) , x_{a,b,N} \bigr)
}

b.) The fields $\widehat{K}_{a,b,N,D}$ are ray class fields
for $K_D$.  Moreover,
all finite abelian extensions of $K_D$
are subfields of some $\widehat{K}_{a,b,N,D}$.

\bigskip

A corollary of the above discussion and theorem
7.2.1 is that the $\CN=4$ attractor varieties
$S_{2 Q_{p,q}}\times E_{\tau(p,q)}$ are
arithmetic. If the attractive K3 surface is
 Kummer then the variety is defined over a finite extension of
$K_{D_{p,q}}$ related to the  ray
 class fields of $K_{D_{p,q}}$.

Similar statements hold for the
  case of the $\CN=8$ attractor.
Firstly, since the attractor is an abelian
variety of CM type it follows from
very general arguments that it is
defined over a number field. (See
\shimtan\ Prop. 26, sec. 12.4 for the
rather abstract argument.)
In our case we can be
more specific since the variety is
isogenous to
$E_\gamma \times E_\gamma \times E_\gamma$,
where
$E_\gamma$ is defined over $\widehat{K}_D$ for
$D=-I_4(\gamma)$.
Isogenous varieties will in general be defined
over different fields. For an elliptic curve
the relation between the fields can be
deduced from the modular equation
$\Phi_N(j(\tau), j(N\tau))=0$. In general,
the isogeny defines a collection of points
of order $N$ (where $N$ is the degree of the
isogeny), so the field of definition of the
attractor variety will be related to the field
of definition, $\tilde K$, of  all
the points of order $N$ of
$E_\gamma \times E_\gamma \times E_\gamma$.
We believe  the attractor variety is
defined over $\tilde K$, and no
further extension is required, but we have not
proved this.

\newsec{Attractors for general CY 3-folds}

In this section we make some remarks on general
aspects of the attractor equations for CY
3-folds, and study some examples.

\subsec{Attractor points of rank one and rank two}

Suppose that two charges
$\gamma_1, \gamma_2\in H_3(X;\IZ)$
have a common attractor point $X_*$ with
holomorphic $(3,0)$ form $\Omega$. Then
there are constants $C_1, C_2$ so that
\eqn\doublattr{
\eqalign{
 \im (2 \bar C_1 \Omega) & =  \hat \gamma_1 \cr
 \im (2 \bar C_2 \Omega) & =  \hat \gamma_2 \cr}
}
It follows by simple algebra that
\eqn\dbltto{
\langle \hat \gamma_1, \hat \gamma_2 \rangle =
2 \im (\bar C_1 C_2) i \langle \Omega,
\bar \Omega \rangle .
}
Thus, if $\langle \hat \gamma_1, \hat \gamma_2 \rangle\not=0 $
are mutually nonlocal charges   we can invert the
equations and write:
\eqn\dblttoo{
\Omega = {1 \over  2  \im (\bar C_1 C_2) } \biggl( C_1 \hat \gamma_2 - C_2 \hat
\gamma_1 \biggr)
}
On the other hand, if
$\langle \hat \gamma_1, \hat \gamma_2 \rangle =0 $
then $C_1 = \lambda C_2$ for some {\it real}
constant $\lambda$, and hence $\hat \gamma_1 = \lambda \hat \gamma_2$ for some
real constant $\lambda$. Thus we
have proved the simple

\bigskip
\noindent
{\bf Proposition 8.1.1.} Suppose a complex structure satisfies
the attractor equation for two   vectors
$\gamma_1, \gamma_2 \in H_3(X;\IZ)$. Then either

a.) $\gamma_1 = \lambda \gamma_2$ for $\lambda\in \IQ$, or,

b.) $\langle \gamma_1 , \gamma_2 \rangle \not=0$

\bigskip
\noindent
{\bf Definition.}
If an attractor complex structure satisfies $(b)$
for some pair of charges $\gamma_1, \gamma_2$ then
we will refer to it as an attractor point of rank two.
If an attractor point is not of rank two we call
it an attractor point of rank one.

\bigskip
{\bf Remarks.}

\item{1.} We have found some examples
of exact CY attractors. They are
 described in section 8.3.
They are all of rank two.

\item{2.} The physical interpretation of
the rank two attractor is that two mutually nonlocal
BPS states take their minimum mass at the same
complex structure.

\item{3.} At an attractor point of
rank two, $H^{3,0} \oplus H^{0,3}$ is the complexification
of a rank two submodule of $H^3(X;\IZ)$.
Conversely, if $H^{3,0} \oplus H^{0,3}$ is the complexification
of a rank two submodule so that
$\Omega= \hat \gamma_1 + \xi \hat \gamma_2$, $\xi\in \IC, \gamma_1, \gamma_2
\in H_3(X;\IZ)$,
then for any real numbers $n,m$:
\eqn\dblaat{
2 \im (\bar C_{n,m} \Omega) = n \hat \gamma_1 + m \hat \gamma_2
}
for
\eqn\dblaatt{
C_{n,m} = {(m-\xi n) \over  2 \im \xi}
}
In particular, taking $m,n$ integral we see that
such a point is an attractor point of rank two.

\item{4.} Attractor points can
be characterized as follows. Since $\hat \gamma = \hat \gamma^{3,0} + \hat
\gamma^{0,3}$, if $\CJ$ is a complex
structure on $H^3(X;\IR)$
diagonal in the Hodge decomposition then
$\CJ\cdot \hat \gamma =-i  \hat \gamma^{3,0} + i \hat \gamma^{0,3}$. Thus,
\eqn\complx{
H^{3,0} \oplus H^{0,3} = {\rm Span}_{\IR}\{ \hat \gamma, \CJ \cdot \hat
\gamma\} \otimes \IC
}
Moreover, the change of variables from a symplectic
to a complex basis:
$\hat \gamma = p^I \hat \alpha_I - q_I \hat \beta^I =
z^I f_I + \bar z^I \bar f_I$ is given by
\eqn\cplxchg{
z^I = {i \over  2} (\im \tau)^{-1,IJ}(q_J + \bar \tau_{JK} p^K)
}
since $\hat \gamma^{3,0} = -i \bar C \Omega$ we can
take periods to obtain the suggestive expression
\eqn\perds{
\bar C X^I = - {1 \over  2} (\im \tau)^{-1,IJ}(q_J + \bar \tau_{JK} p^K)
}
This is valid for {\it both} the Griffiths and Weil complex
structures on $H^3(X;\IR)$. Note that it does
{\it not} mean that $H^{3,0} \oplus H^{0,3}$ is the complexification of an
integral lattice since in the
basis $\alpha^I, \beta_I$ it is not {\it a priori} obvious
that the matrix elements of
the operator $\CJ$  are integral or even arithmetic.

\subsec{The   attractor conjectures}

\bigskip
\noindent
{\bf Attractor Conjecture 8.2.1:}
Suppose $\gamma\in H_3(X;\IZ)$ for
a polarized CY 3-fold $X$ defines an attractor point
$z_*(\gamma)\in
\widetilde{\CM}$ in the Teichmuller
space of complex structures.
Then the period  vector is   valued in a number
field $\IP^{h^{2,1}}(K(\gamma))$. In particular,
the special coordinates $t^i$, $i=1,\dots, h^{2,1}$
 associated to this
point are valued in a number field $K(\gamma)$.

\bigskip
\noindent
{\bf Attractor Conjecture 8.2.2:} Suppose
$\gamma\in H_3(X;\IZ)$ for a polarized CY
3-fold $X$ defines an attractor point  $z_*(\gamma)\in
\widetilde{\CM}$ in the
Teichm\"uller
space of complex structures. Then the corresponding
variety $X_{\gamma}$  is arithmetic, and defined over
a number field $\widehat{K}(\gamma)$.
More precisely, the moduli space of
complex structures for the polarized variety
$X$ has an embedding  \viehweg\
\eqn\viehweg{
f: \CM(X) \rightarrow \IP^N
}
to a quasiprojective variety. We conjecture that
there is an embedding (independent of $\gamma$)
such that complex
structures corresponding to attractor points
$\gamma \in H_3(X;\IZ)$ map to   $\widehat{K}(\gamma)$-rational
points in $\IP^N$, where $\widehat{K}(\gamma)$ is a number field.

\bigskip
\noindent
{\bf Attractor Conjecture 8.2.3:} The embedding
\viehweg\ can be chosen so that the compositum
$\widehat{K}(\gamma) K(\gamma)$ is a Galois extension
of $K(\gamma)$,

\bigskip

In section 12 below we will put these conjectures
in a broader context showing how they generalize
``Kronecker's Jugendtraum.''

\bigskip
\noindent
{\bf Remarks.}

\item{1.} There are two ``moral reasons'' behind the attractor
conjectures. First, they are a natural generalization of
the phenomena we have found for $d=4, \CN=4,8$
compactifications. Second, we expect that the
entropy of $d=4, \CN=2$ Calabi-Yau black holes will
be obtained from special conformal field theories
along the lines of  \sv.  We hope these conformal
field theories are related to RCFT's. It is known
that for RCFT's the values of $c$ and $h$ (the
dimensions of Virasoro primaries) are rational
numbers \andermoore. Rationality of $c$ implies
rationality of the black hole discriminant, and
this would most naturally follow if the periods
are algebraic. Of course, this is hardly a proof.

\item{2.} The attractor conjectures
 can be made much more
explicit and concrete by considering
families of CY 3-folds constructed using
toric geometry as in, e.g.,  \batyrev\agm\hkt\hkty.
Let us adopt the notation of \hkty, section 3, for
definiteness. Working around a
large complex structure limit (LCSL)
the periods are obtained from Picard-Fuchs
differential equations $\CL_s \varpi=0$ which
themselves are obtained from a
   GKZ system of hypergeometric equations.
These hypergeometric equations are
written in the algebraic coordinates for
toric complex structure deformations.
(They are denoted by $z_s$ in eq. (3.6)
of \hkty.)  It is shown in \dmlcsl\delignelcsl\hkty\
that $h^{2,1}$ of the periods have $\log$-singularities
(and not $\log^2, \log^3$) so we may use
these to define special coordinates:
$t^i(z) = \varpi_i(z)/\varpi_0(z)$, where
$\varpi_0(z)$ is a distinguished period with
no logarithmic singularities.
Conjecture 8.2.1 above states that we can
choose an integral basis of periods so that
if $\gamma$ admits solutions $t^i_\gamma$ to
\hodgede\ within the region of convergence
of the series expansion for
$\varpi_0(z)$ then $t^i_\gamma$ are valued in
a number field $K(\gamma)$. Conjecture  8.2.2,
states that the
corresponding values of $z_s$ are also
valued in a number field $\widehat{K}(\gamma)$, and
conjecture 8.2.3 states that they are related in
the way suggested by the theory of complex
multiplication of elliptic curves.
Thus, in this context the conjecture becomes a
conjecture about the arithmetic values of certain
special generalized hypergeometric functions.

\item{3.} Because there are two different kinds of
attractor varieties, namely, those of rank 1 and rank 2,
the above conjectures can be stated in a weak and
a strong form. The weak form of the conjectures
only makes the assertion for attractor points of rank 2.
The strong form makes the assertion for both ranks 1 and
2.  We thank P. Deligne for stressing the importance
of this distinction: The weak form is far more likely
to be true than the strong form. On the other hand,
it might well prove to be the case
that attractor points of rank 2 are rare.

\item{4.} There are several extensions of the definition
of CM-type to $\dim > 1$
Calabi-Yau varieties in the literature. A definition
of K3 surfaces of CM type appears in a
paper of Piatetski-Shapiro and Shafarevich
\psshaf.
%
%(Closely related matters are discussed
%in \deligne.)
This class of K3 surfaces differs from the
attractive K3 surfaces discussed in this paper
because there are K3 surfaces of CM type
in the sense of \psshaf\  with $\rho<20$.
A definition of  CM type for CY 3-folds was
proposed by Borcea in \borcea. This class of
CY 3-folds differs from the attractor varieties
because, as we mention below, the Fermat
quintic is of CM type in the sense of Borcea,
but is {\it not} an attractor variety. Finally,   in
independent work,
A. Todorov has proposed that a natural
extension of definition of CM type to
CY 3-folds is the requirement that
$H^{3,0} \oplus H^{0,3}$ is defined over
$\IZ$. In our language, these are the attractor
points of rank 2.

\subsec{Examples of exact arithmetic CY 3-fold attractors}

As mentioned above, a
prediction of the attractor conjectures,  taken
together with the explicit expressions
for periods of toric hypersurfaces, is that the
generalized hypergeometric
functions $ {}_{N+1} F_{N}$ should have
special arithmetic properties at special values of
the monodromy parameters. Unfortunately
these  do not seem to follow (at least, do
not follow easily)
from the known results about these hypergeometric
functions such as those listed in
\slater. However, in some examples the
periods and mirror maps reduce to special
cases of ${}_2 F_1$ or to modular functions, and
in these cases we can verify the above conjectures.

\subsubsec{Example 1}

Mirror symmetry for
one parameter families of Calabi-Yau's
generalizing   the
famous analysis of the quintic
\cdgp\ were studied in
\berglund. As shown in
these papers, at a Fermat point,
after a rational
change of basis the periods
are given by roots of unity. Thus,
the existence of charges satisfying
the attractor
equations at a Fermat point reduces
to a question about the ring of integers in a
cyclotomic field $\IQ[\beta]$, with
$\beta= e^{2\pi i /k}$.

Of the four
examples in \klemmtheis\ with
$k=5,6, 8,10$ one finds that the Fermat
point is {\it not} an attractor point
for $k=5,8,10$.
On the other hand, the case $k=6$
does give an example. The  Fermat point
is defined by
\eqn\finmonvi{
M: \qquad \{ x: 2 x_0^3 + x_1^6 + x_2^6+ x_3^6+ x_4^6 = 0 \} \subset
\IP^{2,1,1,1,1}
}
with $W=M/G$  constructed via the
 Greene-Plesser  construction with
$G=\IZ_3 \times \IZ_6 \times \IZ_6$.
The attractor charges are most easily
found using the basis:
\eqn\finmonvii{
\pmatrix{R_2 \cr R_1 \cr R_0\cr R_5} =
\pmatrix{-6 & 3 & 1& -2\cr
0& -1 & 0 & 1\cr
0 & -1& 0 & 0 \cr
-3 & 3 & 1& -1\cr}
\pmatrix{q_0 \cr q_1 \cr p^0 \cr p^1\cr}
}
since in this basis the periods are
$\varpi = K(\beta^2, \beta, 1, \beta^5)$ for
some constant $K$ and $\beta=\exp[2\pi i/6]$.
The attractor equations   have a solution
iff
\eqn\finmonviii{
R_5=-R_2=R_0-R_1
}
While $R_1,R_0$
are arbitrary, and hence the attractor point is
of rank two.
The minimal BPS mass for such a charge is given by
\eqn\finmonvix{
M^2(p,q;p,q) = {2 \sqrt{3} \over  3} (R_0^2 - R_0 R_1 + R_1^2)
}
This is positive definite, and once again we see
that, up to an overall constant,
 at an attractor point the mass-squared
spectrum of BPS states is integral.

\ifig\ceezero{The complex structure
moduli space for the two-parameter
family of \twop, as depicted in
\lookingglass. We find attractor points
on the divisor $C_0$. In section 9 we compare
with attractor points near the LCSL. }
{\epsfxsize2.5in\epsfbox{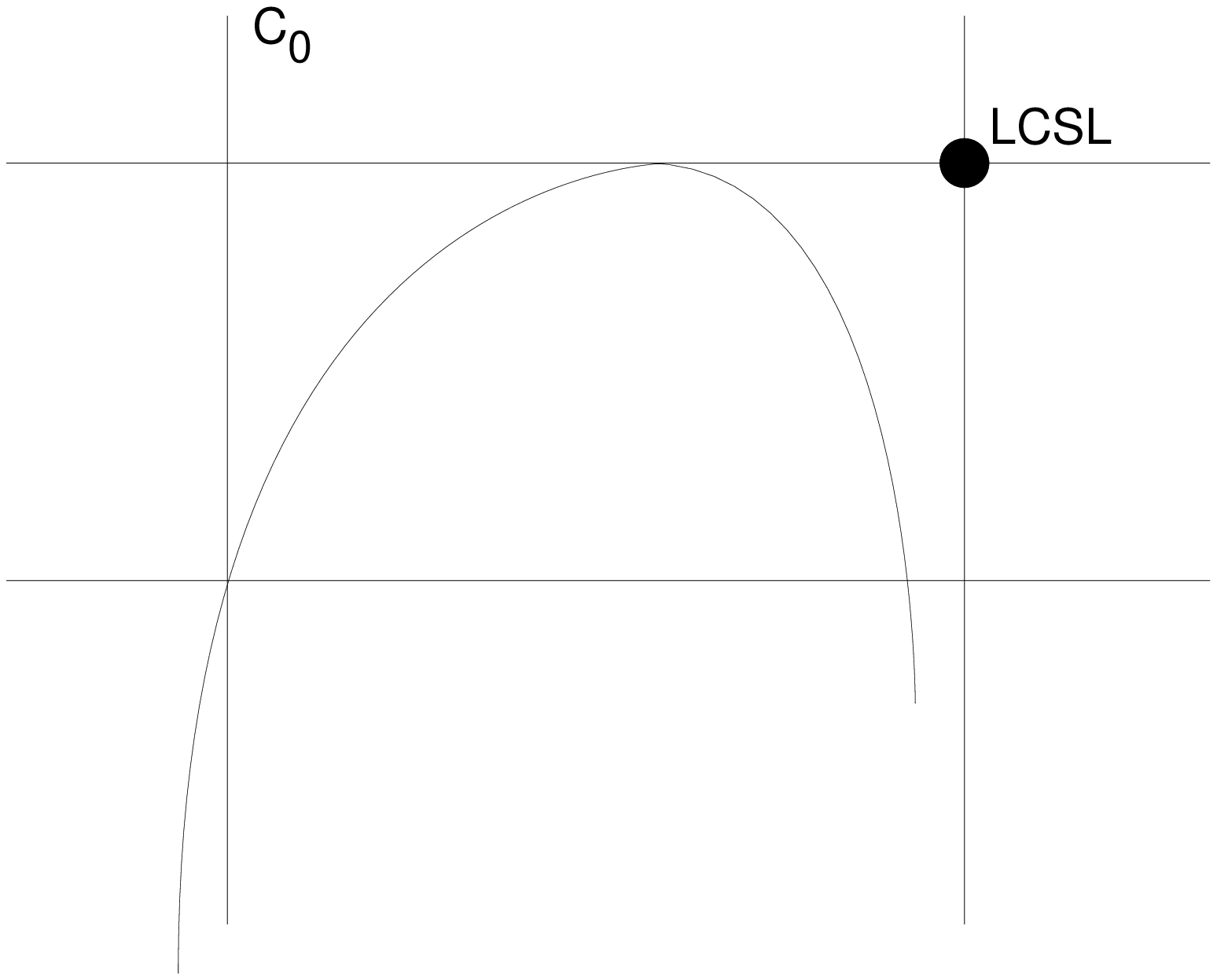}}

 \subsubsec{Example 2}

We now consider the 2-parameter family
studied in detail in  \twop.
For $M$ we take the family of surfaces:
\eqn\familyi{
x_1^8 + x_2^8 + x_3^4 + x_4^4 + x_5^4
- 8 \psi x_1 x_2 x_3 x_4 x_5 - 2 \phi x_1^4 x_2^4 = 0
}
in  $\IP^{1,1,2,2,2}[8]$.
Resolving the singularities of the weighted projective
space introduces a second Kahler parameter.
For $W=M/G$ we take a $G= \IZ_4^3$ quotient
using the Greene-Plesser construction.

We examine the periods of $W$ on the locus $C_0$
where $\psi =0$, and
shown in \ceezero. The monodromy around this
divisor is finite, with values in $8^{th}$ roots of $1$.
%
%\foot{This is the divisor given by the vertical line
%on the left hand side  of figure 1 of
%\lookingglass. In particular, note that it is far from
%the LCSL.}
%
The explicit
periods of $\psi^{-1} \varpi_j(\psi,\phi)$
at $\psi=0$ follows from eq. (6.18) of \twop.
\eqn\valpers{
\eqalign{
\psi^{-1} \varpi_j(\psi,\phi) \vert_{\psi=0}
& = K \alpha^j u_{-1/4}(\phi)  \qquad j ~ {\rm even} \cr
& = K \alpha^j u_{-1/4}^D(\phi)  \qquad j ~ {\rm odd} \cr}
}
where $\alpha = e^{2 \pi i /8} = (1+i)/\sqrt{2}$,
$K=-2 \Gamma(1/4)/(\Gamma(3/4))^3$,
and $u_\nu, u_\nu^D$ are related to Legendre
functions.

Define $\varpi' = \varpi/ (K  \psi)
=\pmatrix{\varpi_0 &\varpi_1 &\varpi_2 &\varpi_3 &\varpi_4
&\varpi_5 \cr}/(K\psi)  $. Then
the attractor equations can be written in the basis:
\eqn\aeqdiv{
2 \Im \bar C' \varpi'
= \pmatrix{ R_0 & R_1 & R_2 & R_3 & R_4 & R_5 \cr}^{tr}
}
The $R_i$ are related to the integral basis by
a rational matrix called $m$ in \twop.
(See also eq. (9.23) below.)
In the basis \aeqdiv\ the attractor equations
become the equations:   $R_5=-R_1, R_4 = - R_0$, and
\eqn\adqdv{
{ \alpha u^D_{-1/4}(\phi_*) \over  u_{-1/4}(\phi_*)} =
  {R_3 + i R_1 \over  R_2 + i R_0} .
}
Thus, up to an overall constant the attractor periods
are in the field $\IQ[i]$. This verifies
conjecture 8.2.1.

 It is useful to define rational
numbers $a,b$:
\eqn\defabee{
a+ i b \equiv (1+i) {R_3 + i R_1 \over  R_2 + i R_0}
}
then:
\eqn\aeqdvi{
\pmatrix{ R_3 \cr R_1 \cr} = \pmatrix{ \half(a+b) & \half (a-b) \cr -\half(a-b)
& \half (a+b) \cr} \pmatrix{ R_2 \cr R_0 \cr}
}
Using the matrix $m$ we find that $\Omega$ is
proportional to
\eqn\omegprop{
\eqalign{
\Omega_{a,b} & \equiv \hat \gamma_1 + i \hat \gamma_2 \cr
\hat \gamma_1 & = 2 \hat \alpha_0 - \hat \alpha_1 + (a+1) \hat \alpha_2 - (a +
b-2) \hat \beta^0 - 2(b+1) \hat \beta^1 - 4 \hat \beta^2 \cr
\hat \gamma_2 & = \hat \alpha_1 + (b-1) \hat \alpha_2 - (b-a) \hat \beta^0 -
2(1-a) \hat \beta^1 \cr}
}
In particular all attractor points are of rank two.

We would now like to verify that the corresponding
Calabi-Yau variety is indeed arithmetic.
Note that $x_i \rightarrow \alpha x_i$
takes $\phi \rightarrow - \phi$, so we only
give the value of $\phi^2$ at an attractor point.
In order to
understand the arithmetic nature
of $\phi^2$  we must examine the Legendre functions
$u_\nu, u^D_\nu$ in more detail.
The ratio $u_{-1/4}/u^D_{-1/4}$ is related to
a Schwarz triangle function for the triangle
group of type $(2,4,\infty)$. This is one of the
arithmetic triangle groups
\takeuchi\ and so we may expect special
arithmetic properties. We will show explicitly
this is the case for $\vert \phi \vert > 1$.

Choosing the system of cuts described in
\twop\  one can derive the values of the periods
in the region  $\vert \phi \vert >1,\Im \phi>0 $:
\eqn\youknew{
\eqalign{
u_\nu(\phi) & = (2 \phi)^\nu {}_2 F_1(-{\nu \over  2} , {1-\nu \over  2} ; 1;
{1 \over  \phi^2} ) \cr
u^D_\nu(\phi) & = { \sin \pi \nu \over  \pi} (2 \phi)^\nu
\sum_{n\geq 0} {(-{\nu \over  2})_n ({1-\nu \over  2})_n \over  (n!)^2}{1 \over
 \phi^{2n}} \cdot \cr
 \cdot \biggl[ - \log[-\phi^2] +
&
\psi(-{\nu \over  2}+n) + \psi({1-\nu \over  2}+n) - 2 \psi(n+1) \biggr] \cr
& - \cos \pi \nu  (2 \phi)^\nu {}_2 F_1(-{\nu \over  2} , {1-\nu \over  2} ; 1;
{1 \over  \phi^2} ) \cr}
}
where the log is to be evaluated with
$\vert \arg(-\phi^2) \vert<\pi$, and
$\psi(x)$ is the dilogarithm function.

Now make two quadratic transformations on the
hypergeometric functions, let
\eqn\quadtrani{
 \phi^{-2}   = {16 z (1-z) \over  (1+ 4z(1-z) )^2}
}
and define the automorphic function:
\eqn\quadtranii{
z = {\vartheta_2^4(\tau) \over  \vartheta_3^4(\tau) },
}
in terms of standard Riemann theta functions.
One  finds the  result:
\eqn\quadtraniii{
\eqalign{
u_{-1/4}(\phi) & = (2 \phi)^{-1/4} (1+4 z(1-z) )^{1/4} \vartheta_3^2(\tau) \cr
u_{-1/4}^D(\phi) & = -{i \over  \sqrt{2}} (\tau+1) (2 \phi)^{-1/4}
(1+4 z(1-z) )^{1/4}   \vartheta_3^2(\tau) \cr}
}
where $-1 < \re(\tau+1) < 1 $.

The attractor equations now say that $\tau +1 = a+ ib$.
Note that $b>0$ and $a,b$ must be mapped
to $\vert \phi \vert > 1, \im\phi>0$. This covers a region
near the cusp at infinity. We  can now use the standard
theory of complex multiplication to check
 conjectures 8.2.2 and 8.2.3.
Using \quadtranii\ $z$ may be expressed
in terms of the value of $\wp(x,\tau)$ at   2-torsion points,
so it follows from the second main theorem of complex multiplication that
$\phi^{2}$
is valued in a ray classfield of $K=\IQ[i]$.
We can be more precise about this field.
Let $\tau = p_1/q_ 1+ i p_2/q_2$ be in lowest
terms so that $a \tau^2 + b \tau + c =0$ with
$D=b^2 - 4 a c = -(2 q_1^2 p_2 q_2)^2$.
$\phi^2$ is   related to the Weber function
$f(\tau) = (\vartheta_3/\eta)^{1/2}$ by:
\eqn\weber{
\phi^2 = {f^{24} \over  2^8} + {2^8 \over  f^{24}} + \half
}
and $(f^{24}-16)^3 - j(\tau) f^{24}=0$, so $\phi^2$  is
in a cubic extension of the ring class field of the
order $\CO(D)$. In particular, it is an algebraic
number of degree approximately
$\vert 2 q_1^2 p_2 q_2\vert$ when this number is
large.

To summarize, in this section
we have given a small confirmation of the weak
attractor conjectures
along the divisor $C_0$
in complex structure moduli space.

The BPS mass may be calculated from
\eqn\entropy{
\vert Z \vert^2 = -i \vert C \vert^2 \Pi^\dagger \pmatrix{ 0 & 1 \cr -1 & 0
\cr}  \Pi =
 + \half b (R_0^2 + R_2^2) = \half (\Im \tau) (R_0^2 + R_2^2)
}
where $\Pi$ denotes the period vector in an
integral basis, as in  \twop.

Unfortunately, we cannot easily use the exact results
on the periods along LCSL divisors (such as the
divisor $q_1=0$ in \twop) since the attractor
equations do not appear to have a good limit on
these compactification divisors.

\subsubsec{Other examples}

Other examples of exact $\CN=2$ attractors are provided
by the FHSV model (section 5 above), other Voisin-Borcea
Calabi-Yau manifolds, and orbifolds of attractor
tori.

\newsec{Attractor equations in the large
complex structure limit}

Several interesting aspects of the attractor
equations become apparent in the ``large complex
structure limit'' (LCSL), which may be characterized
by maximal unipotent monodromy on the
period vector. (For a more precise description
see \dmlcsl\delignelcsl.)    In this
section we begin with a review of previous
work on the subject and then give three applications:
First, the attractor points are dense.
Second, a given charge can
lead to several distinct attractor points with
different values of minima.
Third,
the arithmetic attractor conjectures imply
interesting identities between  $\zeta(3)$ and
polylogarithms.

\subsec{Review of the approximate
general  solution}

In this section we review the general solution
of the attractor equations given in
\schmakova\cardi. In IIB theory we work
around a large complex structure limit.
Choose a vanishing  cycle $\alpha^0$ to
separate out a period $X^0$,
 and write special coordinates $t^a = X^a/X^0$,
 $a=1,\dots, n_v= h^{2,1}(X)$.
The attractor equations   now become the
system:
\eqn\chosgag{
\eqalign{
\Im (\bar\CC ) & = p^0 \cr
\Im(\bar \CC t^a) & = p^a \cr
\Im (\bar \CC \CF_a) & = q_a \cr
\Im (\bar \CC(2 \CF -t^a \CF_a) )& = q_0 \cr}
}
where $\bar \CC = 2 \bar C X^0$,
$\CF$ is the inhomogeneous prepotential, and
$\CF_a = {\p \over  \p t^a} \CF$. In our
conventions the
  BPS mass-squared is given by:
\eqn\masssq{
\eqalign{
M_*^2 & =   \vert C \vert^2 e^{-K} \cr
& = +{i \over  4} \vert \CC \vert^2   \biggl(
2 \CF - 2 \bar \CF - (t^a-\bar t^a)(\CF_a + \bar \CF_a) \biggr) \cr}
}

Now we use the general form of the inhomogeneous
prepotential near a point of maximal unipotent
monodromy \hkty\hkt:
\eqn\maxuni{
\eqalign{
\CF = {1 \over  3!}
&
D_{abc} t^a t^b t^c + \half A_{ab} t^a t^b
+ B_a t^a  - \half { i \zeta}  + \Sigma\cr
\Sigma&\equiv
{1 \over  (2\pi i)^3} \sum_r n_r Li_3(e^{2 \pi i r \cdot t}) \cr}
}
The various terms are most simply interpreted by considering
instead the IIA compactification on the mirror $\widetilde{X}$
in
the large radius limit. Then, $-D_{abc}$ are
intersection numbers, the sum
on $r$ is over rational curves in $\tilde X$,
\eqn\defzet{
\zeta = {\zeta(3) \over  (2\pi)^3} \chi(\tilde X) ,
}
$B_a$ is   given by:
\eqn\beea{
B_a = - {1 \over  24} \int J_a c_2(\tilde X) ,}
and $A_{ab}$ is a symmetric integral matrix.
In our conventions, dropping $\zeta,\Sigma$ leads to
\eqn\convens{
e^{-K} = - {4 \over  3} \vert X^0 \vert^2 D_{abc} (\im t^a)
(\im t^b)(\im t^b) + \cdots
}
which must be positive. The point of maximal
unipotent monodromy is approached when
$\im t^a \gg 1$.

Using these facts it is convenient to rewrite the
equations \chosgag\ as:

\eqna\aerenorm
$$
\eqalignno{
\Im \bar \CC & = p^0  & \aerenorm a \cr
\Im \bar \CC t^a & = p^a & \aerenorm b \cr
\Im \bar \CC \biggl( \half D_{abc}t^b t^c + \CI_a(t) \biggr) &
= \tilde q_a & \aerenorm c \cr
\Im \bar \CC \biggl(
-{1 \over  3!} D_{abc} t^a t^b t^c + \CI_0(t) \biggr) &
= \tilde q_0& \aerenorm d \cr}
$$
where we have defined the shifted charges:
\eqn\shifchrg{
\eqalign{
\tilde q_a & = q_a - A_{ab} p^b - B_a p^0 \cr
\tilde q_0 & = q_0 - B_a p^a\cr}
}
and defined instanton sums:
\eqn\instsum{
\eqalign{
\CI_a(t) & \equiv \p_a \Sigma = {1 \over  (2\pi i)^2} \sum_r r_a
n_r Li_2(e^{2 \pi i r \cdot t}) \cr
\CI_0(t) & \equiv  - i \zeta  +   {2 \over  (2\pi i)^3} \sum_r
n_r \widetilde{Li_3}(e^{2 \pi i r \cdot t}) \cr
\widetilde{Li_3}(e^{2 \pi i r \cdot t})& = Li_3(e^{2 \pi i r \cdot t})-
\half (2\pi i ) t \cdot r  Li_2(e^{2 \pi i r \cdot t})\cr}
}

The equations  \aerenorm\
were explicitly solved by Schmakova, in
the approximation that one drops $\CI_0, \CI_a$,
as follows \schmakova.
Define:
\eqn\chrgdefs{
\eqalign{
\Delta_a(p) & \equiv \half D_{abc} p^b p^c - p^0 \tilde q_a \cr
\Delta(p) & \equiv D_{abc} p^a p^b p^c \cr
\CM(p,q)  & \equiv \Delta(p) - 3 p^0(   p^0 \tilde q_0  +  p^a  \tilde q_a)
\cr}
}
If $p^0 \not=0$ a solution to \aerenorm{a,b}  is   given by:
\eqn\trivsolve{
\eqalign{
\bar \CC & = \xi_0 + i p^0 \cr
t^a & = {1 \over  p^0}
\biggl( p^a - \CC {x^a \over  \vert \CC \vert} \biggr) \cr
\im t^a & = {x^a \over  \vert \CC \vert} \cr
\re t^a  & = {1 \over  p^0} (p^a - {\xi_0\over  \vert \CC \vert} x^a ) \cr}
}
where $x^a$ is  real. Moreover, if $x^a$ is a real
solution  to the system of
quadratic equations:
\eqn\system{
 \half  D_{abc} x^b x^c    = \Delta_a(p)
}
then \aerenorm{c} is solved. It is useful
to define
$ \Delta(x) \equiv D_{abc} x^a x^b x^c
= 2 \Delta_a(p) x^a$, as well as
\eqn\defcurlyell{
\CD(x,p,q) \equiv \Delta(x)^2 - \CM(p,q)^2.
}
Then \aerenorm{d}\
implies $\xi_0/\vert \CC \vert = \CM/(2 x^a \Delta_a)=
\CM/\Delta(x) $. This leads to a quadratic
equation for $\xi_0$ and we choose the root:
\eqn\trvslvii{
\xi_0 =   - \vert p^0 \vert {\CM(p,q) \over  \sqrt{
\CD(x,p,q)} }
}

The solution must satisfy the consistency conditions:
\eqna\consis
$$
\eqalignno{
\CD(x,p,q)= ( \Delta(x) )^2 -\CM(p,q)^2 & > 0  & \consis a\cr
x^a & > 0 & \consis b\cr
\Delta(x) & <  0  &\consis c\cr}
$$
Here \consis{a}\ is needed for reality of $\xi_0$. The BPS
square-mass is given by: \eqn\zeesqr{ \vert Z \vert^2 = -{ \Delta(x)
\over  3 \vert \CC \vert } = {1 \over  3 \vert p^0 \vert }
\sqrt{\CD(x,p,q) } } hence \consis{c}, and this governs the choice
of root in \trvslvii.

An important special case arises for $p^0=0$.
The solution is given by:
\eqn\specci{
t^a = w^a + i {p^a \over  \xi_0}
}
where $w^a$ is a solution to
\eqn\speccii{
D_{ab}(p) w^b = \tilde q_a , \qquad D_{ab}(p)\equiv D_{abc}p^c
}
and $\xi_0$ is given by:
\eqn\specciii{
\xi_0^2 = {\Delta(p) \over  6 \tilde q_0 + 3   w^a \tilde q_a } .
}
The solution exists if the RHS of \specciii\ is positive
and there is a choice of root so that $p^a/\xi_0 > 0 $
for all $a$.

A corollary of this calculation is:

\bigskip
\noindent
{\bf Proposition 9.1.1}  The attractor points of rank one
are dense in the neighborhood of a point of maximal
unipotent monodromy.

This follows already from   the special solutions
\specci\speccii\specciii. It is natural to conjecture
that they form a dense set of points throughout
Teichmuller space $\widetilde{\CM}$.

{\bf Remarks}

\item{1.} {\it Domain of validity.}
In order for this solution to be valid we must have
$\im t^a \gg 1$ in order to justify neglecting
the sum over rational curves. (Neglecting the term involving
$\zeta$ is more subtle and discussed below.)
Under a rescaling of charges
$p \rightarrow \lambda p, q \rightarrow \lambda q$,
the quantities
$x^a, \CC $ scale like $\lambda$ so $t^a$ is
invariant. Thus, one must always take a ``skew limit''
of charges. From \zeesqr\ one may read off an
approximate expression for the discriminant and
hence distinguish charges of type a,b,c, in the
language of section 2.6.

\item{2.} {\it On the choice of branch in
\system.} There are $2^{n_v}$
complex solutions to \system.
If the charges $p^a$ are all positive and the
RHS of \specciii\ is positive then there is
a branch of solutions to \system\ that is
distinguished in the limit $p^0/p^a \rightarrow 0$.
This is given by:
\eqn\branch{
\eqalign{
x^a & = p^a - p^0 w^a - (p^0)^2 w^a_2 + \cdots \cr
D_{ab}(p) w_2^b & = \half D_{a bc} w^b w^c \cr}
}
with $w^a$ defined by \speccii.
One finds
\eqn\lmit{
\CD(x,p,q) = (\Delta(x))^2 - \CM(p,q)^2 \rightarrow
(p^0)^2 (6 \tilde q_0 + 3 w^a \tilde q_a)
\Delta(p) + \CO((p^0)^3)
}
so
we recover \specci\specciii\ from \branch,  provided
the solution exists, i.e., provided $\xi_0$ is real.

\item{3.} {\it Charge shifts}. The shifts in the charges
\shifchrg\ from $B_a$ are related to anomalous D-brane
charges. See, for example, \msw. The term $A_{ab}$
can be removed by a monodromy transformation, but
is important if one is careful about the relation
of $\varpi, \Pi$, as in \twop\dmlcsl.

\subsec{Multiple solutions and area codes}

In this section we show that a single charge
$\gamma$ can
lead to several different attractor
complex structures in $\widetilde{\CM}$,
not related by the duality group.

Let us note at the outset that this is
{\it not} in contradiction with
the minimization
principle described in section 2.5.
Theorem 2.5.1 alone cannot be used to deduce
that there is a unique
global minimum of the BPS mass in Teichmuller
space. For example, the simple function on $\IR^2$
given by
\eqn\simple{
f(x,y) = e^{-2 x} - e^{-x -y^2}y^2
}
has only two stationary points, and both
are minima. \foot{Thanks to K. Rabe for help finding this
example.}

An example of  charges leading
to multiple attractor points is provided by the
exact solutions given in section 8.3.2 above,
when
combined with the analysis near a LCSL given
in the previous subsection.
The charges leading to the attractor points
studied in section 8.3.2, when referred to the
integral basis $\Pi$ of \twop, are given by:
\eqn\charges{
\eqalign{
\pmatrix{q_0 \cr q_1 \cr q_2 \cr p^0\cr p^1 \cr p^2\cr}
=
&
\pmatrix{ -1 & 1 & 0 & 0 & 0 & 0 \cr
1 & 0 & 1 & -1 & 0 & -1 \cr
3/2 & 0 & 0 & 0 & -1/2 & 0 \cr
1 & 0 & 0 & 0 & 0 & 0 \cr
-1/4 & 0 & 1/2 & 0 & 1/4 & 0 \cr
1/4 & 3/4 & -1/2 & 1/2 & -1/4 & 1/4\cr}
\pmatrix{ R_0 \cr R_1 \cr R_2 \cr R_3 \cr R_4 \cr R_5\cr} \cr
&
= \pmatrix{
 \half (R_2 (b-a) + R_0 (a+b-2))\cr
R_0(b+1) - R_2(a-1)\cr
 2 R_0\cr
R_0\cr
\half ( R_2-R_0)\cr
\half( R_0(b-1) - R_2(a+1)) \cr}\cr}
}

By construction, these charges lead to attractor points
on the divisor $C_0$ far from the LCSL.
If we trivialize $H^3(X;\IZ)$ over
$\widetilde{\CM}$  we can
  consider the same charges in the equations
near a LCSL on the Teichmuller space.
(Equivalently, we can choose a path from $C_0$
to the LCSL so that $\Pi = m \varpi$, where $m$ is the
matrix in \charges.)  The prepotential
is \twop:
\eqn\twoprep{
\CF = - {1 \over  6} (8 (t^1)^3 + 12 (t^1)^2 t^2) -2 t^1 t^2 -{11 \over  3} t^1
- t^2  +  i {\zeta(3) \over  2 ( 2\pi)^3}168 + \cdots
}
One finds \system\ has a unique acceptable solution
under the  conditions:
\eqn\deltwoo{
\eqalign{
- \Delta_2 & = \half (R_2^2 + 5 R_0^2 ) > 0 \cr
2 \Delta_2 - \Delta_1 & = (b-1) R_2^2 + (b+2/3) R_0^2 > 0 \cr}
}
The expression $\CD(x,p,q)$
in \defcurlyell\  is a complicated $6^{th}$
order polynomial in $R_0,R_2$. If $R_0 \gg R_2$
then we have:
\eqn\curlyell{
\CD \cong { 9 R_0^6 \over  4} ( 5 b^2 + 40 b -16a(a-4) + 16)
+ \CO(R_0^5 R_2)
}
which is positive for suitable $a,b$
for $R_0 \gg R_2$, while the
attractor points are approximately
\eqn\attpts{
\eqalign{
\im t^1 & \cong {\sqrt{5}\over  8} {\sqrt{5 b^2 + 40 b -16a(a-4) + 16} \over
 2-a} \cr
\im t^2 & \cong  {3 b + 2 \over  15} \im t^1 \cr}
}
in this limit. Thus, by choosing suitable $a,b$ and
large $R_0$ we get a consistent solution in the
neighborhood of a point of maximal unipotent
monodromy. The BPS mass in this limit is given by
\eqn\bpsmass{
\vert Z \vert^2 \cong 2 R_0^2 \sqrt{5 b^2 + 40 b -16a(a-4) + 16}
}
Comparison with \entropy\ raises  the interesting question of
which expression should be explained by the Dbrane
model. (Since \msw\ explicitly refers to a large
radius (IIA) limit, and the other attractor point is deep within
moduli space we expect \bpsmass\ to be the preferred answer. )

\bigskip
{\bf Remarks}

\item{1.} Because of the attractor mechanism, the parameters which
specify the near-horizon geometry of a black hole are a
proper subset of the ``hair,'' that is, the minimal
data needed to specify uniquely the entire black hole geometry outside
the horizon. Naively, the horizon geometry is solely
determined by $\gamma$. However, our example shows that the
dynamical system \dynsys\ defined by a charge $\gamma$
can have several basins of attraction $\CB_\alpha$,
$\alpha=1,2,\dots$. The data $(\gamma, \CB_\alpha)$ uniquely
specifies the near-horizon geometry and should be referred
to as the {\it area code} of the black hole.

\item{2.} In the opposite
limit,  $R_0 \rightarrow 0$ one finds that
\eqn\opplimit{
\CD \rightarrow - {R_0^2  \over  4} (3b-1)(9b+19) R_2^4
}
so the above charges give an example of acceptable
charges for   solutions of the attractor equations
which do not admit a physical branch with $p^0 \rightarrow 0$.
Indeed, it is interesting to note that our example requires
$p^0\not=0$. In the setup of \msw\ this means there
is  large
magnetic   (D6) charge in the IIA description,
or a large $N$ multi-Taub-NUT space in the M-theory
description. The existence of multiple minima should
be related to some interesting physics of these
multiply-wrapped D6-branes.

\item{3.} The approximate solution
described in the    previous section can only
lead to multiple solutions if $p^0\not=0$ or if
$p^0=0$ and $D_{ab}(p)$ is noninvertible. The latter
situation is excluded if $p^a$ represents a divisor
in the NEF cone of $\tilde{X}$ as in \msw. We have
searched for   examples of multiple {\it acceptable}
solutions of \system, but without success. However, if one includes the   $\zeta(3)$ correction, but continues to drop
the instanton sum then the equations are still
tractable.  Indeed, \trivsolve\ and
\system\ still remain valid and solve \aerenorm{a,b,c}.
The final equation \aerenorm{d} is modified to
\eqn\zetrhee{
(\xi_0^2 + (p^0)^2)\bigl( \CM -
{3 \over 2}(p^0)^2 \zeta \xi_0\bigr)^2
= (\Delta(x))^2 \xi_0^2
}
(Here we take the case $p^0\not=0$. If $p^0=0$ there
is a cubic equation and similar remarks apply.)
Equation \zetrhee\ is quartic and simple examples show that
one can choose charges for which it admits {\it two} real
solutions $\xi_0$. Unfortunately such examples have
$\im t^a \cong \CO(1)$, and one can show that it is
impossible to find sequences of charges for which both
solutions have $\im t^a \rightarrow \infty$.
Thus these examples certainly {\it suggest} there are
multiple solutions, but  are inconclusive.

\item{4.} {\it Comment on flop transitions}. It is interesting
to consider the role of flop-induced spacetime
topology change \wittenphase\agm\ in the context
of the attractor mechanism. See \fvdeeii\blsii\gmms\
for related discussion. We have examined several
examples of partially enlarged Kahler cones of
Calabi-Yau 3-folds and found {\it no} examples of
charges with different basins of attraction containing
different LCS/large radius limits.
This is consistent with the
example studied in \fvdeeii. It also has the
interesting consequence that the radial supergravity
flow starting out at the ``wrong'' large radius
limit must lead to a topology change, as suggested
in \blsii. (The flow through the wall of the K\"ahler
cone, discussed in \blsii\gmms, needs further study.)

\item{5.} A consequence of multiple solutions is that
the ``Weinhold metric'' discussed in \fgk\
must have negative eigenvalues for some points
in moduli space. Conversely, if the Weinhold metric
defines a positive definite metric then there can be
no multiple solutions.
\foot{We thank G. Gibbons for a discussion on this
point.}

\subsec{Digression on   trilogarithms - a moral point}

The strong attractor conjecture is extremely strong,
as one can see by examining the implications for the
infinite sums of trilogarithms entering the
expansion of the prepotential
near a point of maximal unipotent monodromy.

In order to appreciate how outrageous
the strong form of the conjecture is, consider the
so-called ``axion free case'' with
charges $\gamma = q_0 \alpha^0 - p^a \beta_a$
\cardi\cardii. All of the attractor equations
except that for the period $F_0$ are solved
by the simple expression $t^a = i p^a/\lambda$ where
$\lambda\in \IR$ and
the $p^a$ all have a common sign so that
$\im t^a>0$. Moreover, $\im t^a$ must
be    sufficiently large that the
sum over rational curves converges.
The last equation in \chosgag\
becomes a complicated
transcendental equation for $\lambda$:
\eqn\lambde{
{\Delta(p) \over  6 \lambda^2} -\tilde q_0 =
\lambda \Biggl[ \zeta - {2 \over  (2\pi)^3} \sum_r n_r {\rm Li_3}  (e^{-2\pi r
\cdot p /\lambda}) \Biggr]
}
and $\Delta(p) \equiv D_{abc} p^a p^b p^c$.
The strong attractor conjecture asserts that
the solutions to this equation are algebraic
numbers!

This assertion  might look
completely  unreasonable. As weak
evidence in support of the conjecture we
note that fixed points of the action of
the duality group $\Gamma$ on
$\widetilde{\CM}$ lead to lots of
nontrivial identities between infinite
sums of trilogs and $\zeta(3)$. To quote
one simple example we consider the perturbative
prepotential of the ``STU model.''
This involves the function
$\CL(T,U)$ in \hmi\ defined by:
\eqn\ellfun{
\eqalign{
\CL(T,U) & \equiv
 \sum_{r>0 } c(-r^2/2) Li_3(e^{2 \pi i (k T + \ell U )}) \cr
& =
Li_3(e^{2 \pi i  (T-U)}) +
 \sum_{k,\ell \geq 0} c(k\ell) Li_3(e^{2 \pi i (k T + \ell U )}) \cr}
}
where
\eqn\solprpi{
F(q) = \sum_{n=-1}^\infty c(n) q^n =  {E_4 E_6 \over  \eta^{24}} \quad .
}
One easily derives the monodromies of the
prepotential under $T\rightarrow T, U \rightarrow -1/U$.
\eqn\curident{
\eqalign{
\sum_{r>0} c(kl ) Li_3(e^{2 \pi i (k T + l U') })
=
& U^{-2} \sum_{r>0} c(kl ) Li_3(e^{2 \pi i (k T + l U) })
+ \half \zeta(3) c(0)( U^{-2}-1) +\cr
&
+ {4 \pi^3 i \over  3} (U - 5 U^{-1} + U^{-3}) \cr}
}
where $U' = -1/U$ and we are in the chamber
$\Im T > \Im U, \Im U'$.
Evaluating \curident\ at the fixed point $U=i$ we get
two curious identities:
\eqn\zetsum{
\eqalign{
{\zeta(3) \over  2 (2 \pi)^3} & = {7 \over  2880}- {1 \over  (2\pi)^3}
\sum_{j=1}^\infty Li_3( e^{-2 \pi j})   \cr
(2.42301....) \times 10^{-3} & = 2.43056 \times 10^{-3}  -  (7.53....) \times 10^{-6} + \CO(10^{-8})  \cr}
}
and
\eqn\secident{
Li_3(e^{-2 \pi(\beta-1) } ) = - \sum_{k>0, l\geq 0} c(kl)
Li_3(e^{-2 \pi(k \beta + l ) } )\qquad \beta > 1
}

We quote the
identities \curident\zetsum\secident\ because they establish the
``moral point'' that there are {\it lots} of identities
relating $\zeta$, infinite sums of polylogarithms,
and algebraic expressions in flat coordinates,
because there are {\it lots} of fixed points of
$\Gamma$ on $\widetilde{\CM}$.

{\bf Remarks.}

\item{1.}  Incidentally,
taking $\Im T \rightarrow +\infty$ \curident\
becomes an identity of Ramanujan:
\eqn\ramang{
\sum_{k=1}^\infty Li_3(e^{-2 \pi k y} ) =
-{1 \over  x^2} \sum_{k=1}^\infty Li_3(e^{-2 \pi k x} )
-\half \zeta(3) (1+ 1/x^2) + {4 \pi^3 \over  3 (240)} (x + 1/x^3+ 5/x)
}
for $xy =1$, and both positive.

\item{2.} The fixed points of the Siegel modular
group $Sp(4;\IZ)$ acting on the Siegel
upper half-plane $\CH_2$ have
been classified in \ueno. Combining this with
the results for prepotentials in the ``STUV models''
\lustline\lustrev\ one can find many more
explicit examples of   identities like \curident.

\item{3.} In a systematic expansion starting
with the solution of the attractor equations
near a LCSL the next term in the expansion
corrects the entropy by:
\eqn\correctentr{
\vert Z \vert^2 = - {4 \Delta(x) \over 3 \vert \CC\vert}
\Biggl(1 - {3 \over 4} \vert p^0\vert^3
{\CM^2 \over (\Delta(x)^2-\CM^2)^{3/2}}\zeta+
\CO((p^0/p^a)^6) \Biggr)
}
It has been suggested in \cardii\ that the appearance of the
transcendental object $\zeta(3)$ will be an obstacle to a
microscopic derivation of the entropy via D-branes.
One proposal \cardii\ is that $\zeta(3)$ is related to
the statistical entropy of membranes rather than
strings. The existence of the class of
identities discussed above
suggests an alternative
explanation: One cannot include the effects of
$\zeta$ without  the infinite sum of instanton
corrections, and  these differ from $\zeta$
by an algebraic expression in the flat coordinates.

\newsec{Attractor points and rational conformal field theory}

The attractor
mechanism singles out the attractive K3
surfaces.  These surfaces are
elliptically fibered (in many ways), so it is natural to
ask what   8-dimensional heterotic theories
correspond to these attractive K3 surfaces
in the $F$-theory dual \vafa\vfmr.
In this section we show that the corresponding
heterotic theories are built from the rational conformal field
theories on $T^2$. The masses of BPS states
are given by norms of ideals  in quadratic imaginary
fields and the BPS generating functions  may be
related to  ray class theta functions. Moreover,
this relation to the heterotic theory provides a
new physical interpretation of the existence of a
Mordell-Weil (MW) group in the F-theory compactification:
The MW group is an even integral lattice to which
we may associate a vertex operator algebra.
This vertex operator algebra is the enhanced
left-moving chiral algebra of the rational conformal
field theory in the heterotic dual. It acts as an
automorphism algebra on the algebra of (Dabholkar-Harvey)
BPS states. We establish these
results in detail in the following sections.

\subsec{Review: $F$-theory/Heterotic duality in 8D}

\subsubsec{Complex structure
moduli of elliptically fibered K3 surfaces}

The moduli space of complex structures of
marked elliptically fibered K3 surfaces with
a section may be described as follows.
We fix a marking $\gamma_I$,
$I=1,\dots, 22$, on a K3 surface $S$,
so that the intersection matrix
$  \gamma_I  \cdot \gamma_J   $ is the
standard intersection form for the K3 lattice
\eqn\ktlatt{
\Lambda_{K3} \equiv
H(1)^3 \oplus (E_8(-1) )^2.
}
Here $H(1)\cong II^{1,1}$ stands for the
even unimodular lattice with the
intersection form
\eqn\hform{
\pmatrix{ 0 & 1 \cr 1 & 0 \cr}.
}
Choose null vectors
$e,e^*$ generating $H(1)$ so that the class of the fiber is
$e$ and the class of  the section is $e^* - e$.
We take $\gamma_1 = e, \gamma_2 = e^*$.
These are algebraic cycles and must be orthogonal
to the holomorphic $(2,0)$ form $\Omega$.
By the global Torelli theorem
the space of marked complex structures on $S$
is given by
\eqn\fmdspce{
\CB^{18,2} \cong
\{
\Omega = \sum_{J=3}^{22} z^J \gamma_J\  : \
   \Omega\cdot \bar \Omega >0,
\quad    \Omega \cdot \Omega  =0\}/\IC^*
}
where
we identify $\Omega \sim \lambda \Omega$ for
$\lambda\in \IC^*$.

The comparison to the heterotic string is facilitated
by identifying $\CB^{18,2}$ with a space of
projection operators.
Giving  a class $\Omega$ in \fmdspce\
 is equivalent to
specifying the oriented positive 2-plane
$\langle \re \Omega, \im \Omega \rangle_{\IR} \subset
II^{18,2}\otimes \IR$. Hence the moduli space of
marked complex structures is identified with
\eqn\dfncrlbee{
\CB^{18,2} \equiv Gr_2^+(II^{18,2}\otimes \IR).
}
Explicitly,
 $\Omega$ determines projection
operators of $II^{18,2}\otimes \IR$ onto
definite-signature  subspaces:
\eqn\projection{
\eqalign{
\Pi_\Omega:
II^{18,2}\otimes \IR & \rightarrow \IR^{18,0}\perp \IR^{0,2} \cr
\Pi_\Omega^+(x) &   \equiv
{  x\cdot  \Omega  \over
  \Omega\cdot \bar \Omega } \bar \Omega +{  x\cdot \bar \Omega  \over   \Omega \cdot  \bar \Omega }   \Omega \cr}
}
To match with heterotic string notation we will denote $\Pi_\Omega^+(x) =x_R$.
We let $x_L \equiv x- x_R\equiv \Pi_\Omega^-(x)$. In our conventions, $x_L^2\leq 0$.
Many authors identify $\IR^{0,2} \cong \IC$
by   setting $x_R = e^{K/2}   x \cdot \Omega  $
where $K = -\log[  \Omega \cdot \bar \Omega  ]$.

\subsubsec{The Neron-Severi lattice
and the Mordell-Weil group }

The Neron-Severi lattice
$NS(S)$ consists of
vectors $\gamma\in H^2(S;\IZ)$ such that
$\int_\gamma \Omega=0$. Given the structure
of an elliptic fibration we have a decomposition:
\eqn\nsvcts{
\gamma = a e + a^* e^* + \tilde \gamma
}
where $\Pi_\Omega^+(\tilde \gamma) = \tilde \gamma_R =0$,
so $\tilde \gamma^2\leq 0$.
Effective divisors have $\gamma^2=-2$ or
$\gamma^2 \geq 0$. If they are represented by an
irreducible curve of genus $g$ then $\gamma^2 = 2 g -2$.

The general elliptic K3 fibration
$\Phi: S \rightarrow \IP^1$ in the family
\dfncrlbee\  has
a rank 2 NS lattice and hence $\tilde \gamma=0$
for all $\gamma\in NS(S)$.  On special
subvarieties of $\CB^{18,2}$   the
image of $\Pi_\Omega^-$ contains integral vectors
and the Picard number $\rho(S)$ jumps. This jump
occurs in one of two very
different ways. First, there can be
rational curves in the singular fibers of
the fibration $\Phi$. These vectors correspond to
roots of the enhanced
gauge symmetry in the F-theory compactification.
Second,  there can be a  jump in the rank of the
MW group.
All this is summarized by
the formula \shiodaef:
\eqn\rankform{
\rho(S) = r(\Phi) + 2 + \sum_\nu (m_\nu-1)
}
where $r(\Phi)$ is the rank of the
MW group, the sum is over the
singular fibers of $\Phi$, and  $m_\nu-1$ is the rank
of the associated simple factor in the enhanced
gauge symmetry in the IIB theory.
For $I_1$ fibers (corresponding to $U(1)$ factors
in the gauge group) we have
$m_\nu=1$.

\subsubsec{$F$-theory/heterotic correspondence in 8D}

The $F$-theory/heterotic correspondence in 8D
is simply obtained by identifying the moduli space
$\CB^{18,2}$ with the universal cover of Narain
moduli space and the lattice
$\langle e, e^* \rangle^\perp \cong II^{18,2}$
with the Narain lattice. The right-moving
component of a Narain vector $p\in II^{18,2}$
is identified with the period via
\eqn\pright{
p_R =e^{K/2}  \int_p \Omega
}
with $K = - \log [ \Omega \cdot \bar \Omega ]$.

In order to relate points in
$\CB^{18,2}$ to geometrical compactification
data in the heterotic string it is useful to introduce a
tube domain realization of $\CB^{18,2}$.
We make a  choice
 of integral lightlike vectors $w,w^* \in \IR^{s,2}$
in a Minkowski space of signature $(-1)^{s}, (+1)^2$
(in our case $s=18$).
$w,w^*$ have intersection form \hform, and hence:
\eqn\tdu{
\IR^{s,2} \cong \IR^{s-1,1} \perp \langle w, w^* \rangle_\IR
}
The tube domain realization is then
an isomorphism:
\eqn\tdii{
\Psi_{w,w^*} : \CB^{s,2} \rightarrow \IR^{s-1,1} + i C^+
\subset \IC^{s-1,1}
}
where $C^+$ is the forward light-cone in  $\IR^{s-1,1}$.
The (inverse) map is defined by:
\eqn\tdiii{
   y
\quad \mapsto\quad \Omega \cdot \IC^* = (y + w -\half y^2
w^*) \cdot \IC^*
}

Geometrically, on the IIB side,  we choose  a point of
maximal unipotent monodromy and
a vanishing cycle $w$ with $w^2=0$. This
represents a
nonalgebraic torus. In a model for K3 based
on a product of elliptic curves $Km(E_1 \times E_2)$
we could choose $w$ to be a product of
``$a$-cycles.''  On the heterotic side, if
we choose a 1-cycle on $T^2$ then we
canonically isolate a hyperbolic sublattice
$\langle w , w^*  \rangle_{\IZ} \subset II^{18,2}$
\eqn\cusp{
\langle w , w^* \rangle_{\IZ} \cong H(1)
}
given by the momentum/winding lattice for that
cycle. Thus, a choice of cusp in $\CB^{s,2}$
is a choice of decompactification of the
heterotic string to 9 dimensions.

Having phrased things this way
it is natural to choose  a basis
of 1-cycles in $H_1(T^2;\IZ)$.
  The momentum/winding lattice for the
torus $T^2$ becomes
\eqn\twodecompact{
\langle w_1, w^*_1 \rangle_{\IZ} \oplus
\langle w_2, w^*_2 \rangle_{\IZ} \cong
H(1) \oplus H(-1)
}
Normalize the period point to be $\Omega =
(y + w_1 - \half y^2 w^*_1 ) $. Now,
using $w_2, w_2^*$
 we can give $y$ the physical
interpretation in terms of standard heterotic moduli:
\eqn\physintep{
y = \int_a \vec A + i \int_b \vec A + T w_2 + U w_2^*
= \vec y + T w_2 + U w_2^* \in \IC^{0,s-2}\oplus \IC^{1,1}
}
Here $T,U$ are the   Kahler and complex
moduli, respectively. The moduli $\vec y$
represent the Wilson lines
scaled by appropriate metric factors. (For
explicit formulae see \clm.)

\subsec{Review: Toroidal Rational conformal field theories}

We now recall some standard facts about
rational conformal field theories. The RCFT's
of a single boson $X$ on a circle of radius
$R$ are well-known \difrancesco.  They
occur when $R^2 = p/q$ is rational.
Let $\CA(\Gamma)$ be the chiral vertex operator
algebra associated to an even integral lattice $\Gamma$.
Then the  chiral algebra of the rational
circle is  $\CA(\Gamma)$ for  $\Gamma = \sqrt{N} \IZ$
with $N = 2 pq$. The representations of the
chiral algebra are labelled by
$\Gamma^*/\Gamma \cong \IZ/N\IZ$. The
different gluings of left- and right-moving
theories with chiral algebra
$\CA(\Gamma)_{\rm left}
 \otimes \widetilde{\CA}(\Gamma)_{\rm right}$
are given by the
automorphisms of the fusion algebra
$Aut(\Gamma^*/\Gamma) \cong (\IZ/N\IZ)^*$
\ms\dv.

These statements generalize readily to the
rational torus compactifications with
Narain lattice $II^{d+8s,d}$ with signature
$(-1)^{d+8s},(+1)^{d}$. We regard a point in
the universal cover of Narain moduli space as
a projector $\Pi:  II^{d+8s,d}\otimes \IR \rightarrow
\IR^{d+8s,0} \perp \IR^{0,d}$. The rational
conformal field theories which occur are
based on enhanced chiral algebras
$\CA(\Gamma_L) \otimes \widetilde{\CA}(\Gamma_R)$
associated with  even integral
lattices $\Gamma_L, \Gamma_R$
of signatures $(-1)^{d+8s}$ and $(+1)^d$, respectively.
The irreducible representations
of $\CA(\Gamma)$ are labelled by the discriminant
group $\CD(\Gamma) = \Gamma^*/\Gamma$
and the group algebra of $\CD(\Gamma)$
is the Verlinde algebra. The gluing of left-
and right-moving chiral rational conformal field theories
is determined by a choice of glue vectors $g_i$ such
that:
\eqn\gluevectors{
II^{d+ 8s, d} \cong \amalg_i \biggl( \Gamma_L \perp \Gamma_R + g_i \biggr)
}
The map $g_i^L \rightarrow g_i^R$ provides an
isomorphism of the discriminant groups
$\CD(\Gamma_L) \rightarrow \CD(\Gamma_R)$.
and such   isomorphisms are in 1-1 correspondence
with the automorphisms of the fusion rules.

\bigskip
{\bf Remark.}
The isomorphism $g_i^L \rightarrow g_i^R$ of discriminant groups  is the same as that  used in the Nikulin embedding theorem \nikulin\dolgachevi. Indeed, proposition 1.6.1 of
\nikulin\ corresponds to the
the theorem that different left-right gluings of the rational
conformal field theories are obtained
by using an automorphism of the Verlinde algebra,
as in \ms\dv.

\subsec{The RCFT's corresponding to attractor varieties}

We now put together the facts reviewed in sections
10.1, 10.2 in the context of the attractor mechanism.

In the case of Narain compactification on a
$II^{18,2}$ lattice the criterion for an RCFT is
that the right-projection $\Pi_R$ contains a rank
two lattice. Equivalently, there are two independent
vectors $p,q$ with $p_L =q_L=0$. On the other hand,
as we have seen, attractive K3 surfaces  (compatible
with a fixed elliptic fibration)
are defined by $p_L=q_L=0$ for
vectors $p,q\in \langle e,e^* \rangle^\perp \subset
H^2(K3;\IZ)$. As we have seen, these
 determine the complex structure
\eqn\attrccpl{
\Omega  = \CC (q - \bar \tau   p)
}
where $\tau = \tau(p,q)$ as defined in \solvtau.
Therefore, under F-theory duality heterotic
compactification on RCFT's
corresponds to IIB compactification on
attractive K3 surfaces.

It follows immediately that the purely
right-moving momentum lattice is the
transcendental lattice $T_S$ so
the right-moving chiral algebra is
$\widetilde{\CA}(T_S)$. Indeed,
the possible right-moving RCFT's are
classified by even   positive
definite rank 2 lattices.

Let us now consider the left-moving momenta.
The attractive $K3$ surfaces
have Neron-Severi lattice $NS(S)$ of signature $(1,19)$
and we may write:
\eqn\mwgp{
 \langle e, e^* \rangle_{\IZ} \perp \widehat{NS}(S)
\subset NS(S)
}
where $\widehat{NS}(S)$ is a negative definite
even lattice of rank $18$. For generic attractor points
all the singular fibers are   of Kodaira type $I_1$,
and hence by \rankform\  $r(\Phi) = 18$.
Thus, at the generic attractor point we should
identify $\widehat{NS}(S)$ with the torsion free
part of the MW group of the elliptic
fibration $\Phi: S\rightarrow \IP^1$.

Under $F$-theory duality the
purely left-moving momentum
lattice is $\widehat{NS}(S)$ so the left-moving
chiral algebra includes the chiral vertex algebra
$\CA(\widehat{NS}(S))$.  For generic attractor
points, therefore, {\it the  left-moving chiral algebra is
generated by the Mordell-Weil group}. This gives a
physical interpretation of the appearance of
that group in $F$-theory.

\bigskip

{\bf Remarks.}

\item{1.}   We identified the
right-moving chiral algebra with $\widetilde{\CA}(T_S)$.
A comparison  with the Shioda-Inose
theorem leads to an interesting subtlety.
Given a chiral algebra $\CA(\Gamma_L) \otimes
\tilde \CA(\Gamma_R)$ one can always obtain
another RCFT based on $\CA(\Gamma_L') \otimes
\tilde \CA(\Gamma_R)$ if $\Gamma_L'$ is in the
same genus as $\Gamma_L$. Thus, several different
rational conformal field theories are associated to
a given choice of $\Gamma_R$, and the enumeration
of these compactifications will again be expressed
in terms of class numbers.  Presumably these
correspond to different elliptic fibrations of the
same K3 surface.

\item{2.}
 Every $\rho=20$ surface $S$ admits an
embedding of $(E_8(-1))^2 \hookrightarrow NS(S)$
\morrpicard.
Since attractor points are dense this appears to
lead to a paradox since enhanced symmetry
points with $E_8 \times E_8$ gauge symmetry
cannot be dense! The condition
for enhanced gauge  symmetry is that there is
an element
$\tilde \gamma\in \langle e, e^* \rangle^\perp$
with $\tilde \gamma^2=-2$. Geometrically
this corresponds to rational curves in
singular fibers. In general the embedding
of $(-E_8)^2$ involves
vectors of type \nsvcts\ but with $a,a^*$ nonzero.
Thus, these divisors give rise to enhanced gauge
symmetry with respect to {\it some} elliptic
fibration, but not with respect to the fixed one
used in defining the $F$-theory moduli space.

\item{3.} There are several other physical
interpretations of the MW group which have
appeared in the literature.
Unbroken $U(1)$ factors were
associated with  the torsion-free part of the MW
group in  \vfmr. In the present discussion
these  correspond to the
vertex operators $- i \p X^I$ of the compact bosons.
This group also plays a key
role in recent work on hypermultiplet moduli
\asphm.
Finally some very interesting recent papers
\berpan\bkmt\amrp\ investigate the
physics of the torsion parts of
the MW group.

\subsubsec{Arithmetic and the mass spectrum}

Let us now consider the physical spectrum
of the theory.
The transcendental cycles give right-moving
vertex operators  in the chiral algebra of
the conformal field theory
$\sim e^{i p_R \cdot \tilde X(\bar z)}$.
They can be combined
with left-moving chiral algebra elements to
give DH-BPS states. In particular, the mass-squared
of these BPS states is
\eqn\msssqr{
M^2 = \half (np + m q)_R^2 =
\half\bigl(n^2 p^2 + 2mn p\cdot q + m^2 q^2 \bigr) ,
}
where $m,n\in \IZ$. As in the FHSV model, these
are norms of  ideals in the ideal class
associated to $2 Q_{p,q}$. On the other hand,
the only chiral vertex operators
 in $\CA(\widehat{NS}(S))$
which survive in the BRST cohomology are
those vertex operators related to rational
curves in the singular fibers. This is the
standard correspondence between enhanced
gauge symmetry in F-theory and in heterotic
theory. Nevertheless, the
  full chiral algebra $\CA(\widehat{NS}(S))$
is not devoid  of physical meaning since
it gives automorphisms of the algebra of
BPS states in 8D compactifications and,
in a related IIA compactification to 6D, it would
be a vertex subalgebra of the algebra of
BPS states.

Finally, we note that the mass generating functional
of the BPS states is just
\eqn\mssgenfurn{
\sum_{P\in II^{18,2}}  d(P) e^{-4 \pi \im \tau M^2(P)}
=\oint d \tau_1 {\sum_{P\in II^{18,2}} q^{\half P_L^2} \bar q^{\half P_R^2}
\over  \eta^{24}}
}
and at an attractor point we have a
decomposition into characters of the RCFT
following from:
\eqn\mssgenii{
\sum q^{\half P_L^2} \bar q^{\half P_R^2}
= \sum_{g_i} \vartheta_{\widehat{NS}(S) + g_{i,L}}(\tau)
\overline{\vartheta_{T(S) + g_{i,R}}(\tau)}
}
where $g_i$ are the glue vectors
and the coset theta functions
$\vartheta_{T(S) + g_{i,R}}(\tau)$ can be expressed
in terms of ray-class theta functions for the
quadratic field $K_D$ using Theorem 3.4 of
\taormina. Therefore, $L$-functions
of $K_D$ can be interpreted as
BPS mass counting functions
$\sim \sum M^{-s} $ where we sum
over certain subsets of BPS states.

\subsubsec{Relation to rational values of $G,B,\vec A$ }

We now study the attractor points in a tube-domain
realization of $\CB^{18,2}$, using section 10.1.3.
We will show   that the standard
compactification data $G,B,\vec A$ are rational
(in string units).

As in section 10.1.3 we choose integral $w, w^* \in
 \langle e, e^* \rangle^\perp\subset H^2(S;\IZ)$.
Then we can write
$p = p_+ w + p_- w^* +\tilde p $,
$q = q_+ w + q_- w^* +\tilde q $
with $\tilde p, \tilde q \in \langle w, w^* \rangle_{\IZ}^{\perp}
\in \Gamma^{s-1,1}$. Note
$\Gamma^{s-1,1}$ need not be unimodular.
Using the standard normalization $\int_{w^*} \Omega=1$ we
have:
\eqn\stannorm{
\Omega = y + w - \half y^2 w^* =
{1 \over  p_+ - \bar \tau q_+} \bigl( q - \bar \tau p \bigr)
}
and therefore can solve for $y$:
\eqn\solvewhy{
y = {\tilde q - \bar \tau(p,q) \tilde p \over  p_+ - \bar \tau q_+}
}
Going one step further as in  \twodecompact\
we can
read off the
physical moduli $\vec A, T, U$
by decomposing the vectors
\eqn\vectrs{
\eqalign{
p & = p_+^1 w_1 + p_-^1 w^*_1
+ p_+^2 w_2 + p_-^2 w^*_2+ \vec p  \cr
q & = q_+^1 w_1 + q_-^1 w^*_1
+ q_+^2 w_2 + q_-^2 w^*_2+ \vec q
\cr}
}
The moduli are:
\eqn\moduli{
\eqalign{
T & = { q_+^2 - \bar \tau p_+^2 \over  q_+^1 - \bar \tau p_+^1} \cr
U & =  { q_-^2 - \bar \tau p_-^2 \over  q_+^1 - \bar \tau p_+^1} \cr
\vec A & = { \vec q  - \bar \tau \vec p  \over  q_+^1 - \bar \tau p_+^1} \cr}
}
In particular, they are  quadratic imaginary.

It is useful to be even more explicit. Consider,
for simplicity,  the subspace $\CB^{2,2}\subset
\CB^{18,2}$ corresponding to enhanced
$E_8 \times E_8$ or $Spin(32)/\IZ_2$ symmetry.
Let us parametrize $\CB^{2,2}$ by
$T,U$, $Im(T) >0, Im(U) >0$
and identify the Narain lattice
with  $II^{1,1} \oplus II^{1,1}$ for
momentum and winding around two circles in a basis for
$H_1(T^2)$.  $T,U$ are related to the
compactification data $G,B$ through the standard
expressions:
\eqn\standexp{
\eqalign{
T& = B + i \sqrt{\det G} \cr
U & = {G_{12} + i \sqrt{\det G} \over  G_{11}} \cr}
}
in terms of which the rightmoving projection of
a momentum vector
\eqn\pvctr{
p   =  k_1 w_1+  \ell_1 w_1^*+   k_2 w_2 +  \ell_2 w_2^*
}
is
\eqn\standii{
p_R = {1 \over   \sqrt{2 \im T \im U} } (
k_2  + \ell_1  U + (k_1  + \ell_2  U)  T )
}
The moduli of the  attractor point   defined by
electric and magnetic vectors:
\eqn\hetatt{
\eqalign{
q & =  (k_1^e, \ell_1^e) \oplus  (k_2^e, \ell_2^e)\cr
p & =  (k_1^m, \ell_1^m) \oplus  (k_2^m, \ell_2^m)\cr}
}
is obtained from the vanishing of the left-moving
projections:
\eqn\hetatti{
\eqalign{
k_2^e + \ell_1^e U + (k_1^e + \ell_2^e U)   \bar T & = 0 \cr
k_2^m + \ell_1^m U + (k_1^m + \ell_2^m U)   \bar T & = 0 \cr}
}
Equation \hetatti\   can be rewritten as
\eqn\hetattii{
\pmatrix{ \ell_2^m & - \ell_2^e \cr -k_1^m & k_1^e\cr}
\pmatrix{k_2^e & \ell_1^e \cr k_2^m & \ell_1^m\cr}
\pmatrix{1\cr U\cr} = - \bar T (k_1^e \ell_2^m - k_1^m \ell_2^e)
\pmatrix{1\cr U\cr}
}
Comparison with \cmplxi\ shows that the elliptic curve
$E=\IC/(\IZ + U \IZ)$ has complex multiplication by
$\bar T (k_1^e \ell_2^m - k_1^m \ell_2^e)$. In particular,
$T$ and $U$ are quadratic imaginary.
Conversely, from \standexp\ it follows that if
$G,B$ are rational then $T,U$ are quadratic imaginary
and $\Gamma_R$ contains a rank two lattice.

Computing the complex numbers $p_R$ corresponding
to purely right-moving vectors one finds, up to
an overall constant, the elements of the integral
ideal $\underline{\bf a}_{p,q}$
defined in  \coridel.
%
%$p_R\in {1 \over  \sqrt{p^2}} \underline{\bf a}_{p,q}$.
%The projections $p_R$ of arbitrary
%Narain vectors are in the fractional ideal
%$({p^2 \over  \sqrt{D} })\underline{\bf a}_{p,q}$.
%

\subsec{Brief remarks on 4D and 6D compactifications}

Given the connection between attractor
varieties and rational conformal field theory in
8D compactifications it is natural to ask if
the correspondence somehow extends
to 6D and 4D compactifications of the heterotic string.

In order even to formulate such a correspondence
we need to say something about hypermultiplet
moduli.  Mirror symmetry allows us to identify
``attractor hypermultiplet moduli'' as mirror to
attractor vectormultiplet moduli, at least at
weak string coupling (where we can isolate
a special K\"ahler submanifold of the
quaternionic space).
With this understood it is   natural to conjecture
that the   compactifications
of the heterotic string on RCFT's are dual to type II
compactifications at  attractor points.

In the case of $Het[T^4]  = IIA[K3]$ one can
confirm this guess combining the above results
with  the $IIA$ geometrical interpretation of
Narain moduli described in \aspinwall\ (and
reviewed in appendix C):
RCFT points in the Narain moduli
space $\CN^{20,4}$ map to K3 surfaces
with quadratic imaginary $\Omega$, and
quadratic imaginary complexified K\"ahler class.

\newsec{Arithmetic of the $F$-map}

\subsec{Definition of the $F$-map}

There are two standard ways to
parametrize the moduli space of
8D $F$-theory compactifications.
One was described in the previous
section based on  period points
and Grassmannians. This is the description
in flat coordinates. Alternatively
one can give an algebro-geometric
description of the family of elliptically
fibered K3 surfaces. We will loosely refer
to the map between these two descriptions
as the $F$-map.

More precisely, since
 we are describing elliptically fibered $K3$'s
with a section we can give a Weierstrass model
of the elliptic fibration $\Phi: S \rightarrow \IP^1$:
\eqn\weiermodel{
\eqalign{
Z Y^2 = & 4X^3 - f_8(s,t) X Z^2 - f_{12}(s,t) Z^3 \cr
f_8(s,t) = & \alpha_{-4} s^8 + \cdots + \alpha_{+4} t^8 \cr
f_{12}(s,t) = & \beta_{-6} s^{12}+ \cdots + \beta_{+6} t^{12}\cr}
}
where $(s,t)\not=0$ are homogeneous coordinates
on the base $\IP^1$, $ (X,Y,Z)\not=0$ are homogeneous
coordinates for the elliptic fiber, and we divide by
the $\IC^* \times \IC^*$ action:
\eqn\globcor{
\eqalign{
(s,t; X, Y,Z) & \sim (s,t; \lambda X, \lambda Y , \lambda Z) \cr
& \sim (\mu s, \mu t; \mu^4 X, \mu^6 Y, Z) \cr}
}

The $F$-theory
moduli space in algebraic coordinates is then:
\eqn\algemduli{
\CM_{\rm algebraic}   = \biggl[ \{ (\vec \alpha, \vec \beta)\} -
\CD\biggr]/GL(2,\IC)
}
where $\CD$ is an appropriate discriminant locus
and we divide by $GL(2,\IC)$ because Mobius
transformations on $s/t$ define equivalent surfaces.

Choosing a cusp for $\CB^{18,2}$ as in
section 10.1.3  the  $F$-map is then defined to
be the map
 $\Phi_F: y \rightarrow [(\vec \alpha, \vec \beta)]$
identifying the moduli spaces.
Combining the Shioda-Inose theorem with
the theory of complex multiplication we see
that the $K3$ mirror map should behave
analogously to the elliptic functions in the theory of
complex multiplication. Unfortunately, the results
of \shioda\ do not refer to any specific projective
model, such as we have in \weiermodel,  so we formulate the
\foot{I would like to thank E. Diaconescu for pointing out
a subtlety in the statement of this conjecture in version 1.
The examples below already indicate that the relevant
finite extension in general is not a ray class field of $K_D$,
but involves, at least, and abelian extension of such ray
class fields.}

\bigskip
{\bf Conjecture 11.1} Under the $F$-map, the
  quadratic imaginary periods  $y^i\in K_D$
 map to $  \alpha_i, \beta_i $ in some
finite  extension of $K_D $, for an appropriate
choice of coordinates $X,Y,Z,s,t$.

Of course, since $(\vec \alpha, \vec \beta)$
can be redefined by $GL(2,\IC)$ we can
only expect the class $[(\vec \alpha,\vec \beta)]$ to have a
representative in the field.
We have not given a mathematical
proof of this statement but we will check it in several
examples below. We do not attempt to find the minimal
field of definition.

\subsec{Digression on the F-map and K3 mirror symmetry }

The F-theory map  described in the previous section
is the same as the mirror map for K3 mirror symmetry.
This is tautological if one defines the mirror
map as a map between algebraic and flat
coordinates as in, for example, \lianyau\klm.
However, the $F$-map
can also be related to the description of
K3 mirror symmetry given in \aspmorr\dolgachev.
A post-duality-revolution description of \aspmorr\
proceeds as follows.

Recall the standard string dualities:
\eqn\dualities{
Het[T^4_{6789}]
= F[K3_{89} \times T^2_{67}] = M[K3_{78910} \times T^1_6] = IIA[K3_{6789}]
}
where the subscripts indicate ``coordinate directions.''
This is a useful notational device to keep track of the
various spaces involved.

Consider a family of compactifications
so that the flat triple $(G, B, \vec A)$ in
the Narain moduli space $\CN^{20,4}$
respects a product structure
$T^4_{6789}= T^2_{67}\times T^2_{89}$ but is
otherwise generic. This means that the
 Wilson lines on  $T^2_{89}$ leave
a rank $s$ sub-group
$H_{89}\subset E_8 \times E_8\times U(1)^4$ {\it unbroken},
while the Wilson lines on $T^2_{67}$,
leave a rank $20-s$ $H_{67} \subset E_8 \times E_8 \times U(1)^4$
unbroken. The Wilson lines on $T^2_{89}$
break $H_{67}$ to its Cartan subgroup, and
similarly for those on  $T^2_{67}$.

The family of compactifications described above
is given by:
\eqn\family{
\CB^{18-s,s}_{\rm 89 moduli} \times
\CB^{2+s,2}_{\rm 67 moduli}
}
forming a 20-complex parameter subspace of
the family $\CN^{20,4}$.

Given a family $\CB^{\rho,2} \times \CB^{20-\rho,2}$
we define the {\it mirror family} to be
 $\CB^{\rho',2} \times \CB^{20-\rho',2}$ for
$\rho + \rho' =20$.
There is an obvious mirror map between these
families obtained by involution
\eqn\involution{
(x,y) \rightarrow (y,x).
}
It simply corresponds to exchanging the roles of
the compactification data of $T^2_{89}$ and $T^2_{67}$.

The relation of the involution \involution\
to what is usually understood by the mirror
map is explained by the geometrical interpretation
of the Narain moduli space explained in
\aspinwall (and reviewed in
appendix C). This provides the detailed
map between the moduli spaces for
$Het[T^4]$ and $IIA[K3]$. Consider the
interpretation of  \involution\ under this map.
As explained in appendix C.1, the duality map
$\Phi_{e,e^*}$ requires a choice of a hyperbolic plane
spanned by $e,e^*$.
Under the map $\Phi_{e,e^*}$ the moduli space
$\CB^{18-s,2}_{\rm 89 moduli} $ is mapped to the
moduli of complex structures of the elliptic
fibration and thence to the complex moduli
of the IIA K3 surface. Following through this
map on $T^2 \times T^2$ compactification
one finds that the
heterotic Wilson line moduli  $\vec A_6  + i \vec A_7 $
essentially determine the complexified
Kahler class of the dual IIA K3 surface.
Restricting to the family \family,
this only involves the Wilson lines of
the rank s group  $H_{89}$ unbroken
by the ``first''  torus $T^2_{89}$. Thus,
$\CB^{2+s,2}_{\rm 67 moduli}$ is mapped to the
complexified Kahler moduli of the IIA K3
surface, and the involution \involution\ is just
K3 mirror symmetry.

The complex structure moduli are more
naturally described in terms of algebraic
coordinates and the complexified Kahler
moduli are more naturally described in
flat coordinates. For this reason, the
$F$-theory map above coincides with
the K3 mirror map in the sense of
\lianyau.

\subsec{Examples of the arithmetic nature of the
$F$-map}

Conjecture 11.1 about the arithmetic nature of the
$F$-theory map can be checked for several
families of K3 surfaces.

\subsubsec{Equations from the Riemann relations}

One can write explicit families of
Kummer surfaces as quartic surfaces in $\IP^3$
using the Riemann theta   relations.
Specifically, let $A_\tau$ be a principally
polarized abelian variety with polarization
$L$. Suppose that    $\tau\in\CH_2$ is a period
matrix for $A_\tau$ and  define \mumford:
\eqn\thetfuns{
f_{\vec a}^{(n)}(\vec z) \equiv
{\vartheta\biggl[{\vec a/n  \atop  \vec 0 }\biggr](n\vec z \vert n \tau)}
}
where $n$ is a positive integer,
$\vec a\in \IZ^2/(n \IZ^2)$, and
$\vartheta$ is a Riemann theta function with
characteristics (we use the notation of \mumford).
The four functions
$X_{\vec a}\equiv f_{\vec a}^{(2)}$ are a basis
for  $H^0(A_\tau; L^2) $.
Using \mumford,  p.222, equation 6.6,  with
$n=2, z_1 = z_2= 2 \vec z$ we can express
6 even theta functions
in $H^0(A_\tau; L^4) $
as quadratic expressions in $X_{\vec a}$.
To be precise, we take: $Y_{\vec a}(\vec z)= f_{\vec a}^{(4)}$
with $\vec a\in \IZ^2/(4 \IZ^2)$
and find the relations:
\eqn\quadrels{
\eqalign{
\pmatrix{ X_{00}^2 \cr X_{01}^2\cr X_{10}^2\cr X_{11}^2\cr}
&=
\pmatrix{
Z_{00} & Z_{20} & Z_{02} & Z_{22} \cr
Z_{02} & Z_{22} & Z_{00} & Z_{20} \cr
Z_{20} & Z_{00} & Z_{22} & Z_{02} \cr
Z_{22} & Z_{20} & Z_{02} & Z_{00} \cr}
\pmatrix{ Y_{00} \cr Y_{20}\cr Y_{02}\cr Y_{22}\cr}
\cr
& \cr
\pmatrix{X_{00} X_{10}\cr X_{11} X_{01}\cr}
& =
\pmatrix{ Z_{10} & Z_{12} \cr Z_{12} & Z_{10}\cr}
\pmatrix{ Y_{10}+ Y_{30} \cr Y_{12}+Y_{32}\cr}
\cr}
}
where $Z_{\vec a} \equiv Y_{\vec a}(0)$ are
thetanullwherte.
Now, using the Riemann relations
(\mumford,  p. 223 with $n=4$)
we can find a quadratic relation on
the $Y$'s:
\eqn\quadrelsii{
\eqalign{
\lambda_1^2 (Y_{00}^2 + Y_{20}^2 + Y_{02}^2 + Y_{22}^2)
& + 2 \lambda_2^2 (Y_{00} Y_{20} + Y_{02} Y_{22})\cr
 =
\lambda_1 \lambda_2 \biggl( (Y_{10}+Y_{30})^2
&  +( Y_{12}+Y_{32})^2\biggr) \cr
\lambda_1  = f^{(8)}_{(2,0)}(0)
 + f^{(8)}_{(2,4)}(0) & + f^{(8)}_{(6,0)}(0)+ f^{(8)}_{(6,4)}(0)\cr
\lambda_2   = f^{(8)}_{(0,0)}(0)
 + f^{(8)}_{(0,4)}(0) & + f^{(8)}_{(4,0)}(0)+ f^{(8)}_{(4,4)}(0)\cr}
}
Combining \quadrels\ and \quadrelsii\ we obtain
 a quartic relation
on $X_{\vec a}$ embedding $A_\tau/\IZ_2$ into
$\IP^3$.  For further discussion of such equations see
\birklang.

As $\tau$ degenerates:
\eqn\degentau{
\tau \rightarrow \pmatrix{ \tau_1 & 0 \cr 0 & \tau_2 \cr}
}
the quartic equation degenerates to the
Segre embedding: $Q_2(X)^2\equiv (X_{00} X_{11} - X_{01} X_{10})^2=0$ and thus
fails to give an embedding
of the Kummer surface $Km(E_{\tau_1} \times E_{\tau_2})$.
There is a standard way to handle this difficulty.
We can write the quartic relation as:
\eqn\segtoo{
Q_2(X)^2 - t^2 F_4(X) = 0
}
Here $F_4$ is a quartic polynomial in
$X$. Moreover,   $t^2$ and the
coefficients of $F_4$
are rational expressions in thetanullwherte of
levels 4 and 8 such that $t^2$ vanishes
on the locus \degentau\ and $F_4$ does not.
(The actual expressions are rather complicated.)
We now introduce
a new (affine) coordinate $y$ and replace \segtoo\
  with the   equations:
\eqn\dblcvr{
\eqalign{
t y - Q_2(X) & =0 \cr
y^2 - F_4(X) & =0  \cr}
}
These are equivalent to \segtoo\ for $t\not=0$
and when $t=0$ they express the surface
as a  double cover of $\IP^1 \times \IP^1$ branched
on a degree $(4,4)$ curve.
The first equation is the Segr\'e embedding
$\IP^1 \times \IP^1 \hookrightarrow \IP^3$, so,
if we consider the fibration  say,  to the
first $\IP^1$ factor: $\pi:(y,X) \rightarrow
X_{00}/X_{10}\in \IC$ then
the fiber is an elliptic curve. Moreover, on the locus
\degentau\ the  coefficients  of $F_4$ are expressed
in terms of level $4$ and level $8$ genus one
theta functions of $\tau_1, \tau_2$.
Using the values of $\tau_1, \tau_2$ given in
\shodii\ and the theory of complex multiplication we
finally get an arithmetic Weierstrass model with
coefficients in a ray class field of $K_D$, as claimed.

\subsubsec{Commensurability relations on the
mirror map}

In \lianyau, Lian and Yau studied the K3
mirror map for certain one-parameter families
of K3 surfaces.
They showed that the mirror map satisfies a
comensurability relation:
\eqn\commens{
P(j,x) =0
}
where $P$ is a polynomial with integral coefficients
and  $x= x(q)$ is the mirror map. See also
\klm. By CM theory $x$
is therefore algebraic over $\widehat{K}_D$ at the attractor
points.

\subsubsec{The Morrison-Vafa $E_8 \times E_8$ family}

Morrison and Vafa  considered an
interesting 2-parameter family of Weierstrass
models with two $E_8$ singularities \vfmr:
\eqn\twoeate{
y^2 = x^3 + \alpha z^4 x + (z^5 + \beta z^6 + z^7)
}
The explicit
map to flat coordinates was found in \cardcurl\lerchstie :
\eqn\twopmap{
\eqalign{
j(T) j(U) & =- \bigl( 48 \alpha \bigr)^3\cr
(j(T)-1728)(j(U)-1728) & =
+  \bigl( 864 \beta \bigr)^2 \cr}
}
Once again we can apply class field theory to conclude that
$\alpha^3, \beta^2$ are in certain extensions of a
quadratic imaginary field. As we have seen in the
previous section, the attractor points
have $T,U\in K_D$ for some $D$.
Consider the field extensions:
\eqn\compositi{
\matrix{
  &   &  L= K_D(j(T), j(U)) &   &   \cr
   & \swarrow &   & \searrow & \cr
K_D(j(T)) &  &  &  & K_D(j(U)) \cr
  &  \searrow & & \swarrow & \cr
 &     & K_D &  &  \cr}
}
Now by a general theorem of Galois theory
(See, for example \langi\
ch. 8.1, Theorem 5)
$Gal(L/K_D) \hookrightarrow Gal(K_D(j(T))/K_D )
\times Gal(K_D(j(U))/K_D )$ is an injection so
$Gal(L/K_D) $ is abelian and by the second main
theorem of complex multiplication
$L$ is a ray class field.
(Because of  scaling of $x,y$ we are only
considering $\alpha,\beta$ as defined up to
cubic and square roots of one. )

\subsubsec{$F$-map for stable degenerations}

The most complete results available
for the $F$-map follows from the results of
Friedman, Morgan, and Witten \fmw. (
See also  \amiii\donagi\ for descriptions.)
These works give
a construction of the $F$-map in the
limit  $\im T \rightarrow \infty$.
The remaining moduli are the complex structure
of $T^2$, commonly called
$U$, and the $E_8 \times E_8$ Wilson
line moduli. It turns out that we can treat
the $E_8$ factors separately so we focus on a
single $E_8$
 and take  $\vec A \in \liet\otimes\IC  /\bigl[(\Lambda + U
\Lambda)\sdtimes W_{E_8}\bigr] $ where
$\liet$ is the Cartan subalgebra,  $\Lambda$ is
the root lattice, and $W_{E_8}$ is the Weyl
group  of $E_8$.

As shown in \fmw\amiii\ the F-theory dual of the
limit $\im T \rightarrow \infty$ is the stable
degeneration of the K3 surface to a union of
two rational elliptic surfaces
\foot{We abbreviate ``rational elliptic surface''
to $dP_9$, since they are almost del Pezzo.}
$S_1, S_2$
glued along a common elliptic curve. This
curve is identified with the heterotic torus
$E_U = \IC/(\IZ + U \IZ)$. The two $dP_9$'s
may be written in the Weierstrass form:
\eqn\dpnine{
\eqalign{
S_1 \qquad  y^2 & = 4 x^3 - \biggl(\sum_{i=0}^4 \alpha_i s_1^i t_1^{4-i}
\biggr) x - \biggl(\sum_{i=0}^6 \beta_i s_1^i t_1^{6-i} \biggr)\cr
 S_2 \qquad  y^2 & = 4 x^3 - \biggl(\sum_{i=0}^4 \alpha_{-i} s_2^i t_2^{4-i}
\biggr) x - \biggl(\sum_{i=0}^6 \beta_{-i} s_2^i t_2^{6-i} \biggr)\cr}
}
These are glued along the common elliptic curve,
\eqn\common{
Z Y^2 = 4X^3 - \alpha_0 t^4 X Z^2 - \beta_0 t^6 Z^3
}
over $[(s_1=0,t_1)]\sim [(s_2=0,t_2)]$,
where $(X:Y:Z)$ are homogeneous coordinates in
$\IP^2$.
The two sections of the $dP_9$'s  meet in a
point over $[(s_1=0,t_1)] \sim [(s_2=0,t_2)]$.
Suppose $U\in K_D$. As we have seen, we may
``choose a gauge,''  $t$, so that $\alpha_0,\beta_0$ are in
the ring class field $\widehat{K}_D$.

Following \fmw\ we can also account for the bundle
moduli. One key point is the mapping between   flat
gauge fields on $E_U$  and collections of points
on $E_U$ written in the Weierstrass form. For a $U(1)$
gauge field the map is standard. Let $z \sim z+1 \sim
z+U $ be a flat coordinate on $E_U$. Let
$\phi,\theta$ specify the holonomies around a basis of
one-cycles on $T^2$, with $0 \leq \phi, \theta \leq 1$.
Then we map
\eqn\flatwei{
A_{\bar z} = {\phi + U \theta \over  \im U}
\quad \rightarrow \quad (x=\wp(\phi + U \theta), y= \wp'(\phi+ U \theta) ).
}
By the second main theorem of complex multiplication if
$A_{\bar z}$ is in $K_D$ then after suitable rescaling
$(x,y) \rightarrow (t^2 x, t^3 y) $ the Weierstrass
coordinates are in a ray class field of $K_D$.
An appropriate rescaling is given by
\eqn\weber{
t^2 = {g_2(U) g_3(U) \over  \Delta(U)}
}
With this choice the $x$-coordinate becomes
a Weber function.

Now consider the full $E_8$ bundle moduli.
As explained in section 4 of \fmw, the moduli space
of flat $E_8$ bundles on an elliptic curve $E_U$
may be identified with the moduli of complex
structures on a corresponding del Pezzo surface
$\bar S$ together with a fixed anticanonical
divisor, identified as $E_U$.
One maps the flat gauge field to a collection of
$8$ points on $E_U$. These 8 points are then
identified with the intersection between  $E_U$ and a
collection of divisors forming a $\IZ$-basis for
the component $E_8(-1)$ of $H^2(\bar S;\IZ)$.
We now view $\bar S$ as a blow-down of
the  $dP_9$ called  $S$ above. From the complex structure of
$S$ we obtain the ``algebraic moduli coordinates''  $\alpha_i,\beta_i$  in
\dpnine,

The procedure of the previous paragraph can
be implemented quite explicitly following appendix A
of \gms, and we do so here. We choose a basis for
the Cartan subalgebra of $E_8$ so that a
flat gauge field may be considered as an
8-tuple
$\vec w = (w_1, \dots, w_8)$ with certain identifications
listed in \gms. We define nine points on
$E_U$ in flat coordinates by:
$w_i = z_i-z_0$, $i=1,\dots, 8$,
with $z_0 + z_1 + \cdots + z_8 =0$.
These determine nine points $e_i$ on the curve
\common. Their homogeneous coordinates are written:
\eqn\ninpts{
e_i = \bigl( X_i ,Y_i , Z_i \bigr)
}
with $X_i = t^2 \wp(z_i)   Z_i$,
$Y_i = t^3 \wp'(z_i)   Z_i $.
Equations for the surface $S$ can be written by
expressing it as a blowup of $9$ points in  $\IP^2$
sitting at the intersection of \common\ with a
second elliptic curve $Q(X,Y,Z)=0$. The curve
$Q$ is determined as follows. We have the equation
for $S$:
\eqn\dpnineii{
u (4 X^3 - g_2 X Z^2 - g_3 Z^3 - Y^2 Z) + Q(X,Y,Z)=0
}
where $u$ is an affine parameter for the base of the
elliptic fibration of $S$.
The coefficients of $Q$ are compleletely deteremined,
up to an overall scale, by enforcing linear equations
 requiring that the nine
points \ninpts\ lie on the cubic $Q=0$ \gms.

By choosing $t^2$ as in \weber\ and $Z_i=1$ we find
that $X_i, Y_i$ are in an abelian extension of
a ray class field, call it
$\tilde K_{\vec w, D}$,  of $K_D$ and consequently,
so are the coefficients of $Q$. Finally, the
cubic \dpnineii\ in $\IP^2$ can be put in Weierstrass form,
hence allowing us to extract $\vec\alpha,\vec\beta,$ by a
standard method described in
\clemens. This procedure can be implemented over
any field of characteristic zero so we conclude that
$\alpha_i, \beta_i \in \tilde K_{\vec w, D}$.
This finally completes our verification of
Conjecture 11.1  in the limit of a stable degeneration.

\newsec{Kronecker's Jugendtraum, Hilbert's Twelfth problem,
and Mirror symmetry}

Perhaps the most interesting implication of the
attractor mechanism from the purely mathematical point
of view is the possible relation between mirror symmetry
and generalizations of complex multiplication which
it suggests.

Consider the problem of finding all {\it abelian}
extensions of a number field $K$.
The Kronecker-Weber theorem states that
all abelian extensions of $K=\IQ$ are subfields of
some cyclotomic extension $\IQ[e^{2 \pi i /n}]$.
Thus, there is a ``magical''  transcendental
function, $x \rightarrow f(x) = \exp [2 \pi i x]$
whose values on torsion points in the
circle $f(x), x \in \IQ$  generate all
abelian extensions of $\IQ$.

What mathematicians refer to as
``Kronecker's Jugendtraum''  is an extension of
this phenomenon to quadratic imaginary
fields. Namely, an analogous situation
holds with $K=\IQ$ replaced by   imaginary quadratic
fields  $K=K_D$, and $f(x)$ replaced by
the elliptic functions: $j(\tau),\wp(z,\tau), \wp'(z,\tau)$
evaluated on   torsion points of elliptic curves $E_\tau$
of CM type. Kronecker's Jugendtraum is true:
This is the theory of complex
multiplication, outlined  above.

In his famous address to the ICM in 1900
Hilbert posed his $12^{th}$ problem  \hilbert
\foot{See especially \langlands\ in \hilbert. I thank
G. Zuckerman for bringing this reference to my
attention.}  in which he
encouraged mathematicians to:

\bigskip
\ndt
{\it `` ... succeed in finding and discussing those functions which play the
part for any algebraic number field corresponding to that of the
exponential function in the field of rational numbers and of the elliptic
modular functions in the imaginary quadratic number field.'' }

This has been partially solved by the Shimura-Taniyama
theory of abelian varieties of CM type \shimtan\langii\gshimura.
However, in  view of the relations we have found
between   complex multiplication,
K3 mirror symmetry, and the attractor points of supersymmetric
black holes, we cannot help speculating that
the transcendental functions provided by the mirror map
are just the functions which Hilbert was seeking.
Indeed, we can now recognize Hilbert's $12^{th}$
problem as the proper context for the attractor
conjectures of section 8.2.  Schematically the proposed
generalization may be summarized in the following table:
\eqn\generalize{
\eqalign{
{\rm elliptic\quad  curve} \qquad & \rightarrow \qquad  {\rm Calabi-Yau \
d-fold}\cr
\tau \in \CH_1 \qquad & \rightarrow \qquad t^a\in \widetilde{\CM}\cr
{\rm   discriminant}\  D\qquad  & \rightarrow \qquad \gamma\in H^d(X;\IZ) \cr
a \tau^2 + b \tau + c =0 \qquad & \rightarrow  \qquad 2 \im \bar C \Omega =
\hat \gamma\cr
\IQ[\sqrt{D}] \qquad & \rightarrow \qquad K(\gamma)\cr
\tau \mapsto j(\tau) \qquad & \rightarrow \qquad
{\rm The\  mirror\  map} \cr
({a + b \tau \over  N} , \tau) \mapsto (y_{a,b,N} = \wp', x_{a,b,N} = \wp)
\qquad
& \rightarrow \qquad {\rm Coordinates\ of\  }\cr
& \qquad\qquad  {\rm \widehat{K}(\gamma) \ rational \
points} \cr  }
}

Of course, if the attractor conjectures turn out to
be correct (or even ``approximately correct'') then
one would want to know precisely which number
fields $K(\gamma)$ and which extensions
$\widehat{K(\gamma)}$  actually appear.

\newsec{Possible applications of arithmetic to
the study of BPS states}

This section is {\it even more speculative}
 than the previous
sections. In the above  sections  we have shown
that, as complex varieties, the
attractor varieties are arithmetic.
Indeed they are defined over ``class fields'' of
$K_D$ or abelian extensions of class fields of
$K_D$.  In this section we
sketch a few consequences of this result.

\subsec{Galois action on  attractor vectormultiplet moduli}

One of the main points of class field theory
is that $\widehat{K _D}$ is Galois over $K_D $,
and $Gal(\widehat{K _D}/K_D)$ is in
fact isomorphic to the class group $C(D)$ of
the field $K_D$ itself.
In the case of ring class fields of an
order $\CO(D)$ the isomorphism can
be expressed very  concretely:
\eqn\clssfield{
\eqalign{
[\tau] & \rightarrow \sigma_{[\tau]}\in Gal(\widehat{K _D}/K_D)\cr
j([\bar \tau_i] * [\tau_j]) & = \sigma_{[\tau_i ] }(j[\tau_j]) \cr}
}

\bigskip
\noindent
{\bf Example.} Continuing our example
\explclss\   with $D=-20$
we have seen that $\widehat{K}_{D=-20} =K_{D=-20}(\sqrt{5}) = \IQ(\sqrt{-1},\sqrt{-5})$. See \explii. Note that indeed
the Galois and class groups are isomorphic:
${\rm Gal}(\widehat{K _D}/K_D) \cong \IZ/(2\IZ)\cong C(-20)$.
The isomorphism \clssfield\ can be verified explicitly:
\eqn\explcfim{
\eqalign{
 (50 - 26 \sqrt{5})^3=    j({1+ i \sqrt{5} \over 2} )= j([\tau_2]*[\tau_1]) & =
 \sigma_{[\tau_2]}(j([\tau_1]))   = \sigma_{[\tau_2]}\bigl[ j(i \sqrt{5})\bigr]
\cr
 =  & \sigma( (50 + 26 \sqrt{5})^3)   =
(50 - 26 \sqrt{5})^3\cr}
}

Using \clssfield\ it
 follows that  ${\rm Gal}(\bar{\IQ}/\IQ)$ acts on the (complex structure)
attractor IIB vectormultiplet
moduli, and permutes the attractor moduli
at fixed discriminant.
Thus, the Galois group extends the $U$-duality
group and ``unifies'' the different attractor points at
discriminant $D$ enumerated in section 3.

The physical use of this group action is at present
unclear, although it is aesthetically pleasing that
a group of some mathematical importance does
unify the ``accidental degeneracy'' of $U$-duality
inequivalent classes with fixed horizon geometry,
as described above. One error we should guard
against: the Galois action is {\it not} a symmetry
of string theory. This can be seen simply by
examining the mass-spectrum of BPS
states at the attractor points.
As we showed in section 5 above, in the case of the
FHSV model the BPS spectrum is
completely disjoint at the different attractor points
permuted by ${\rm Gal}(\bar{\IQ}/{\IQ})$.

\subsec{Galois action on positions of D-branes}

It is also worth noting that in string duality
moduli space and spacetime are often
identified with one another. Since there can
be distinguished coordinates in moduli
space there can be distinguished coordinates
in spacetime (somewhat at odds with
general covariance) so it might make sense
under some circumstances to ask about
the physical meaning of $K$-rational points
on an arithmetic variety.

An interesting example of this is provided by
the position of D7 branes in 8-dimensional
F-theory compactifications. It follows from
the results of section 11 that attractive
K3 surfaces can be given arithmetic Weierstrass
presentations. In these coordinates the
  D7 branes are located at arithmetic points
of the variety. For example, in the
Morrison-Vafa family
\twoeate\ the discriminant of the cubic is
\eqn\cudis{
\Delta = s^{10} t^{10} \bigl[ 4 \alpha^3 (st)^2
+ 27(t^2 + \beta st + s^2)^2 \bigr]
}
There are two $E_8$ singularities at $s=0$ and $t=0$,
each containing 10 7-branes.
There are 4 other singularities at the solutions of
\eqn\foroth{
z^4 + 2 \beta (z^3+ z) + (4 (\alpha/3)^3 + \beta^2 +2) z^2 + 1 = 0 . }
So the positions of the D7's, in $z$ coordinates,
 are at algebraic integers
in  a quartic extension of $\widehat{K}_{D}(\beta)$.

In general, if Conjecture 11.1 is correct
the 7-branes will be located at arithmetic points,
and their positions will be permuted by
${\rm Gal}(\bar{\IQ}/\IQ)$. The study of this
and related actions on D-brane positions
might turn out to be very interesting.

\subsec{A calculation indicating a possible
relation of $\dim \CH_{BPS}^\gamma$ to  heights}

In Diophantine geometry one
can associate a real number (often an
integer)
$H(X,\CL)$ to a polarized variety
$(X,\CL)$ defined over a number field.
This number $H(X,\CL)$ is
called a multiplicative height of $X$.
On the other hand, thanks to
the attractor mechanism, given
an area code $(\gamma,\CB)$ we can associate
an attractor variety
$X_{(\gamma,\CB)}$ and a positive integer,
the dimension $\dim \CH_{BPS}^\gamma$
of the space of BPS states of charge $\gamma$
for IIB compactification on $X_{(\gamma,\CB)}$.
Conjecturally, the attractor varieties
$X_{(\gamma,\CB)}$  are arithmetic.
\foot{For brevity we write $X_\gamma$ below.}
One might wonder whether
$\dim \CH_{BPS}^\gamma$ defines
a notion of a height function
on the varieties $X_{(\gamma,\CB)}$.
If so,
results from Diophantine geometry
might be brought to bear on understanding
the spaces of BPS states (or conversely).

There are many notions of ``height.''
Here we focus on the ``Faltings height''
or ``stable modular height'' of a
Calabi-Yau $X$ defined over a number
field $K$.
Roughly speaking, we form the family
$\pi: {\cal X} \rightarrow S$ with
$S=Spec \CO_K$ together with
the associated line $\CL=R\pi_* \omega_{{\cal X}/S}$,
where $\omega_{{\cal X}/S}$ is the relative dualizing
sheaf. We now define the degree of $\CL$, $\deg \CL$,
by summing a logarithmic norm on
 the fibers of $\CL$ over all the places of $S$.
The norm at the places at
infinity is based on the hermitian form:
$  i \int_X  \Omega \wedge \bar \Omega$.
The choice of normalization of $\Omega$
cancels from the contribution at the finite
places.
See \mazur\silvermanag\langdg\ for precise
descriptions.  It is     more convenient to work
with the logarithmic height $h(X,\CL) = \log H(X,\CL)$.
This height is related to the degree of
$\CL$    by
$ [K:\IQ]  h(X,\CL) = \deg \CL$, where $X, \CL$
are defined over $K$.

In this section we argue that, at least in
some examples,
if $X_\gamma$ is an attractor variety with
discriminant $D$ then
$  h(X_\gamma,\CL)$ has asymptotics for
large $D$ similar to $\log[S/\pi]$ where
$S$ is the black hole
entropy, suggesting that
$\dim \CH_{BPS}^\gamma$ might actually
be related to heights. Specifically, we consider the
Faltings height of the $\CN=8$ attractors,
regarded as principally polarized varieties
defined over a number field. We will
argue (nonrigorously!) that, for certain sequences
of  charges with $I_4(\gamma)\rightarrow \infty$,
\eqn\htbound{
 h(X_\gamma,\CL)
\sim
\kappa  \log[  S/\pi ]
}
where $\kappa$ is a constant of order $1$.
Optimistically one might hope that $\kappa$ is
rational.

In section 6 we showed
that the $\CN=8$ attractor is
isogenous to a product of 3 elliptic
curves with modular
parameter $\tau(\gamma) = i \sqrt{I_4(\gamma)}$.
We denote the elliptic curve by $E_{\gamma}$.
By the definition of
the degree of a metrized line bundle
it is clear that $h(A_1 \times A_2,
 \CL_1 \otimes \CL_2) = h(A_1,\CL_1) + h(A_2,\CL_2)$.
Thus we will concentrate on an  estimate for the height
of the elliptic curve $E_{\gamma}$.
The height of $E_\gamma$ can be studied using
a result of Silverman, but before embarking
on this we must address the issue of the isogeny.

The height of a variety changes under isogeny.
By a result of Faltings
(see, e.g. \silvermanag, Lemma 5, p. 18)
the heights $h(X_\gamma)$ and
$h(E_{\gamma} \times E_{\gamma}\times E_{\gamma})$
differ by logarithmic corrections in
the prime factors of the degree of the
isogeny. Optimistically
these will lead to $U$-duality invariant
corrections $\sim log I_4(\gamma)$, but whether
this is so requires much more study. We will
dodge the issue by taking the case where
the charges are such that
 $I_4(\gamma)\rightarrow \infty$
with $\det R$ held fixed. In this case the
correction for the isogeny will not matter.

Let $D=-I_4(\gamma)$. We will assume that it
is a fundamental discriminant, i.e.,
that $D$ is the
discriminant of $K_D$. This means that
$D=0,1\mod\ 4$, and,
if $D=1 \mod\ 4$, then it  is squarefree, while
  if $D=0 \mod\ 4$ then
$D/4 = 2,3 \mod\ 4$ and is squarefree.
We choose the Weierstrass model for
the curve $E_\gamma$ defined over the
Hilbert class field $\widehat{K}_D$, as described
in section 7.2.  We write it in the form:
\eqn\overarr{
\eqalign{
y^2 & = 4 x^3 - g_2 x - g_3 \cr
g_2& = 27 j (j- (12)^3) \cr
g_3 & = 27 j (j-(12)^3)^2\cr
\Delta & = g_2^3 - 27 g_3^2 = 2^6 \cdot 3^{12}\cdot
j^2 (j-(12)^3)^3\cr}
}
so that the coefficients are in the ring of
integers   $\CO_{\widehat{K}_D}$.

The Faltings height for a curve over a number
field can be computed using
a result of Silverman (Proposition 1.1, p. 254
in \silvermanag). Applying it in the present
example we have:
\foot{There is an unfortunate clash of notation.
Recall that  $h(D)$ stands for the class number.}
\eqn\heightcrve{
\eqalign{
24 h(D) h(E_\gamma /\widehat{K}_D ) & =
\log \vert N_{\widehat{K}_D/\IQ}(\CD_{E_\gamma/\widehat{K}_D })\vert +
\sum_{k=1}^{h(D)} \CR(\tau_k)\cr
\CR(\tau) & \equiv -2 \log\biggl[ (\Im \tau )^6\vert \eta^{24}(\tau)\vert
(2\pi)^{-12}  \biggr]\cr}
}
Here $\CD_{E_\gamma/\widehat{K}_D}$ is the ``minimal discriminant.''
It is a certain integral ideal in $\widehat{K}_D$
which divides the principal ideal
$(\Delta)$ obtained from \overarr.
(For a definition see \silveraec, p. 224.)
The second sum is over representatives for the
ideal class group. We write these
representatives as  $\tau_k = (-b_k + \sqrt{D})/2 a_k$
and assume they are in the fundamental domain
for $PSL(2,\IZ)$.

In order to estimate the growth at large $D$ we
need some estimate of the norm of the minimal
discriminant. While one can probably do better,
we will be content to note that
\eqn\lowbnd{
0< \log [\vert N_{\widehat{K}_D/\IQ}
(\CD_{E/\widehat{K}_D})\vert ]
}
since $\CD_{E/\widehat{K}_D}$ is an integral ideal and that
\eqn\guessi{
\log \vert N_{\widehat{K}_D/\IQ}(\CD_{E/\widehat{K}_D})\vert
< 12 h(D) \log[ 18] + 4 \sum_{k=1}^{h(D)} \log \vert j(\tau_k)\vert + 6
\sum_{k=1}^{h(D)}\log \vert j(\tau_k)-j(i) \vert
}
since $\CD_{E/\widehat{K}_D}$ divides $(\Delta)$.
Putting these bounds together we can estimate
the growth of $h(E_\gamma/\widehat{K}_D)$ in terms
of averages over the class group:
\eqn\avergs{
\langle \CR(\tau)\rangle < 24 h(E_\gamma/\widehat{K}_D)
< 12 \log[18] + 4 \langle \log \vert j(\tau)\vert\rangle
+ 6 \langle \log \vert j(\tau)-j(i) \vert\rangle
+  \langle \CR(\tau) \rangle
}
where
\eqn\defaverg{
\langle f(\tau)\rangle \equiv {1 \over h(D)} \sum_{k=1}^{h(D)}
f(\tau_k)
}
for any function $f$.

In order to estimate these averages we need some understanding
(at least at the heuristic level)
of the distribution of class representatives $[\tau_k]$ in
the fundamental domain for large $D$. This was developed
in collaboration with Steve Miller and will be described
elsewhere. Roughly speaking, the $\tau_k$ are equi-distributed
with a cutoff $\im \tau\leq \vert D\vert^{1/2}/2$
coming from the principal class.
Using this model one finds the rough estimates:
\eqn\logavs{
\eqalign{
\langle \CR(\tau)\rangle & \sim 6 \log\vert D \vert \cr
\langle \log \vert j(\tau)\vert\rangle & \sim 3 \log\vert D \vert \cr
\langle \log \vert j(\tau)-j(i) \vert\rangle & \sim 3 \log\vert D \vert \cr}
}
These estimates have been checked (by Steve Miller) with
extensive numerical calculations.

A consequence of \logavs\ is that
\eqn\bndhet{
{1\over 2}
 \log\vert D\vert \grsim
h(E_\gamma/\widehat{K}_D) \grsim 3  \log\vert D\vert
}
which is our main estimate, leading to the guess
\htbound.

\bigskip
\noindent
{\bf Remarks.}

\item{1.} The first term on the right hand side of
\heightcrve\
can probably be handled much more accurately.
Thanks to results of Gross and Zagier \gzsm\
the minimal discriminant can probably be
calculated explicitly. For example, if
$D=-4m$ and $m\not=0 \mod 3$ then
$j(\tau)$ is the cube of an algebraic
integer in $\widehat{K}$. This allows one
to get a better estimate of the minimial
discriminant.

\item{2.} There are two
  ``moral reasons'' to hope for  a
connection between heights and entropy.
First, according to the mathematicians,
the height ``measures the arithmetic complexity''
of $X_\gamma$. On the other hand   since
arithmetic questions do enter in determining the
existence of BPS states one might hope that
an ``arithmetically  complicated space''
supports  many
BPS states.   Second the entropy is given
in terms of the holomorphic $(3,0)$ form
$\Omega$ on the attractor variety by
\eqn\morres{
\log[S/\pi] = \log\biggl\vert\int_\gamma\Omega\biggr\vert^2-
\log\vert \int_X \Omega\wedge\bar\Omega\vert
}
which bears a remote resemblance to the definition
of heights as a sum of contributions from finite
and infinite places, respectively.

%\MillerAG
\lref\MillerAG{
S.~D.~Miller and G.~W.~Moore,
``Landau-Siegel zeroes and black hole entropy,''
arXiv:hep-th/9903267.
%%CITATION = HEP-TH 9903267;%%
}

\item{3.} For some further discussion along these lines see \MillerAG.

\newsec{Conclusion: Summary of the main results and speculations}

In this paper we have reviewed (and, we hope,
clarified) some of
the literature on the attractor equations. We have
also stated some concrete results and offered
(too) many speculations. Here is a summary of the
main results and   speculations:

\bigskip

\centerline{\it Summary of the main results:}

\bigskip

\item{1.} The number of $U$-duality inequivalent
charges with fixed discriminant (i.e., horizon
area in the supergravity approximation) is given
by class numbers, and grows as a power of the
area for large areas. (Section 3, equations
\clssgrwth,\grwth,\grwthii.)

\item{2.} The attractor varieties for compactification
on $  K3\times T^2,T^6 $ are described explicitly.
They are related to products of three isogenous
elliptic curves with complex multiplication.
(Corollary 4.4.1, Proposition 6.3.1).
Consequently, the attractor varieties
for these compactifications are arithmetic.
(Section 7.2)

\item{3.} The BPS mass-squared spectrum at attractor points
in the FHSV model is integral, and may be described in
terms of norms of ideals in quadratic imaginary fields.
(Section 5)

\item{4.} Examples of exact CY 3-fold attractor varieties
are described in section 8.3.

\item{5.} There can be more than one basin of
attraction for the dynamical system \dynsys.
An example is given in section 9.2.

\item{6.} Under 8D $F$-theory duality, the
attractive K3 surfaces
map to rational conformal field theories. (Section 10).

\item{7.} The mirror map for K3 surfaces has properties
generalizing the arithmetic properties of the $j$-function,
at least in some examples.
(Section 11).

\bigskip
\bigskip

\centerline{\it Summary of the main speculations:}

\bigskip
\bigskip

\item{1.} Attractor varieties are arithmetic, and
define arithmetic values of the mirror map.
(Conjectures 8.2.1, 8.2.2, 8.2.3. Section 12.)

\item{2.} Type II compactifications on attractor
Calabi-Yau varieties are dual to heterotic
compactifications on rational conformal field theories.
(Section 10.4)

\item{3.} Attractor points in vectormultiplet moduli
space form orbits for an action of ${\rm Gal}(\bar {\IQ}/\IQ)$.
(Sections 13.1, 13.2)

\item{4.} The number of BPS states of charge $\gamma\in H_3(X;\IZ)$, $\dim \CH_{\rm BPS}^\gamma$, defines a
height function for the attractor varieties $X_\gamma$.
(Section 13.3)

\bigskip
%\GubserBC
\lref\GubserBC{
S.~S.~Gubser, I.~R.~Klebanov and A.~M.~Polyakov,
``Gauge theory correlators from non-critical string theory,''
Phys.\ Lett.\ B {\bf 428}, 105 (1998)
[arXiv:hep-th/9802109].
%%CITATION = HEP-TH 9802109;%%
}

Many further speculations could be mentioned, but
we spare the intrepid reader. We only mention that
one might suspect that the
Galois group will play a role in the conformal
field theory obtained from brane dynamics as in
\maldacena\wittenholog\GubserBC, since those conformal
field theories are obtained only when the
moduli take their attractor values.

\bigskip
\centerline{\bf Acknowledgements}\nobreak

I would like to thank E. Verlinde for sparking
my interest in attractor points.
B. Lian and G. Zuckerman investigated connections
between some rational conformal field theories and
abelian varieties of CM type in 1993 (unpublished).
I thank them for explanations of their work at the
time. I am particularly indebted to P. Deligne and
D. Morrison for answering many questions and making
some important points.
Finally I would also like to thank R. Borcherds,
P. Candelas, R. Donagi,
S. Ferrara,  H. Garland, A. Gerasimov, G. Gibbons,
J. Harvey, G. Horowitz, R. Kallosh, A. Losev, J. Maldacena,
 M. Mari\~no,
S. Miller, R. Minasian, D. Morrison,
N. Nekrasov, T. Pantev, R. Plesser,
W. Sabra,   A. Silverberg, J. Silverman,
A. Strominger,  A. Taormina,
A. Todorov and G. Zuckerman   for useful
correspondence and discussions. I would
like to thank the Aspen Center for Physics,
and the organizers of the Amsterdam Summer Workshop
on String Theory and Black Holes,
for hospitality during the completion of this
paper. This work is supported by
 DOE grant DE-FG02-92ER40704.

\appendix{A}{Quadratic imaginary fields}

A quadratic imaginary field is a number field of the
form
\eqn\field{
K_D\equiv \IQ[\sqrt{D}] \equiv \{
r_1 + r_2 \sqrt{D}: r_1, r_2\in\IQ\}
}
with $D<0$.
Of course, for any rational number $\ell$ both $D$ and
$D \ell^2$ generate the same quadratic extension.
(This can lead to some confusion.) We usually
reserve $D$ for an integer $=0,1\mod 4$,
not necessarily square-free.

The ring of integers $\CO_K$ in $K$ are the
solutions of monic polynomials with rational integral
coefficients. If $N$ is squarefree and
$K_D = \IQ[\sqrt{N}]$ this ring is,
explicitly:
\eqn\ringint{
\eqalign{
\CO_{K_D} & = \IZ[\sqrt{N}] \qquad \qquad N \not= 1\ \mod 4\cr
& = \IZ[{1 + \sqrt{N}\over  2} ] \qquad N  = 1\ \mod 4\cr}
}

If $\alpha_1, \dots, \alpha_n$ form an integral
basis for the ring of integers
$\CO_K$ in any number field $K$ of degree $n$ then the
discriminant of the field is defined by:
\eqn\discriminant{
d(K) = \det (\Tr(\alpha_i \alpha_j) )
}
In particular, if $N$ is square-free then $\IQ[\sqrt{N}]$
has discriminant
\eqn\discqi{
\eqalign{
d(\IQ[\sqrt{N}]) & = 4 N \qquad N \not=1 \mod 4 \cr
& = N \qquad N = 1 \mod 4 \cr}
}
(Warning: the field $K_D$  is not necessarily of
discriminant $D$.)

An {\it order} $\CO$ in a number field $K$ is a subring of
$K$ containing $1$ which is a $\IZ$-module of
rank   equal to the degree
of $K$. If $\{ \alpha_1, \cdots, \alpha_n \}$ is a
$\IZ$-basis then  since $\CO$ is a ring:
\eqn\multoh{
\alpha_i \cdot \alpha_j = N_{ij}^{~k} \alpha_k
}
so $\alpha_i$, being an eigenvalue of the
integral matrix $(N_i)_j^{~k}=N_{ij}^{~k}$ is an algebraic
integer. Thus $\CO \subset \CO_K$ and we
may think of an order as a finite index sublattice
of $\CO_K$ closed under multiplication.
If $D=0,1\mod 4$ we can construct
an order $\CO(D)$  in $K=\IQ[\sqrt{D}]$:
\eqn\order{
\eqalign{
\CO(D) & = \IZ \oplus \omega \IZ  \cr
\omega & = {D + \sqrt{D} \over  2} \cr}
}
This order has discriminant $D$ as one
easily checks using   the identity:
\eqn\sqrom{
\omega^2 = -{D(D-1) \over  4} + D \omega
}

The index of $\CO(D)$ in $\CO_K$ is called the
``conductor.''  Sometimes it is useful to write:
\eqn\somsu{
\CO(D) = \IZ + f \CO_K
}
If $f=1$ we get the ring of integers $\CO_K$.

An {\it ideal} in $\CO(D)$ is a subring
${\bf \underline{a } }$ of $\CO(D)$
such that for all $x \in \CO(D)$, $x {\bf \underline{a } }
\subset {\bf \underline{a } }$. It can be shown
(\cohen, 5.2.1) that
every ideal in $\CO(D)$ can be put in Hermite normal
form:
\eqn\hnf{
{\bf \underline{a } } = \ell_1 \IZ + (\ell_2 + \ell_3 \omega) \IZ
}
where $0 \leq \ell_2 < \ell_1$, $0<\ell_3$, $\ell_3 \vert \ell_1, \ell_2$
and $\ell_1/\ell_3$ divides
\eqn\idlcondi{
{D \over  4} - \bigl(\ell_2/\ell_3 + D/2\bigr)^2
}
or, equivalently,
\eqn\idlcondii{
\ell_1 \ell_3 \biggl \vert
\ell_2^2 + 2 \ell_2 \ell_3 D + \ell_3^2 {D(D-1) \over  4}
}
The condition \idlcondii\ is necessary and sufficient
for \hnf\ to be an ideal in $\CO(D)$.

Now we may associate certain equivalence classes of
ideals in $\CO(D)$ with certain equivalence classes of
quadratic forms of discriminant $D$.
The map from forms of
discriminant $D$ to ideals is given by:
\eqn\fridl{
\pmatrix{a & b/2 \cr b/2 & c \cr}
 \rightarrow  {\bf \underline{a } }= a \IZ + {-b + \sqrt{D} \over  2} \IZ
}
This map is not 1-1. Indeed, forms related by
``axion shifts''
\eqn\axionshf{
\eqalign{
\pmatrix{a & b/2 \cr b/2 & c \cr}  & \rightarrow
\pmatrix{1 & m \cr 0 & 1 \cr}^{tr} \pmatrix{a & b/2 \cr b/2 & c \cr}\pmatrix{1
& m \cr 0 & 1 \cr} \cr
(a,b,c) & \rightarrow (a, b+ 2am , c + bm + am^2) \cr}
}
map to the same ideal.
For the inverse map we choose an integral basis
 of the ideal
\eqn\intgrlbas{
 {\bf \underline{a } } =
\omega_1 \IZ \oplus  \omega_2 \IZ
}
 and consider the map from ideal-with-basis to
binary quadratic forms ($x,y$ are real):
\eqn\idlfrm{
\omega_1 \IZ \oplus  \omega_2 \IZ \rightarrow
\sqrt{-D} {\vert (x \omega_1 - y   \omega_2) \vert^2 \over
\vert \omega_1 \bar \omega_2 - \omega_2 \bar \omega_1 \vert}
=\sqrt{-D} {\vert x - \tau y \vert^2 \over  2 Im \tau}
= a \vert x - \tau y \vert^2
}

Ideals can be multiplied in an obvious way. A principal
ideal is one of the form $(\alpha)\equiv \alpha \CO(D)$.
An {\it ideal class} is an equivalence class of ideals
under multiplication by principal ideals. One key
result (see, e.g., \cox, theorem 7.7) is that proper
equivalence classes of positive binary quadratic forms
of discriminant $D$ are in 1-1 correspondence with
ideal classes of the order $\CO(D)$.

The norm of an element $\alpha \in K$ in a Galois
extension is
the product over the conjugates. For a quadratic
extension of $\IQ$ this is just $N(\alpha) = \alpha \bar \alpha$.
The norm of the {\it ideal} ${\bf \underline{a } }$,
denoted
$N({\bf \underline{a } })$ is
the order of ${\bf \underline{a } }$ in $\CO(D)$
considered as a sublattice:
\eqn\normideal{
N({\bf \underline{a } })\equiv  \vert \CO(D)/{\bf \underline{a } }\vert
}
This is easily computed since the lattice
with oriented basis
$\langle \omega_1, \omega_2 \rangle$ has
unit cell of volume: $\Im \bar \omega_1 \omega_2$.
For example, the norm of \hnf\   is given by
$\ell_1 \ell_3$. The norm of \fridl\ is
just
\eqn\explnorm{
N( a \IZ + {-b + \sqrt{D} \over  2} \IZ
)=a
}

The norms of the integral ideals in an
ideal class are exactly the integers represented
by the corresponding quadratic form. This set of
integers only depends on the $SL(2,\IZ)$ class of
the form.
Since $f(x,y) = N({\bf \underline{a } }) \vert x - \tau y \vert^2$
we see that this is the norm of the integral ideal
${\bf \underline{a } } ( x - \tau y)$.

\appendix{B}{Lattice embedding and the
Smith-Minkowski-Siegel formula}

\subsec{Some definitions}

In addressing problems of duality symmetry
we often run into the following general
{\it lattice embedding problem}:

Given a lattice $S$ and a unimodular lattice
$L$,

a.) Is there an embedding $S \rightarrow L$?

b.) How many inequivalent embeddings are there?

These problems  can be
addressed using methods described in
\nikulin\dolgachevi\slg, and are solved by
the Nikulin embedding theorem, \nikulin,
Proposition 1.14.1.
\foot{I thank D. Morrison for explaining some of the
key ideas to me.} We collect two definitions and
a few statements useful for understanding the
calculation in the text.

\bigskip
\noindent
{\bf Definition.} If $S$ is an
integral lattice $D(S) \equiv S^*/S$ is a finite abelian
group called the {\it discriminant group} \nikulin\
or {\it dual quotient group} \slg. It comes equipped
with a binary quadratic form:
$q(x+S) = x^2 \mod 2 \IZ$ (for $S$ even).
$q$ is called the discriminant form.

\bigskip
\noindent
{\bf Definition.} Two quadratic forms are in
the same {\it genus} if they have the same
signature and   discriminant form $q$.

One also encounters the definition that
two forms are in the same genus if they
are equivalent over the $p$-adic integers
$\IZ_p$ for all $p$, including $p=\infty$.
As shown in \nikulin\ the above is an
equivalent definition.

In addressing the lattice embedding problem
the following statements are very useful.
They can all be found in \nikulin.

1. Even overlattices are in 1-1 correspondence with
isotropic subgroups of the discriminant group.
(\nikulin, Prop.  1.4.1).

2. An embedding of an even lattice $S$ in a unimodular
lattice $L$ determines
$L$ as  an even overlattice of $ S \oplus K$
where $K \equiv S^\perp$. In this case the
isotropic subgroup of the overlattice is maximal
isotropic. Moreover, there
is an isomorphism $\psi: D(S) \rightarrow D(K)$
and $q_K \cong  - q_S$.

3. Conversely, given a lattice $S$ in the
genus   $(t_+,t_-,q_S)$ and a lattice $K$ in the
genus $(l_+-t_+, l_- - t_-, q_K)$ and an
isomorphism $\psi: D(S) \rightarrow D(K)$
such that $q_S \psi = - q_K$ there exists an
embedding of $S$ into a unimodular lattice
of signature $(l_+, l_-)$.

4. By definition, two embeddings are equivalent if
there is an $O(L)$ transformation taking one to
the other. Two embeddings are equivalent iff there
exists $\phi\in O(K)$ such that
$\psi \bar \phi = \psi'$, where $\bar \phi$ is the
induced map on $D(K)$.
(\nikulin 1.6.1).

\subsec{Proof of $(3.23)$ }

The bound \grwth\  is obtained using the
``Smith-Minkowski-Siegel mass formula.''
For a description, see \slg\csiv.

The positive primitive lattice $L_{p,q}$ determines a
class with discriminant $D$.
For each  $S$-duality class choose a representative
form $Q_{(a_i, b_i, c_i)}$, $i = 1, \dots, h(D)$
and denote the corresponding rank 2 lattice
by $L_i$. One embedding
of $L_i\hookrightarrow II^{10,2}$
is defined by gluing with the orthogonal
complement $K = E_8(-1) \oplus L_i(-1)$,
but there might be other
inequivalent embeddings of
$L_i$ into $II^{10,2}$. We denote the number
of inequivalent embeddings by $\CE(L;II^{10,2})$ so
that
\eqn\nmbrin{
\CN(D) = \sum_{i=1}^{h(D)} \CE(L_i;II^{10,2}) .
}

Now we must estimate $\CE(L_i;II^{10,2})$.
As explained in \nikulin\dolgachevi, in order
to be able to find glue vectors $g_\alpha$ so that:
\eqn\gluvcts{
II^{10,2} = \amalg_\alpha \biggl( L\perp K + g_\alpha \biggr)
}
the
complement $K$ to $L$ must be in a genus
determined by $L$. The genus of $K$
is determined by the signature (in this
case $(-1)^{10}$) and the quadratic form
induced on $K^*/K$ to be $g(K)=(10,0,-q_L)$.
It follows from the discriminant
form technique that \nikulin\mirandai:
\eqn\cembd{
\CE(L_i;II^{10,2}) = \sum_{g(K_j) =(10,0;-q_i) } [O(D(K_j)): Image(O(K_j)
\rightarrow O(D(K_j))]
}
where the sum runs over inequivalent classes of
positive definite lattices $K_j$ in the genus $(-^{10},+^0;-q_i)$,
where $q_i$ is the quadratic form on the discriminant group
$L_i^*/L_i$. Note that $O(D(K_i))$ is a finite group.

In general there is no good formula for the class number
of lattices with dimension $>2$
and the ``right'' thing to do is count the ``mass'' given
by the Smith-Minkowski-Siegel formula \csiv.
The ``mass'' of a genus is defined to be:
\eqn\massfrm{
\CW([\Lambda]) \equiv
\sum {1 \over  \vert Aut(\Lambda) \vert }
}
where we sum over the classes in a genus.

We can use the Smith-Minkowski-Siegel formula
to get a {\it lower bound}  for the number of $U$-duality
inequivalent charges $(p,q)$ at fixed discriminant $D_{p,q}$
because:
\eqn\inequalty{
\eqalign{
\CN(D) & = \sum_{i=1}^{h(D)} \CE(L_i;II^{10,2}) \cr
& =   \sum_{i=1}^{h(D)}   \sum_{K_j: g(K_j) =(10,0;-q_i) } [O(D(K_j)):
Image(O(K_j) \rightarrow O(D(K_j))]  \cr
&\geq  \sum_{i=1}^{h(D)}   \sum_{K_j: g(K_j) =(10,0;-q_i) }  1  \cr
&\geq  \sum_{i=1}^{h(D)}   \sum_{K_j: g(K_j) =(10,0;-q_i) }  {1 \over  \vert
O^+(K_j) \vert   } \cr}
}
The mass formula has the general form:
\eqn\smsform{
\sum_{K_j: g(K_j) =(10,0;-q_i) }  {1 \over  \vert O^+(K_j) \vert   }
= \beta_{\infty} \prod_{p<\infty} \beta_p
}
The factor at the place at $\infty$,  $\beta_\infty$, grows
like $\vert D \vert^{(n+1) /2}$. Here and below
$n=\dim K$. In our case $n=10$
  gives  $\vert D \vert^{11 /2}$
and is independent of the genus
\cassels, Appendix B, 3.6. However, the
factors at the finite places can change the
asymptotics at large $D$, so we must be more
careful.

In order to understand the factors at the finite
places we need to
use the more precise formula of \csiv, eqs. 2,3.
This formula  has the shape:
\eqn\csform{
\eqalign{
\CW([\Lambda]) & = 2 \pi^{-n(n+1)/4} \cdot \prod_{j=1}^n \Gamma(j/2)\cdot \prod_{\rm p\ prime} (2 w_p([\Lambda])) \cr
2w_p([\Lambda]) & = A_p([\Lambda]) B_p([\Lambda]) C_p([\Lambda]) \cr}
}
where $A_p$ is the ``diagonal factor,'' $B_p$ is the
``cross-product'' and $C_p$ is the ``type factor,''
in the terminology of \csiv.

We are interested in the growth of $\CW([\Lambda])$ as
a function of discriminant $D$ for large $D$. To be
precise, we suppose $D$ has a prime factorization
$D = 2^{v_2} p_1^{v_{p_1} } \cdots p_k^{v_{p_k}} $
where $p_i$ are odd primes. We keep the $p_i$   fixed
and take $v_{p_i} \rightarrow \infty, p_i>2$.
Under these circumstances,
the ``type factor,'' while subtle, remains bounded.
Likewise, the diagonal factor,
which  is related to an $L$-function will
neither approach zero nor  infinity in the limit.
Indeed, the only $D$-dependence in the
diagonal factor comes from
$\zeta_D(s) = L(s, \chi)$ where $\chi$ is
a Dirichlet character $\chi(m) = (D/m)$ for
$m$ odd, and $\chi(m)=0$ for $m$ even, and
$s=[(\dim \Lambda + 1)/2]$. Now use
  $\zeta_{p^3 D}(s) = \zeta_{pD}(s) $ for $p$ an
odd prime and any $D$ (\csiv, eq. 14).

The remaining factor
is the ``cross-product.''  Suppose the quadratic form
associated to $\Lambda$ has $p$-adic decomposition:
\eqn\padic{
\oplus_{j=0}^k p^{j} {\bf 1_{r_j \times r_j } }
 }
The ``cross-product'' term is then $p^{\ell/2}$ with \csiv:
\eqn\crsstrm{
\eqalign{
\ell & = \sum_{0\leq i < j \leq k} (j-i) r_i r_j\cr
& = n v_p - \biggl(\sum j r_j^2 + 2 \sum_{1 \leq i<j\leq k} i r_i r_j\biggr) \cr
& \leq (n-1) v_p \cr}
}
clearly, $\ell\geq 1$ so we can certainly get a
lower bound of $D^{1/2}$ in \grwth. But we
can do better. The largest values of $\ell$
are obtained by taking $r_{v_p}=1$ and all
others $=0$. This genus {\it is} realized for
some embedding.  Combining the $p$-factors
we get an estimated growth of  $\CW([\Lambda])$
of $D^{(n-1)/2} $ and  we arrive at \grwth. $\spadesuit$

Since we expect the {\it typical} embedding to realize
the case $r_{v_p}=1$ mentioned in the above
proof we actually expect the growth \grwthest.

\appendix{C}{Grassmannians and the geometrical
interpretation of the attractor equations }

The result \interpi\ is proved by
using the geometrical interpretation of
Narain moduli space in terms of type IIB
compactification  explained by
Aspinwall and Morrison in  \aspmorr\aspinwall.
We begin with a review of their construction,
then we apply it.

\subsec{Geometrical interpretation
of the Narain moduli spaces
for $IIB$ compactification.}

Let $Gr_+(k,V)$ denote
spacelike $k$-planes in a real vector space $V$
equipped with a quadratic form of signature
$( (-1)^{k+16},(+1)^k)$. If we identify $V= II^{k+16,k} \otimes \IR$
then $Gr_+(k,V)$ is the space of projection operators
$P: II^{k+16,k} \otimes \IR \rightarrow \IR^{k+16,0} \perp \IR^{0,k}$.
The moduli spaces $\CN^{k+16,k} $ of  low-energy
supergravity Lagrangians with $\CN>2$ supersymmetry
are {\it locally} of the form $\Gr_+(k,V)$. Taking into
account string $U$-duality the moduli spaces are of the
form $\CN^{k+16,k} = \Gamma\backslash \Gr_+(k,V)$.

\subsubsec{The reduction map}

The key construction is the following.

\bigskip
{\bf Theorem \aspmorr\aspinwall } Suppose
$V$ is a real vector space equipped with form of
signature $((-1)^{q+1},(+1)^{p+1} )$.
Suppose  $\langle e,e^*\rangle_{\IR}$ span a hyperbolic
plane in $V$, with $e^2=0$. We then have an
isomorphism of manifolds:
\eqn\grssmi{
\Phi_{e,e^*}: \Gr_+(p+1, V)\ \rightarrow
\Gr_+(p, e^\perp/e) \times \langle e, e^* \rangle_{\IR}^\perp\times \IR_+
}

{\it Proof}.
We first define $\Phi_{e,e^*} $. Let
$W \in  \Gr_+(p+1, V)$.
We claim that the vector space
\eqn\whyp{
Y' = W \cap e^\perp
}
is a $p$-dimensional spacelike  plane in $V$.
To prove this note that
$Y'$ is spacelike since it is a
subspace of $W$, which is spacelike. Since
$W$ is spacelike, it contains no null
vectors, in particular does not contain the
line $e \IR$. Therefore $(v,e) =0$ defines a nontrivial linear
equation on vectors in $W$ and hence reduces the
dimension by 1.

Furthermore $Y'$ does not
contain any line of the form
\eqn\extrlne{
L = (\alpha e + \beta e^* + \ell)\IR
}
where $\ell \in
\langle e, e^* \rangle_{\IR}^\perp$ and
$\beta \not=0$.
However, since $W$ is $(p+1)$-dimensional
and $Y'$ is $p$-dimensional, $W$
 must contain some spacelike line of the
form $(\alpha e + \beta e^* + \ell)\IR$
where $\ell \in
\langle e, e^* \rangle_{\IR}^\perp$ and
$\beta \not=0$.
We may take this line to be orthogonal to $Y'$,
in which case the line is unique.
We have thus found a decomposition:
\eqn\decomp{
W = Y' \perp L
}
Now, to define the map $\Phi_{e,e^*}$ we
  project
\eqn\prjct{
\eqalign{
\pi: \Gr_+(p, e^\perp)&  \rightarrow \Gr_+(p, e^\perp/e) \cr
Y' & \rightarrow \pi(Y') = Y' + e\IR \cr}
}
We may define:
\eqn\defphi{
\Phi_{e,e^*}(W) = \bigl( \pi(W\cap e^\perp), \ell /\beta, \alpha/\beta\bigr)
}

Let us now describe the inverse.
If we are given a triple $(Y, \ell, \xi)$ we proceed as
follows. We would like to lift $Y$ to $Y'$ in
\prjct.
Note that the fiber of $\pi$ is an $\IR^{p}$-torsor:
Suppose $Y_1'$ projects to $Y$.
Choose a basis $\{ x^\alpha \}$ for $Y_1'$.
If $\pi(Y_2') = Y$ then $Y_2'$ has a basis of the
form: $ \{ x^\alpha  + \eta^\alpha e \} $, $\eta^\alpha\in\IR$. We can
determine
this lift uniquely by first choosing the line
\eqn\reconst{
L = ( \xi e + e^* + \ell) \IR
}
There is then a unique lift $Y'$ of $Y$ which is
orthogonal to $L$. We then define the spacelike
$(p+1)$-plane by \decomp.  $\spadesuit$

\subsubsec{Hyperkahler and complex structures}

We choose a hyperbolic plane
\eqn\chci{
\langle e_1, e_1^*\rangle_{\IR} \subset H^2(K3;\IR)
}
giving a decomposition
\eqn\chcii{
H^2(K3;\IR)\cong \IR^{18,2} \perp
\langle e_1, e_1^*\rangle_{\IR}
}
We may choose $e_1$ to be the class representing a
torus in $K3$.
The reduction map gives
\eqn\reduxi{
\Gr_+(3, H^2(K3;\IR)\bigr) \rightarrow
\Gr_+(2, e_1^\perp/e_1) \times \langle e_1, e_1^* \rangle_{\IR}^\perp\times
\IR_+
}
The LHS may be identified with
the moduli space of volume one
hyperkahler structures \hkmi\hkmii\hkmiii. These are
determined by   a
spacelike 3-plane $\Sigma= \langle \omega^I, \omega^J, \omega^K \rangle_\IR$.
The splitting
\decomp\  becomes
\eqn\cmpkhl{
\Sigma = \Xi' \perp \CK
}
where $\Xi'$ is a spacelike two plane identified
with a complex structure:
$\Xi' = \langle Re \Omega, Im \Omega \rangle_\IR$.
  The spacelike 2-plane
$\Xi' = \Sigma \cap e_1^\perp$ is a complex structure
such that $e_1$ is type $(1,1)$.
With respect to
$\Xi'$ the K3 surface is an elliptic fibration. The
Kahler class lies on the line $\CK$.

\bigskip
{\bf Remark/Warning:} The moduli space
$$
\Gr_+(2, \langle e_1, e_1^* \rangle_{\IR}^\perp)
\cong O(2,18)/O(2) \times O(18)
$$
is the space of spacelike
2-planes in $H^2(K3;\IR)$ which are orthogonal
to $e_1, e_1^*$. If we choose $e_1^*$ so that
$b=e_1^* - e_1$ represents a sphere in $K3$
then we can identify $\Gr_+(2, \langle e_1, e_1^* \rangle_{\IR}^\perp)$ with
the moduli space of
complex structures on $K3$ such that there is an
elliptic fibration with general fiber in the class
$e_1$ and holomorphic section represented by $b$. These
complex structures should be distinguished from
$\Xi'$ in \cmpkhl. There is a unique complex
structure $\Xi'_{e,e^*}$ which is a lift of
$[\Xi']$ and is orthogonal to $\langle e, e^* \rangle$.
This   complex structure   is,  in
general,  {\it not} compatible with the hyperkahler
structure $\Sigma$!

\subsubsec{B-field and hyperkahler structures}

We next choose another hyperbolic plane so that
\eqn\chciii{
H^*(K3;\IR)\cong H^2(K3;\IR) \perp
\langle e_2, e_2^*\rangle
}
Given a spacelike 4-plane $\Pi\in \Gr_+(4, H^*(K3;\IR))$
we define:
\eqn\pihk{
\Pi = \Sigma' \perp (V e_2 + e_2^* + B)
}
with $B\in \langle e_2, e_2^*\rangle^\perp = H^2(K3;\IR)$.
Once again, $\Sigma'$ is {\it not} a subspace in
$H^2(K3;\IR)$, but rather defines an element
$\pi(\Sigma) \in \Gr_+(3, e_2^\perp/e_2)$.
The real number $V$ is interpreted in
\aspinwall\ as the volume of K3.
\foot{As in the case of complex
structures, the lift $\Sigma'_{e_2,e_2^*}$ which is
orthogonal to $\langle e_2, e_2^*\rangle$
defines a hyperkahler
structure, but it is in general distinct from the
spacelike 3-plane   compatible with the ``$\Pi$-structure''
(a generalization of hyperkahler structure).}

\subsubsec{Ramond-Ramond moduli}

We now choose a hyperbolic plane so that:
\eqn\fivedee{
\IR^{5,21} \cong H^*(K3;\IR) \perp \langle e_3, e_3^* \rangle_\IR
}
If we discuss IIB compactification on $K3$ then the
geometrical interpretation of $\Theta\in \Gr_+(5, \IR^{5,21})$
is \aspinwall:
\eqn\fvdeei{
\Theta = \Pi' \perp (\alpha e_3 + e_3^* + C) \IR
}
where $C\in H^*(K3;\IR)$, are the values of the IIB
Ramond-Ramond potentials and $\alpha=e^{-\phi}$
is the dilaton.
As before $\Pi'$ is not a spacelike 4-plane in
$H^*(K3;\IR)$, but does uniquely determine
$\Pi'_{e_3, e_3^*}\in \Gr_+\bigl(4, H^*(K3;\IR)\bigr)$.

Finally, considering $IIB$ on $K3 \times T^2$.
We decompose:
\eqn\fivedee{
\IR^{6,22} \cong \IR^{5,21} \perp \langle e_4, e_4^* \rangle_\IR
}
and the decomposition of $\Upsilon\in \Gr_+(6,\IR^{6,22} ) $
is
\eqn\sixdee{
 \Upsilon = \Theta' \perp (a e_4 + e_4^* + A)\IR
}
Denoting the coordinates on $T^2$ as $x^4,x^5$,
we identify $A\in \IR^{5,21}$   as
Wilson lines around $x^4$ or, alternatively,
as the moduli of the
tensormultiplets reduced  on $T^2_{4,5}$.
The factor $a$ must be interpreted as the
area of $T^2$ \aspinwall.

\subsec{ Proof of $(4.4a,b)$ }

We now collect the above decompositions.
We   choose {\it four} mutually orthogonal hyperbolic planes
so that:
\eqn\orthog{
\IR^{6,22} \cong \IR^{2,18} \perp \langle e_1, e_1^*\rangle
\perp \langle e_2, e_2^*\rangle
\perp \langle e_3, e_3^*\rangle
\perp \langle e_4, e_4^*\rangle
}
and so that $\langle p, q \rangle$ lies in $\IR^{2,18}$.
Using these planes we  define a sequence of spacelike planes
(representing decompactification to succesively
higher dimensions) by:
\eqn\decop{
\eqalign{
\Upsilon & = \Theta' \perp (a e_4 + e_4^* + A) \IR \cr
\Theta & = \Pi' \perp (e^{-\phi}  e_3 + e_3^* + C) \IR
\subset e_4^\perp/e_4 \cr
\Pi & = \Sigma' \perp (V e_2 + e_2^* + B)\IR \subset
\langle e_3,e_4\rangle^\perp/\langle e_3,e_4\rangle \cr
\Sigma & = \Xi' \perp (v e_1 + e_1^* + k ) \IR
\subset
\langle e_2, e_3,e_4\rangle^\perp/\langle
e_2, e_3,e_4\rangle \cr  }
}
The geometrical interpretation
of the data was explained above.
$v,k$ represent  the Kahler class
of the $K3$,

The equations \interpi{b}\ simply follow from
the orthogonality to $p,q$. The equation
\interpi{a}\ follows because $\Xi$ is the
spacelike two-plane given by the span
of the real and imaginary parts of $\Omega$.
This 2-plane must contain the  rank two lattice
spanned by $p,q$.

{}From this decomposition we
see quite explicitly that the
``hypermultiplet moduli'' are not fixed
by the attractor, in agreement with
previous discussions \adfii.

\appendix{D}{A list of some notation}

\medskip
\item{ $\hat \alpha_I, \hat \beta^I$}  A symplectic
basis for $H^3(X;\IZ)$. The dual basis is given
by $ \alpha^I,  \beta_I$.

\medskip
\item{ $\CA(\Gamma)$}  The chiral vertex operator algebra
generated by an even integral lattice $\Gamma$.

\medskip
\item{$\widetilde{\CA}$ }
The rightmoving, or anti-holomorphic chiral algebra. In general
$\tilde {} $ denotes a right-moving quantity.

\medskip
\item{ $\CB^{r,s}$} The Grassmannian of spacelike $s$
planes in $II^{r,s}\otimes \IR$.

\medskip
\item{CM} Complex multiplication

\medskip
\item{ CY} Calabi-Yau.

\medskip
\item{ $D_{abc}$} Intersection numbers for a CY 3-fold.

\medskip\item{
$D(\gamma)$ } The discriminant of a BPS state with
charge $\gamma$. (Sec. 3).

\medskip\item{
$D_{p,q}$ } $(p\cdot q)^2 - p^2 q^2$, for $p,q$ vectors in a lattice.

\medskip\item{
$E_\tau$ } The elliptic curve $\IC/(\IZ+ \tau \IZ)$.

\medskip\item{
$\CF$ } The total electric/magnetic fieldstrength (Sec. 2).

\medskip\item{
$\CF$ } Also, the inhomogeneous prepotential (Sec. 2, Sec. 9).

\medskip\item{
$\gamma$ } The charge of a BPS state. Usually, a vector
in $H_3(X;\IZ)$. The Poincar\'e dual is $-\hat \gamma$.

\medskip\item{
$h(D)$ } The class number of $D$

\medskip\item{
$\CH_r$ } The Siegel upper half plane of $r \times r $ matrices.

\medskip\item{
$K$} A generic number field. Also, a lattice (appendix B).

\medskip\item{
$K_D$ } The quadratic field $K_D  = \IQ[\sqrt{D}]$. In this
paper $D$ is usually $<0$, $=0,1\mod 4$.

\medskip\item{
$\widehat{K}_D$ }  The extension $K_D(j(\tau))$, the
ring class field of $K_D$ for the order $\CO(D)$.

\medskip\item{
$\Lambda$ } The electric/magnetic charge lattice.

\medskip\item{
$L_{p,q}$ } The rank 2 lattice spanned
by $p,q$.

\medskip\item{
LCSL } ``Large complex structure limit,'' or, point of
maximal unipotent monodromy

\medskip\item{
$\langle \cdot, \cdot \rangle $ }  A symplectic bilinear form.

\medskip\item{
$\langle v_1, \dots, v_n \rangle_R $
}
A module over a ring $R$ generated by elements $v_1, \dots, v_n$.

\medskip\item{
$\CM$, $\CM(X)$ } The moduli space of complex
structures on a polarized variety $X$.

\medskip\item{
$\widetilde{\CM}$ } The universal cover of $\CM$.

\medskip\item{MW} Mordell-Weil (sec. 10).

\medskip
\item{$\CN$}  The number of independent 4-dimensional supersymmetries.

\medskip\item{
$\CN^{r,s}$} Narain moduli space

\medskip\item{
$\Omega$ }  A nowhere vanishing holomorphic
$(d,0)$ form on a Calabi-Yau $d$-fold.

\medskip\item{
$\Pi^{p,q}$}  The projection onto type $(p,q)$ in
a Hodge decomposition.

\medskip\item{
$Q$ } A positive integral binary quadratic form,
$a x^2 + b xy + c y^2$. Also denoted as $(a,b,c)$
or $Q_{(a,b,c)}$.

\medskip\item{
$Q_{p,q}$ } A quadratic form defined in \bqform.

\medskip\item{RCFT} Rational conformal field theory.

\medskip\item{$r$} The rank of a gauge group. Also, a
radial coordinate in the black hole ansatz \metanstz.
$r=\infty$ is the flat Minkowskian region.

\medskip\item{$\rho$} $\rho=1/r$ in the black hole ansatz
\metanstz.

\medskip\item{$II^{rs}$} The even unimodular lattice
of signature $(-1)^r, (+1)^s$.

\medskip
\item{$S$} A K3 surface (sec. 4) . Also, a lattice (appendix B).
Also, a
del Pezzo surface (Sec. 11.3.4).

\medskip\item{$T^d$} A torus of $d$ real dimensions.

\medskip\item{$T$} The K\"ahler modulus of a sigma model
with target space a torus $T^2$.

\medskip\item{$t^a$} Flat coordinates near a large radius limit.

\medskip\item{
WPZ } Weil-Peterson-Zamolodchikov

\medskip\item{
$X$ } A Calabi-Yau 3-fold.

\medskip\item{
$\widetilde{X}$ } A mirror to a CY 3-fold $X$.

\medskip\item{
$X^I,F_I$ } Periods of a CY 3-fold. Special coordinates.

\medskip\item{
$Z(\Omega;\gamma)$} The central charge in superselection sector
$\gamma$, \defzee.

\medskip\item{
$Z_*$ } A fixed point value of the central charge determined by the
attractor mechanism.

\listrefs
\bye